\def\F{\Phi}

\def\Fb{{\Phi_b}}
\def\Fobs{{\Phi_{\rm obs}}}

\def\pars{{\cal P}}
\def\shpars{{\cal P'}}
\def\spars{{\cal S}}

\def\eps{\epsilon}

\def\rvec{{\bf r}}
\def\vvec{{\bf v}}
\def\lum{\Lambda}
\def\luma{\Lambda_{\rm app}}
\def\Dop{{\cal D}}

\def\drxn{{\bf n}}
\def\dRdF{{dR\over d\Phi}}
\def\dR{{dR\over d\Phi d\drxn}}

\def\like{{\cal L}}

\def\beq{\begin{equation}}
\def\eeq{\end{equation}}
\def\beqa{\begin{eqnarray}}
\def\eeqa{\end{eqnarray}}
\def\eqn(#1){\label{#1}}
\def\eq(#1){\label{#1}}
\def\ceq(#1){(\ref{#1})}
\def\ceqn(#1){equation~(\ref{#1})}

\documentstyle[11pt,aaspp4]{article}
\slugcomment{Submitted to {\it The Astrophysical Journal}, December 1996}

\lefthead{Loredo and Wasserman}
\righthead{Gamma Ray Bursts: Isotropic Models}

\begin{document}

\title{Inferring the Spatial and Energy Distribution\\
of Gamma Ray Burst Sources. II.  Isotropic Models}

\author{Thomas J. Loredo and Ira M. Wasserman}
\affil{Center for Radiophysics and Space Research,
Cornell University, Ithaca, NY 14853-6801}

\begin{abstract}
We use Bayesian methods to analyze the distribution
of gamma ray burst intensities reported in the {\it
Third BATSE Catalog} (3B catalog) of gamma ray bursts, presuming the
distribution of burst sources (``bursters'')
is isotropic.  We study both phenomenological and cosmological
source distribution models, using Bayes's theorem both to infer unknown
parameters in the models, and to compare rival models.  
We analyze the distribution of the time-averaged peak photon number
flux, $\F$, measured on both 64~ms and 1024~ms time scales, performing
the analysis of data based on each time scale independently.  Several of
our findings differ from those of previous analyses that
modeled burst detection less completely.  In particular, we find
that the width of the intrinsic luminosity function for bursters is unconstrained,
and the luminosity function of the actually observed bursts can
be extremely broad, in contrast to the findings of all previous studies.
Useful constraints probably require observation of bursts significantly
fainter than those visible to BATSE.  We also find that the 3B peak flux data do
not usefully constrain the redshifts of burst sources; useful constraints
require the analysis of data beyond that in the 3B catalog (such as burst
time histories), or data from brighter bursts than have been seen by
BATSE (such as those observed by the {\it Pioneer Venus Orbiter}).  
In addition, we find that
an accurate understanding of the peak flux distributions reported
in the 3B almost certainly requires consideration of
data on the temporal and spectral properties of bursts beyond
that reported in the 3B catalog, and
more sophisticated modeling than has so far been attempted.

We first analyze purely phenomenological power law and broken power law
models for the distribution of observed peak fluxes.  We find that
the 64~ms data is adequately fit by a single power law, but that
the 1024~ms data significantly favor models with a sharp, steep
break near the highest observed fluxes.
At fluxes below the break, the distribution of 1024~ms
fluxes is flatter than that of 64~ms fluxes.  Neither data set is
consistent with the power law distribution expected from a
homogeneous, Euclidean distribution of sources.
Next we analyze three simple cosmological models for burst
sources: standard candles with constant burst rate per comoving
volume; a distribution
of standard candle sources with comoving burst rate proportional to a
power law in $(1+z)$, and 
a bounded power-law burster luminosity function with constant comoving burst
rate but variable power-law index and luminosity bounds.
We find that the 3B data can usefully constrain the
luminosity of a standard candle cosmological population of bursts if
there is no density evolution.
But the 3B data allow strong density evolution and arbitrarily
broad luminosity functions; consequently, they do not usefully constrain
the redshifts or luminosities of cosmological burst sources.  We
elucidate the properties of the models responsible for these results.

For sufficiently flexible models, 
the inferred values for parameters describing the shapes of the distributions
of 64~ms and 1024~ms peak fluxes formally
differ at the 68\%--95\% level.  Because the measurements on these
two timescales are not independent,
it is difficult to ascertain the true significance of this discrepancy;
since many bursts are common to both data sets, it is likely its
significance is larger than these formal values indicate.  In addition,
the inferred amplitude (in bursts per year) of the distribution of 
64~ms peak fluxes is about twice that of 1024~ms peak fluxes.  These results
strongly suggest that a complete understanding of the measured peak
flux distributions requires simultaneous modeling and analysis of
temporal properties of bursts.  We study models that attempt to
reconcile the two data sets by accounting for ``peak dilution,'' the
underestimation of the peak intensity that results from using
data accumulated over a
timescale exceeding the peak duration.  A phenomenological model
strongly correlating peak duration with peak flux is moderately
successful at reconciling the data.  A model that correlates
peak duration with peak flux due to cosmological time dilation and
relativistic beaming is less successful, but remains of interest
in that it is a simple physical model illustrating how one can jointly
model and analyze temporal and spectral properties of bursts with
peak flux data.  A more rigorous accounting for the differences between
the 64~ms and 1024~ms data requires analysis of temporal and spectral 
information about bursts beyond that available in the 3B catalog.
\end{abstract}


%==============================================================================
\section{Introduction}

In the absence of direct measurement of the distances to burst sources
(``bursters''),
or association of bursts with well-localized counterparts, we
must infer the spatial and energy distribution of bursters from the observed
distribution of burst strengths and directions.  The complexity of
the burst data makes this task considerably more difficult than
it might at first appear because we must account for several subtle
biases and selection effects.  So far, analyses of burst data
have relied on largely ad hoc choices of statistics designed
to circumvent some of these effects.  But differing choices of
statistic or analysis method by different investigators have led
to some controversy over the implications of the burst data.

Of particular relevance to this work are the varying conclusions of numerous
previous studies of isotropic models for the burst
source distribution.  On a purely phenomenological level, investigators
differ over whether the logarithmic slope of the distribution of burst
intensities exhibits a significant change in slope (cf.\ 
Loredo and Wasserman 1993; Wijers and Lubin 1993;
Petrosian, Azzam, and Efron 1994).  In the context of cosmological
models, investigators have reached a variety of inconsistent
conclusions about possible characteristics of the burst source
distribution, including the value and uncertainty for the luminosity
of standard candle sources (cf.\ Dermer 1992; Loredo and Wasserman 1993), and the ability of the data to constrain the width of the
luminosity function of bursters (cf.\ Loredo and Wasserman 1993;
Horack, Emslie, and Meegan 1994; and Cohen and Piran 1995) or the
redshift of the faintest bursters (cf.\ Loredo and Wasserman 1993;
Emslie and Horack 1994; Cohen and Piran 1995).  These differences result 
largely from
methodological differences among the published studies.  Without a
vastly larger dataset,
only careful attention to methodological issues can identify the
correct inferences.

In the first paper of this series (Loredo and Wasserman 1995; hereafter LW95), 
we described the Bayesian methodology for
inferring the spatial and energy distribution of burst sources.  
Instead of constructing a customized statistic in an
attempt to circumvent the various biases and selection effects that might
enter inferences, we start from simple models (based on the Poisson
distribution) for burst occurence and detection, and directly calculate
the probability for the observed data:  the likelihood function.
The likelihood function describes how well a model can account for
the {\it joint differential distribution} of observed burst 
strengths and directions, and accounts for biases and selection effects
by construction, rather than trying to circumvent them by a clever
choice of statistic.  
Indeed, from the Bayesian point of view, there is no freedom of
choice regarding what 
statistic to use and how to use it; the data enter Bayes's theorem in the 
likelihood function, and
the rules of probability theory dictate both how to calculate the
likelihood function and how to manipulate it to make inferences.  
This methodology offers several advantages over rival methods:
(1) it does not destroy information
by binning or averaging the data (as do, say, $\chi^2$,
$\langle V/V_{\rm max}\rangle$, and analyses of flux or angular moments);
(2) it straightforwardly handles uncertainties in the measured
quantities; (3) it analyzes the strength and
direction information jointly; (4) it uses information available
about nondetections; and (5) it automatically
identifies and accounts for biases and selection effects, given
a precise description of the experiment. 

In this work we use the Bayesian methodology 
to make inferences about isotropic models for the distribution
of burst sites, using data from the {\it Third BATSE Catalog}
(Fishman et al.\ 1996, hereafter the 3B catalog; the 3B catalog 
inherits some properties of the First BATSE (1B) catalog
described in Fishman et al.\ 1994).  A
companion paper (Loredo and Wasserman 1996) 
uses the same methodology to make inferences about
anisotropic models, including comparisons of isotropic and anisotropic
models.

In the next section, we briefly review the Bayesian methodology and
the form of the likelihood function for burst data.  
The 3B catalog
does not provide all of the information required for a rigorous
analysis, so we are forced to approximate some of the quantities
required for calculating the likelihood.  
In \S~3 we describe the approximations necessary to calculate the
likelihood function based on the data available in the 3B catalog.
Consistency requires that data for some bursts 
be omitted from the analysis when the approximations
fail for those bursts, so we carefully discuss selection of the
analyzable data.  The most serious data cut arises from inaccuracy
of the approximation used to calculate the
detection efficiency reported in the 1B catalog for bursts with
low fluxes.  We describe extensive simulations we performed to
quantify the inaccuracy of the approximation, and find that
a significant fraction of bursts must be omitted from the analysis
to avoid seriously corrupting the inferences drawn. 
Previous analyses of the BATSE
data have not worried about inaccuracies introduced by improperly
including dim bursts; this may account for some of the differences between our
conclusions and those reached in other analyses.

We carry out our analysis in \S~3 and throughout the remainder of
the paper using data based on peak photon number flux measurements taken
on both 64~ms and 1024~ms time scales.  We analyze these data separately; 
they are not
independent, but their dependence is too complicated to quantify with
the data tabulated in the 3B catalog alone.  An important conclusion
of our analyses is that inferences drawn from these two data sets
differ substantially, although the formal significance of the discrepancy
cannot be determined without more information about the bursts
comprising the catalog.  We make
an effort to understand and quantify this discrepancy in the final 
sections of this paper.  Although we cannot conclusively identify
the reason for the discrepancy, we suggest that it is
a result of burst light curves having peak durations that
are often significantly shorter than 1024~ms (and possibly shorter than 64~ms)
and that may be
correlated with burst intensity.  The resulting
errors in peak flux estimates distort the peak flux distribution;
the distortion differs for the two data sets, and can potentially
account for their significantly different shapes.

We begin, however, with models that do not attempt to account for
any effects the different time scales of the data sets may have on
the shape of the observed distribution of burst intensities.
For brevity, we deem such models ``simple'' models.
In \S~4 we present results of analyses of simple
phenomenological models for the differential burst rate (the burst rate
per unit peak flux, $\F$). 
These models help us ascertain what features must be present in the burst
rate without committing us to a particular physical explanation for
these features.  They indicate that the distribution of 64~ms peak fluxes
is adequately fit by a single power law whose logarithmic slope is
very significantly different from the $-2.5$ value associated with the
differential rate for an unbounded homogeneous population of sources.
There is
no significant evidence for steepening of the distribution with
intensity, contrary to thesuggestions of such steepening we found
in the 1B catalog (Loredo and Wasserman 1993).  
In contrast, the 1024~ms peak flux data prefer
models with a broken power law distribution, and the low flux part
of the distribution is significantly shallower than the distribution
of 64~ms peak fluxes.  In \S\ 5
we present results of analyses of three simple cosmological models for
the burst source distribution.  The simplest model presumes that
all burst sources have the same intrinsic luminosity (they are ``standard
candles''), and that the burst rate per unit comoving volume is constant
with redshift, $z$.  Next, we consider standard candle sources with
a burst rate density that varies as a power of $(1+z)$.  Finally, we consider
models with power-law luminosity functions and constant comoving
burst rate density.  We find that our ignorance of the additional parameters
in models with density evolution or a luminosity function greatly
weakens our ability to learn about the spatial distribution and
luminosity of the sources of bursts.  As with the simple phenomenological
models, these models reveal systematic discrepancies between the 64~ms and
1024~ms data.

In \S~6 we discuss how properties of burst light curves might lead to
time scale-dependent distortions of the observed flux distribution qualitatively
capable of reconciling the two data sets.  That such effects might
prove crucial for making inferences from burst peak flux data was
already anticipated in LW95; other authors have also previously
remarked on the importance of these effects for understanding
the flux distribution of bursts (Lamb, Graziani, and Smith 1993; 
Petrosian, Lee, and Azzam 1994).
Following a general discussion, we analyze a more complicated phenomenological
model than those of earlier sections that takes into account the different
measuring time scales of the data sets.  It is only moderately successful
at reconciling the data sets, but remains of interest as an example of
how one can explicitly account for time scale dependent effects in models.

In \S~7 we analyze a final physical model that draws together several of
the lines of thought developed in the preceding sections.  In this model,
burst sources are standard candles and standard clocks in their rest frames,
but undergo relativistic motion with respect to locally comoving
observers.  An isotropic distribution of beaming
angles with respect to the line of sight results in an effective
luminosity function that is a power law if the rest frame emission
spectrum is a power law.  In the absence of time scale effects,
this model is thus identical to the cosmological model with a power law 
luminosity function considered in \S~5, except that the power law index
is a function of the burst spectral index, rather than a free parameter.
Beaming also results in bursts having
observed peak durations that are a function of luminosity, allowing us
to model time scale effects; cosmological redshift additionally correlates
duration and peak flux.  Thus besides being a model of intrinsic
physical interest, this model indicates how both the temporal and spectral
properties of bursts can influence the flux distribution in a manner that
can be straightforwardly modeled.

The final section summarizes our findings 
and their implications.  Two technical appendices describe 
the details of our cosmological models.

%===============================================================================
\section{Review of Methodology}

\subsection{The Differential Burst Rate}

In order to assess candidate hypotheses
about burst sources with Bayesian methods using observed burst directions and strengths, the hypotheses
must specify the {\it differential burst rate}:  the
rate of bursts per unit time per unit peak flux per unit steradian.
We denote it by $dR/d\F d\drxn$, where
$\drxn$ denotes a direction on the sky, and $\F$ denotes the time-averaged
peak photon number flux between 60 and 300 keV (the nominal trigger
range for the 3B catalog), averaged over the trigger time scale $\delta t$
(64, 256, or 1024~ms for BATSE).  The differential burst rate could vary
in time, but in this work we consider only rates that are time-independent.

Phenomenological
models specify $dR/d\F d\drxn$ directly as an ad hoc parameterized function of
flux and direction.
For physical hypotheses about the sites of bursts,
the differential rate is derived
from more fundamental rates such as the burst rate per unit volume in
space, and the burst luminosity function.  For example, let $\lum$ denote
the peak photon number luminosity of a burst source in the spectral
range used for flux measurements, and let
$\dot n(\rvec,\lum)$ denote the {\it burst rate density},
the number of bursts occurring per unit
time, volume, and peak luminosity at position $\rvec$ with peak
luminosity $\lum$.  We presume here that the emission is isotropic.
For models with a small enough length scale that spacetime
curvature can be ignored, we can calculate
the differential burst rate according to
\beq
{dR \over d\F d\drxn} = \int dr r^2 \int d\lum\; \dot n(\rvec,\lum)\,
  \delta\left[\F - \F_{\rm obs}(\rvec,\lum)\right],\eqn(dR-ndot)
\eeq
where $r$ is a radial coordinate and
$\F_{\rm obs}(\rvec,\lum)$ specifies the peak flux observed at Earth due
to a source at position $\rvec$ and luminosity $\lum$.  The observed flux 
follows from the inverse-square law:
\beq
\F_{\rm obs}(\rvec,\lum) = {\lum \over 4\pi r^2}.\eqn(Fobs-Euc)
\eeq
Later in this work we employ a similar integral expression for
cosmological models, generalizing the volume element, burst rate density,
and observed flux function to account for the effects of spacetime
curvature.

%-------------------------------------------------------------------------------
\subsection{Bayes's Theorem}

We compare alternative hypotheses for the differential burst rate by 
calculating their probabilities with Bayes's theorem.  The nature of
the resulting calculations depends on the type of hypothesis being considered.
In practice, we distinguish between two kinds of hypotheses.  
In {\it parameter estimation} we calculate probabilities for
hypotheses about the values of parameters in a particular
differential rate model.  In {\it model comparison} we calculate
probabilities for competing parameterized models, presuming that one
of the models being considered is true.  We now briefly describe these
two types of calculations in turn.
Gregory and Loredo (1992) proved a somewhat more extensive review and further 
references.

The goal of Bayesian parameter estimation is to calculate probabilities
for hypotheses about parameter values (e.g., for statements like
``the burst source luminosity is between $a$ and $b$'').
The probability for any hypothesis about parameter values can be calculated
from the posterior probability {\it density} for the parameters.  Denoting
the information specifying the model by $M$, 
the parameters by $\pars$, and the data by $D$,
Bayes's theorem for the posterior density is,
\beq
p(\pars\mid D,M) = p(\pars\mid M) {p(D\mid \pars,M) \over p(D\mid M)}.
  \eqn(BT-par)
\eeq
The first factor is the prior density for the parameters.  The
numerator in the ratio is called the sampling probability for the data
when its functional dependence on $D$ is of primary interest, or
the likelihood for the parameters when its functional dependence on
the parameters is of primary interest, as it is here.  The term
in the denominator does not depend on $\pars$, and plays the
role of a normalization constant.  It can be calculated simply by
integrating the product of the prior and the likelihood over $\pars$.

Since the dependence of the likelihood on $\pars$ is of central
concern in parameter estimation,
we suppress its dependence on $D$ and $M$, and denote the likelihood
function by $\like(\pars)$.  We discuss this important function
further below.  For the prior density we simply use a constant
function with respect to either the parameter, or its logarithm
for the case of scale parameters.
To indicate which of these two priors we use, we adopt
the convention of plotting the posterior with logarithmic parameter
axes for those parameters with a log-constant prior, and with
linear axes for those parameters with a constant prior.  All of
our priors and posteriors are normalized over the ranges displayed
in the figures.

As long as the prior density does not vary strongly over the width 
of the likelihood function, the details of the prior do not significantly
affect parameter estimates.  We note below those cases where our results
are sensitive to the form of the prior; such behavior simply indicates
that the data are uninformative with respect to the hypotheses under
consideration.

In Bayesian model comparison we presume that one of a specified class
of models is true, and calculate the probabilities
for the various models in order to determine which model is best
supported by the data.  A model can have undetermined
parameters, in which case we seek the probability for the model as
a whole, taking into account parameter uncertainty.  If we denote
the models by the symbols $M_i$, and let the proposition
$I$ specify the set of models being considered, then Bayes's theorem
for the probability for model $M_i$ takes a form very similar to
\ceqn(BT-par):
\beq
p(M_i\mid D,I) = p(M_i\mid I) {p(D\mid M_i) \over p(D\mid I)}.\eqn(BT-mod)
\eeq
Here $p(M_i\mid I)$ is the prior probability for model $M_i$.  The
term in the denominator, which is independent of $M_i$, is simply a
normalization constant.  The
term in the numerator is the global likelihood for $M_i$, or the
prior predictive probability for the data presuming $M_i$ is the
correct model.  Formally, we should write it as $p(D\mid M_i,I)$; but
the $I$ proposition is redundant here (it asserts that one of the
models being considered must be true; but $M_i$ asserts that a particular
model is the true one).  As written, it is clear that the global likelihood
is simply the normalization constant one would calculate for parameter
estimation with model $M_i$ (i.e., the denominator in \ceqn(BT-par)).
Although this quantity is of little intrinsic interest for parameter
estimation, it is the key quantity for model comparison, playing the
same role here as the likelihood function does for parameter estimation.
This is why we call it the global likelihood for the model.

In practice, it is easier to work with ratios of model probabilities than
with the probabilities themselves, since in the ratio the normalization
constant, $p(D\mid I)$, cancels and thus need not be calculated.  Also,
if additional models are later added to the calculation, the probabilities
for the originally considered models will change, but the ratios of these
probabilities to each other will not; this is a further convenience of
probability ratios, which are called {\it odds}.  The odds in favor
of $M_i$ over $M_j$ is
\beq
O_{ij} = {p(M_i\mid I) \over p(M_j\mid I)} \;
   {p(D\mid M_i) \over p(D\mid M_j)},\eqn(Odds-def)
\eeq
where the first factor is the prior odds, and the second factor
is the ratio of global likelihoods, also called the {\it Bayes factor}. 
The prior odds expresses how prior information (perhaps
subjective) distinguishes between the models; the Bayes factor
compares how the models predict the data.  If we assign equal prior
probabilities for the models (as we do throughout this work), 
the odds is given by the data-dependent
Bayes factor.  Interestingly, even when one uses equal prior probabilities,
the odds can significantly favor a simpler model over a more complicated
one (which may include the simpler one as a special case).  Bayesian
model comparison implements an automatic and objective ``Ockham's razor'' that
penalizes models for having excessively large parameter spaces, and in
this way guards against unjustified preference for complicated models.

As noted above, we can calculate the global likelihood for a model simply
by integrating the product of the prior and likelihood for the model's
parameters.  If we denote the parameters for model $i$ by $\pars_i$, then
\beq
p(D\mid M_i) = \int d\pars_i\; p(\pars_i\mid M_i) \like(\pars_i).\eqn(glike)
\eeq
This equation reveals that the global likelihood for a model is
the {\it average} likelihood for the model's parameters,
the averaging weight being simply the prior density for the parameters.
The Bayes factor is thus a ratio of average likelihoods.
In contrast, frequentist model comparison is based on ratios of
{\it maximum} likelihoods.  It is the averaging that takes place
in global likelihood calculations that is responsible for the
Ockham's razor effect in Bayesian model comparison.  A model with
a larger parameter space than a simpler alternative has a smaller
prior density in the vicinity of the likelihood peak, since its
prior density is spread over a larger parameter space.  Its global
likelihood will be larger than that of its alternative only if the
likelihood is large enough to make up for this smaller prior.

A consequence of this behavior is that
the value of the global likelihood for a model
depends much more sensitively on the prior for its parameters---and in
particular, on the width of the prior---than do parameter estimates.
Here we use constant priors; useful model comparison calculations require
that we specify finite prior ranges for any parameters not common to all
models being considered.  The global likelihood is then roughly inversely
proportional to the prior range that we have explored.  All of the
Bayes factors quoted here are ratios of global likelihoods calculated
with priors normalized over the ranges displayed in plots of 
posterior distributions.

%-------------------------------------------------------------------------------
\subsection{The Likelihood Function}

In traditional frequentist statistical methods, it is up to the user to
specify the function of data to be used to draw inferences.  Typical
choices include various moments of the data, counts of data sorted
into bins, some measure of misfit (such as $\chi^2$ or a
Kolmogorov-Smirnov statistic), or some statistic based on the likelihood
function.  In contrast, Bayesian methods offer the user no freedom
of choice:  the data enter Bayesian inferences through the entire likelihood
function.
In LW95 we gave a detailed derivation of the form of the likelihood function
for GRB data such as that provided by BATSE.  It can be written as
\beq
\like(\pars) = \exp\left[-T\int d\F\int d\drxn\, 
         \bar\eta(\F,\drxn)\dR \right]\;
    \prod_i \int d\F \int d\drxn \, \like_i(\F,\drxn)
      \dR.\eqn(full-L)
\eeq
Here $T$ is the duration of the observations;
$\bar\eta(\F,\drxn)$ is the time-averaged detection efficiency for bursts of
flux $\F$ from direction $\drxn$; and $\like_i(\F,\drxn)$ is the
probability for seeing the data for burst $i$, presuming it comes from
a burst with peak flux $\F$ and direction $\drxn$.  We call $\like_i(\F,\drxn)$
the individual burst likelihood function; it is the function one would use
to infer the properties of a particular burst.
LW95 derive expressions for $\bar\eta(\F,\drxn)$ and $\like_i(\F,\drxn)$ in 
terms of raw photon count data in the eight BATSE detectors.

%-------------------------------------------------------------------------------
\subsection{Marginal Distribution for Shape Parameters}

Most models for the differential burst rate have among their parameters
an overall {\it scale factor}, $A$, such that we can write
\beq
{dR \over d\F d\drxn} = A\, \rho(\F,\drxn;\shpars),\eqn(scale-def)
\eeq
where $\shpars$ denotes the remaining parameters, which we call 
{\it shape parameters}.  The scale factor is typically some measure
of the burst rate per unit volume, while the shape parameters define
the geometry of the burst distribution and the physical parameters of
individual burst sources (such as their characteristic luminosity).  Many 
questions about burst sources refer only to shape parameters.  Thus
it is useful to eliminate the scale factor from the analysis.  

An important advantage
of Bayesian methods over their frequentist counterparts is that they
allow straightforward inferences about an interesting subset of model
parameters in a manner that fully accounts for the uncertainty in
the neglected parameters.  Such inferences are obtained by
{\it marginalizing}:  integrating the full joint posterior distribution
with respect to the uninteresting parameters.  Marginalization of
the amplitude parameter (using a log-constant prior) can be performed
analytically.  LW95 show that the resulting marginal likelihood for the 
shape parameters is
\beq
\like(\shpars) = \prod_i\;
  {  \int d\F \int d\drxn \, \like_i(\F,\drxn)\,\rho(\F,\drxn) 
  \over  \int d\F \int d\drxn\, \bar\eta(\F,\drxn)\,\rho(\F,\drxn)}.
         \eqn(full-mL) 
\eeq
%Note that the detection efficiency does {\it not} appear in the numerator.
As mentioned above, we use flat prior densities for shape parameters
(or their logarithm) throughout this work;
thus the posterior distribution for the shape parameters
(or their logarithm) is simply the
marginal likelihood, normalized with respect to its arguments.

%-------------------------------------------------------------------------------
\subsection{Likelihood Functions for Isotropic Models}

In this work we consider only isotropic differential rates, for
which
\beq
\dR = {1 \over 4\pi} \dRdF,\eqn(dR-iso-def)
\eeq
where $dR/d\F$ denotes the burst rate per unit flux from {\it all}
directions.  If the differential rate is derived from an isotropic physical
model for the burst rate density, the counterpart to \ceqn(dR-ndot)\ is
\beq
{dR \over d\F} = 4 \pi \int dr r^2 \int d\lum\; \dot n(r,\lum)\,
  \delta\left[\F - \F_{\rm obs}(r,\lum)\right],\eqn(dR-iso)
\eeq
where $\dot n$ and $\Fobs$ are now functions only of $r$ and not
additionally of direction.

The likelihood function can be simplified for isotropic models by
performing the integrals over direction once for all.  The full likelihood
function becomes
\beq
\like(\Theta) = \exp\left[-T\int d\F\, 
         \bar\eta'(\F)\dRdF\right]\;
    \prod_i \int d\F \, \like_i(\F)
      \dRdF,\eqn(iso-L)
\eeq
where $\bar\eta'(\F)$ is the direction and time averaged detection
efficiency given by
\beq
\bar\eta'(\F) = {1 \over 4\pi} \int d\drxn \, \bar\eta(\F,\drxn),
     \eqn(etab-def)
\eeq
and $\like_i(\F) = \int d\drxn\, \like_i(\F,\drxn)$.  This is the
likelihood function used below when the amplitude parameter is of
interest.

Similarly,
writing $dR/d\F = A \rho(\F)$,
the marginal likelihood for the shape parameters becomes
\beq
\like(\pars) = \prod_i\;
  {  \int d\F \, \like_i(\F)\,\rho(\F) 
  \over  \int d\F\, \bar\eta'(\F)\,\rho(\F)}.\eqn(shape-L)
\eeq
This is the likelihood function used below when only the shape parameters
are of interest.

%===============================================================================
\section{Approximations}

As noted above, LW95 derive expressions for $\bar\eta(\F,\drxn)$ and 
$\like_i(\F,\drxn)$ in terms
of raw photon count data in the eight BATSE detectors.  However, these
functions are not directly reported in the 3B catalog, so we must approximate
them using the reported information.  The approximations in turn require
us to omit from our analyses data for which the approximation would be poor.

%-------------------------------------------------------------------------------
\subsection{Individual Burst Likelihood Functions}

The $\like_i(\F,\drxn)$ functions could be well 
approximated by functions proportional to $\exp[-\chi^2(\F,\drxn)/2]$
for each burst,
where $\chi^2(\F,\drxn)$ is the familiar goodness-of-fit measure, given
as a function of peak flux and direction.  The 3B catalog does not
provide $\chi^2$ functions for each burst, but instead
provides simple summaries of the behavior of $\chi^2$ near the 
best-fit $\F$ and $\drxn$ consisting of best-fit values and simple
measures of the widths of independent confidence regions for $\F$
and $\drxn$.  We thus approximate $\like_i(\F,\drxn)$ as a product of
a Gaussian about the best-fit $\F$ value (with width given by the reported
standard deviation), and a Fisher distribution (a spherical generalization
of the Gaussian) about the best-fit $\drxn$ with a width derived from
the reported direction errors.  Thus we write
\beq
\like_i(\F,\drxn) \propto  \exp\left[-
       {(\F - \F_i)^2 \over 2\sigma^2_i}\right]\, \Theta(\F) \,
        e^{\kappa_i\mu}.\eqn(Li-def)
% {1 \over s_i \sqrt{2\pi}}           {\kappa_i \over 4\pi \sinh \kappa_i}
\eeq
The first factor is the Gaussian distribution for the flux
uncertainty, with $\F_i$ the best-fit flux for burst $i$ and $\sigma_i$
the standard deviation for the peak flux; both quantities are reported in
the 3B catalog (one set for each of the three trigger time scales). 
The Heaviside function, $\Theta(\F)$, merely restricts
the Gaussian to positive flux values.  A more rigorous likelihood might
truncate more smoothly, but the
best-fit fluxes of the 3B bursts are all sufficiently positive that 
this truncation occurs at least a few
standard deviations away from $\F_i$ and thus has little effect on the
results.

The last exponential factor is the Fisher distribution describing the direction
uncertainties.  Although it is irrelevant for the analysis of isotropic
models, it is important for analyses of anisotropic models that
we will present in subsequent papers, so we discuss it here.
It is azimuthally symmetric about the burst direction;
we have written it in terms of spherical coordinates with the
polar axis aligned with the best-fit burst direction, so that
$\mu=\cos\theta$, with $\theta$ the polar angle.  If
we work in  Galactic coordinates, so that 
$\drxn = (l,b)$ with $l$ the Galactic longitude and $b$ the
Galactic latitude, then
\beq
\mu = \sin b \sin b_i + \cos b \cos b_i \sin(l-l_i),\eqn(mu-def)
\eeq
where the best-fit
Galactic longitude and latitude for the burst are $l_i$ and $b_i$.
The concentration parameter, $\kappa_i$, specifies the width of
the Fisher distribution.  The 3B catalog reports the angular size, 
$\delta\theta_i$, of a 68\% confidence circle for each burst.
The corresponding value of $\kappa_i$ is that which makes the 3B
confidence circle contain 68\% of the probability according to the
Fisher distribution.  This value can be found by solving the
following nonlinear equation for $\kappa_i$:
\beq
{\kappa_i \over 2 \sinh \kappa_i} 
      \int_{\cos\delta\theta_i}^1 d\mu\, e^{\kappa_i\mu} = 0.68.
  \eqn(kappa-def)
\eeq
For small values of $\delta\theta_i$ (corresponding to large values
of $\kappa_i$), this gives $\kappa_i \approx 2.3/(\delta\theta_i)^2$;
we use the exact (numerical) solution in our analysis.  Also, the
value of $\delta\theta_i$ we use combines the statistical uncertainty,
$\delta\theta_{i,{\rm stat}}$, reported in the 3B tables with the systematic
uncertainty, $\delta\theta_{\rm sys}$, 
estimated by the BATSE team to be  $1.6^\circ$.
We combine these uncertainties in quadrature, so that
\beq
\delta\theta_i = \left[(\delta\theta_{i,{\rm stat}})^2
         + (\delta\theta_{\rm sys})^2\right]^{1/2},\eqn(dtheta)
\eeq
as recommended in the 3B catalog.  We note that Graziani (1995) has found
evidence both that the average size of the systematic error may be 
understimated, and that there are
significant correlations between the size of the systematic error and
the size of the statistical error, although the limited calibration
data available prohibits careful measurement of such correlations
over the entire dynamic range of bursts.  As we discuss further
in the companion paper, these finding appear to have negligible
import for the analysis of models with large scale anisotropy, such
as those we analyze in this series of papers.
However, they may prove quite crucial in analyzing models
with small scale structure (such as models with repeating burst sites).

As already noted, \ceqn(Li-def)\ is an approximation to the actual
individual event likelihood function.  The principal weakness of
this approximation is probably omission of correlations between the
three arguments of the event likelihood (peak flux and two angles).
As noted in LW95, the correlations between the flux and either angle
variable are probably not strong because of compensating correlations
in the likelihood factors for detectors on opposite sides of the
spacecraft; Pendleton et al.\ (1992) reach a similar conclusion
empirically.  But correlations between inferred values of the two angles
needed to specify the burst direction
can be fairly strong; they are exhibited by noncircular contours in the $\chi^2$
maps for the burst directions.  A superior approximation would take
into account ellipticity in the contours (e.g., by replacing the
Fisher distribution with a Kent distribution).  Accounting for such
correlations is probably crucial for assessing hypotheses with
small angular scales, such as models invoking repeating burst sites
or angular correlations between burst sites.  But the axisymmetric
approximation we use should be entirely adequate for the analysis of models
with only large-scale angular structure, and in any case has no
effect on analyses of isotropic models such as those reported here.

Another possibly important weakness is that the behavior of the
true likelihood away from its peak may be non-Gaussian in $\F$
or non-Fisher in $\drxn$.  For example, the true likelihood may decrease
less rapidly than do these exponential distributions. 
This possible shortcoming is noted in the 3B catalog, but no information
is provided (such as the sizes of error circles or ellipses at confidence
levels greater than 68\%) that would allow us to quantitatively assess the 
accuracy of our approximation in this regard.

% We cannot
%assess the accuracy of our approximation in this regard without
%significantly more information than is available in the 3B catalog.

%A better approximation would
%result from using the full $\chi^2$ function from which the summaries
%presented in the 3B catalog were obtained; or one could calculate the
%exaclt likelihood from Poisson distributions for the counts in the
%various detectors.  These more complicated functions would be too
%unwieldy to use in the full likelihood calculation.  

%-------------------------------------------------------------------------------
\subsection{Detection Efficiency}

Approximating the detection efficiency is problematical because several
important elements of a proper efficiency calculation were omitted from the
3B catalog calculations.  Rather than report
$\bar \eta(\F,\drxn)$ as a joint, three-dimensional
function of $\F$ and $\drxn$, the 3B catalog instead reports
two simple functions
indicating the detection sensitivity as a function of flux and of declination.
Fortunately, the correlations between $\F$ and $\drxn$, and the dependence
on right ascension,
are expected to be quite weak at all but the lowest fluxes.  

More seriously,
the 3B calculations ignore Poisson fluctuations in the count rate and
the contribution to to the count rate from atmospheric scattering 
of gamma rays, which increases the effective areas of the eight BATSE
detectors.
Both of these effects are important for properly calculating the
efficiency for detecting weak bursts.  

We have performed extensive simulations
of an idealized BATSE instrument to determine when the neglected
terms become important.  The simulated instrument consisted of eight
detectors arranged in the same octagonal geometry as the eight BATSE
detectors.  The simulated orbit was circular and equatorial, at a
fixed altitude of 400~km.  During the course of the simulated
observations, the satellite spent equal time in each of 15 different
pointings (the first 15 pointings of BATSE's observing plan).  Each
detector had an area equal to the nominal area of a BATSE detector
(1500 cm$^2$), and a purely geometric angular response function
in the outward hemisphere (i.e., proportional to the cosine of the
angle between the burst direction and the outward normal).  In addition,
an atmospheric scattering component was included by isotropically
scattering half of the gamma rays incident at each point on Earth in the 
outward hemisphere of that point (the other half we presume to be
absorbed).  
Each detector had a background rate equal to the nominal
background rate in a BATSE detector (2255~ct~s$^{-1}$), and a trigger
threshold set at a number of counts 5.5
standard deviations above that expected from the background in 
the trigger interval $\delta t$ (the nominal BATSE trigger criterion). 

We simulated burst observations by picking a burst direction from an 
isotropic distribution, and by picking a burst peak flux from a
smooth broken power law distribution that had
a differential slope of 1.5 below a peak flux of 7~ct~cm$^{-2}$~s$^{-1}$
and 2.5 (the slope expected from a homogeneous isotropic burst site
distribution) above that flux.  This purely phenomenological flux
distribution is similar to that expected from a variety of physical
models, and is discussed further below.  Once a burst was chosen, 
we calculated the expected counts in the eight detectors (presuming
a constant peak flux throughout $\delta t$) and
chose actual values for the detected counts from Poisson distributions
that took into account the background rate.
The burst was detected only if the counts equaled or exceeded the
threshold value in two or more detectors.  For detected bursts, we
calculated exact individual event likelihood functions as described
in LW95, and derived from them summaries corresponding to those
published in the 3B catalog.

We simulated 10 sets of 400 bursts using a 64~ms trigger
time scale (each set thus had a number of bursts
comparable to the number of 3B bursts with sufficient information to
perform a Bayesian analysis).  For each set, we calculated the likelihood
for the smooth broken power law model with its parameters
fixed at the true values.  Then we maximized the likelihood by allowing
the break flux and the slope below the break to vary, and calculated
the logarithm of the ratio of the maximum and true likelihoods, 
$\Delta L$ (this corresponds to the analysis we perform in \S~4, below).  
We performed this calculation using the actual detection
efficiency, calculated as specified in LW95, and also using an
approximate detection efficiency that ignores Poisson fluctuations
and atmospheric scattering, as does the efficiency reported in the 3B
catalog.  We denote these values $\Delta L_{\rm true}$, and 
$\Delta L_{\rm approx}$, respectively.
A comparison of the $\Delta L$ values from analyses based on the
true and approximate efficiency
indicates how accurately the likelihood function based on the 
approximate efficiency is; by using
the ratio, we concentrate on the shape of the likelihood function, since its
overall normalization is irrelevant.

Figure 1 shows the the true and approximate average
efficiencies when the simulated detector is operated with a 64~ms
trigger time scale.  The dotted curve shows the approximate result,
ignoring counting uncertainties and atmospheric scattering (corresponding
to the procedure used in the 3B catalog).  The dashed curve incorporates
counting uncertainties; they broaden the region over
which the efficiency falls, allowing the efficiency to be nonzero at all 
positive fluxes (although it is very small at fluxes near zero).
The solid curve is the true efficiency, incorporating both counting
uncertainties and atmospheric scattering.  Atmospheric scattering shifts
the low-flux cutoff to smaller values.

Figure 2a shows the $\Delta L$ values found for each of the 10 simulated data
sets, displayed as a scatterplot showing the approximate $\Delta L$
versus the true $\Delta L$.  To set a useful scale to this plot, note
that an approximate 68\% credible region for a single parameter
is bounded by a likelihood contour with a log likelihood one less than
the maximum.  We thus need $\Delta L$ to be accurate to within
an additive error $\ll 1$ if our inferences are to be accurate.
It is clear from Figure 2a that the approximate efficiency seriously corrupts
inferences, leading to $\Delta L_{\rm approx}$ values unacceptably different 
from the true values.

Figure 1 reveals that
counting uncertainties and atmospheric scattering affect the detection
efficiency most strongly at low fluxes.  We therefore
studied how well the approximation performed when bursts dimmer than
a specified cutoff flux were
omitted from the analysis and the efficiency function was truncated at
that flux.
Such truncation of the data and efficiency is not completely
self-consistent, since $\bar\eta'(\F)$ is the probability for
detecting a burst whose {\it true} peak flux is $\F$, but we can
only truncate the data based on the {\it estimated} peak flux values.
We hoped that the flux uncertainties would be small enough that
the inconsistency would not corrupt the analysis.  That this hope is
realizable is made clear in Figures 2b and 2c.  These Figures show 
$\Delta L$ scatterplots when the data and efficiency are truncated
at increasing values of $\F$.  Once the threshold flux
reaches 1.5~cm$^{-2}$~s$^{-1}$, analyses using the approximate 64~ms
efficiency accurately duplicate the results of analyses using the
true efficiency.  Similar calculations based on a 1024~ms trigger
time scale indicate that the 1024~ms efficiency and data must
be truncated at $\F=0.4$~cm$^{-2}$~s$^{-1}$.

In LW95 we noted that the detection efficiency is also a function
of the burst spectrum and peak duration.  The efficiency reported
in the 3B catalog was calculated for a single burst spectrum
(a power law falling with photon energy $\eps$ like $\eps^{-1.5}$);
the catalog reports that the efficiency changes significantly at 
low fluxes if one steepens the spectrum to $\propto \eps^{-2.5}$,
a spectrum not atypical of some bursts (as we note in \S~5).  No
effort was made to quantify the dependence on peak duration.  Without
further information, we cannot ascertain the effect of these omissions
on our results.  Hopefully, these effects are minor for bursts with
fluxes above our cutoff values.

%We find that use of the reported detection
%efficiency below $\F=1.5$~cm$^{-2}$~s$^{-1}$ for the 64~ms trigger
%time scale significantly corrupts inferences.  Similarly, use of the
%reported efficiency below $\F=0.4$~cm$^{-2}$~s$^{-1}$ for the 1024~ms
%trigger time scale is unacceptable.  We thus truncate the reported
%detection efficiency at these values, and omit bursts with best-fit
%fluxes below these values from our analyses.

%-------------------------------------------------------------------------------
\subsection{Data Selection}

Having settled on approximations for the $\like_i$ and $\bar\eta'$
functions, we can analyze only those data for which the approximations
are acceptable.  Here we summarize all of the resulting selections,
some of which were mentioned above.

First, although the 3B catalog contains information about 1122 bursts,
many of the bursts triggered the BATSE instrument on only one or two
of the three trigger time scales.  Only 453 of these bursts triggered
the instrument on the 64~ms time scale; 557 bursts triggered the
instrument on the 1024~ms time scale.  We can hope to model
selection effects only for these subsets of the full catalog.

A complication in BATSE's triggering criterion results in a further cut.
Once a burst triggers BATSE, it takes roughly 90 minutes to transmit the
burst data to Earth.  During this readout period, further triggering
is disabled on the 256 and 1024~ms time scales, and the threshold for
64~ms triggers is increased to the peak value seen in the detected burst.
The reported efficiency does not take into account BATSE's
change in threshold upon detection of a burst, and so does not accurately 
describe BATSE's 64~ms detection efficiency for the readout period.
Thus all data for bursts detected during such periods (even if the
previous trigger was later identified as a solar flare or some other
non-GRB event) must be omitted from the analysis.  In the parlance of the 3B
catalog, these are bursts that ``overwrote'' a previous trigger.

For a few bursts observed during the second year of operation of
the BATSE instrument, data gaps caused by
failure of the on-board tape recorder resulted in these bursts having
insufficient data for estimating a peak flux or a reliable direction
uncertainty.  Similar limitations may also have arisen for particularly
weak bursts.  In principle, what little information that is available
could be used to specify broad individual event likelihood functions
for these bursts (for example, we may know that the peak flux for
a particular burst is above some value, and could construct an event
likelihood function that reflects this constraint).  In practice, the
3B catalog reports no useful information about the individual event
likelihood functions for these bursts, so they must be omitted from the 
analysis.  After omitting overwrites and bursts without sufficient
direction or flux information, 407 bursts remain available on the 64~ms
time scale, and 554 on the 1024~ms time scale.

Finally, as already noted, the approximation used to calculate the
detection efficiency reported in the 3B catalog fails at low flux
values, so we must truncate the efficiency below some critical
flux and omit bursts with best-fit fluxes below that value from
the analysis.  The critical flux values are different for data
based on different trigger time scales and are identified above.
Unfortunately, this requires that we omit nearly a third of
the remaining 64~ms bursts, and about 16\% of the remaining 1024~ms
bursts.

As a result of these cuts, the portion of the 3B catalog suitable for a 
consistent statistical analysis contains
279 bursts triggered on the 64~ms time scale and 463 bursts triggered
on the 1024~ms time scale.  Data from each time scale must be analyzed
separately, and the results of such analyses are not independent.  
We present here
only the results of analyses of the 64~ms and 1024~ms data (i.e., we
present no results based on the intermediate 256~ms time scale).  We
will find that analyses of these two data sets often lead to different 
conclusions 
about the shape of the burst distribution, indicating that modeling the
temporal properties of bursts is probably important for fully understanding
the distribution of burst strengths and directions; we describe
preliminary modeling along these lines below.
In several cases we have additionally studied the 256~ms data, and
found that its implications are always intermediate between the 64~ms
and 1024~ms data (for example, if a power law index inferred from the
64~ms data is larger than that inferred from the 1024~ms data, then
that inferred from the 256~ms data lies between the two).  This may
be further evidence that the trends we find are real; but this is
difficult to ascertain because of the lack of independence
of the data sets and the limited information provided about each
burst in the 3B catalog.

We emphasize that our data selection is based simply on internal
consistency among elements of the published catalog.  The
resulting subsets of data are thus the usable subsets for {\it any}
analysis of the peak flux data, not just for our Bayesian analyses.

Finally, we note that although the number of bursts in the 1024~ms 
subset is much larger than that in the 64~ms subset, the number
of bursts in the 1024~ms subset with estimated peak fluxes
above 1.5~cm$^{-2}$~s$^{-1}$ (the cutoff for 64~ms data) is only 156,
far below the number in the 64~ms data set (279).  
The published detection efficiencies for both timescales
are nearly identical above 1.5~cm$^{-2}$~s$^{-1}$.  It must then
be the case that data taken with different triggering time scales
represent significantly different samplings of the intrinsic
peak flux distribution.  Lamb, Graziani, and Smith (1993) have
previously pointed out that a particular burst can have very different
64~ms and 1024~ms peak intensities; Mao, Narayan, and Piran (1993) have
made similar observations.  As a result, bursts spanning
a certain range of peak intensities in the 64~ms data may span
a very different range in the 1024~ms data.
We will see below that
this ``shuffling'' of peak intensities distorts the shape of the
observed intensity distribution, so that inferences based on data
from different timescales can differ significantly.  All previous
analyses of burst distribution models have used data from only one timescale 
(usually 1024~ms), and have thus overlooked this effect.

%===============================================================================
\section{Simple Phenomenological Models}

We begin by analyzing two phenomenological models to get a sense of what
information is in the data, independent of a particular physical model
for the burst distribution.  These models simply specify a
parameterized functional form for $dR/d\F$ directly, rather than deriving
the differential rate from an hypothesized physical burst distribution.  The 
forms
we explore involve power laws.  Despite being purely phenomenological, 
they have the
potential to offer some insight into the physical burst distribution,
because power-law differential distributions arise naturally from
consideration of simple physical models.

As an example, consider the differential flux distribution resulting from
sources that are standard candles distributed uniformly throughout an
infinite Euclidean space.  The rate of bursts from within a radius $r$
grows like the volume, so that the cumulative rate from bursts
closer than $r$ obeys $R(<r) \propto r^3$; thus
the differential rate per unit radius follows $dR/dr \propto r^2$.
We can change variables from $r$ to flux by noting that
$\F\propto 1/r^2$, which also implies that $dr/d\F \propto \F^{-3/2}$.  
Thus the differential rate
obeys $dR/d\F \propto \F^{-5/2}$, and the cumulative rate obeys
the familiar $3/2$ law, with $R(>\F) \propto \F^{-3/2}$.  
Inferences about the power-law index therefore 
can be used to quantify acceptance
or rejection of a homogeneous distribution.

More generally, suppose that the rate of bursts from within $r$ grows
like $r^a$, so that
\beq
{dR \over dr} \propto r^{a-1}.\eqn(dRdr-a)
\eeq
The homogeneous Euclidean case just described corresponds to $a=3$.
For burst sources distributed throughout a thin disk, $a=2$, and for
burst sources distributed throughout an isotropic $1/r^2$ halo, $a=1$.
For a cosmological population of bursters it proves convenient
to replace the radial coordinate $r$ with the redshift $z$ ($r\propto z$
when $z\ll1$), writing $dR/dz \propto z^{a-1}$.  
If the cosmological population has a comoving burst
rate density varying with $z$ like $(1+z)^{-\beta}$, then
$a=3$ (the Euclidean value) for bursts at small redshift (i.e., the
brightest bursts); but for sources at large $z$ we show in Appendix A that 
$a = -({3\over 2} + \beta)$.  From these examples it is clear that if we
could infer $a$, we could potentially make important inferences about 
the geometry of the burst distribution.

Similarly, generalize the flux-distance relation so that
\beq
\F(r) \propto r^{-b}.\eqn(F-r)
\eeq
For bursts in Euclidean space, $b=2$.  For cosmological burst sources
emitting $\gamma$ rays of energy $E$ with a power-law photon number spectrum
proportional to $E^{-\alpha}$, the flux-distance relation has this
same form for bright bursts originating from sources at low redshift.
For sources at large redshift, it again is most convenient to
switch from $r$ to $z$, writing $\F \propto z^{-b}$.  In this
case we find in Appendix A that $b=\alpha$.  If we could
infer $b$, we could potentially identify a uniquely cosmological
aspect to the burst data.

We can calculate the differential burst rate implied by these
generalized laws by using \ceqn(F-r)\ to change variables from $r$
to $\F$ in \ceqn(dRdr-a).  The result is a differential rate that obeys
\beq
{dR \over d\F} \propto \F^{-\left({a\over b}+1\right)},\eqn(dRdF-ab)
\eeq
with the corresponding cumulative rate obeying
$R(>\F) \propto \F^{-a/b}$.  It is important to note that these power
laws depend only on the ratio, $a/b$.  Information about
the radial behavior of the burst rate and the flux-distance relation
are inextricably combined in the power-law slope of the differential
burst rate.  This simple result presages an unfortunate conclusion
of our analyses below:
that the flux distribution provides little constraint on cosmological
scenarios for bursters once one considers source distributions more
complicated than standard candles with a constant comoving burst
rate (for which $\beta=0$ so that the behavior of $a$ is fixed a priori).

%Table 1 summarizes the power law behavior of the models just discussed.

The cosmological scenario distinguishes itself among the possibilities
just discussed in that its power-law exponents change with $\F$.
They take on the simple Euclidean values at large $\F$ (corresponding
to sources at redshifts $z\ll 1$), but change to other values for
the dimmest bursts.  This change indicates the presence of characteristic
distance and luminosity scales in the burst distribution.  Quite
generally, if the logarithmic slope of the differential rate changes 
in the vicinity of a
characteristic flux $\F_b$, we can write
\beq
\F_b = {\lum_c \over 4\pi r_c^2},\eqn(Fb-def)
\eeq
where $\lum_c$ is a characteristic luminosity, and $r_c$ is a
characteristic distance.  Detection of a break in the logarithmic
slope thus has the potential of revealing information about scales for
both the spatial and luminosity distribution.  In
cosmological models, the Hubble distance, $c/H_0$ 
(with $c$ the speed of light and $H_0$ equal to Hubble's constant) provides
a fixed characteristic length scale, so detection of a break allows
measurement of the characteristic luminosity of cosmological bursters.

The Euclidean examples discussed above presume unbounded distributions,
and thus have constant power law indices.
If the burster distribution is bounded and bursters are standard candles,
the differential rate will be bounded as well, falling to zero at the flux
value associated with a burst viewed from the farthest boundary.  But if
bursters have a distribution of luminosities, information about that distribution
can be extracted from the shape of the differential rate.  As an
illustrative example, consider the case of an isotropic distribution
of sources with a power law luminosity function, so that
\beq
\dot n(\rvec,\lum) = \dot n(r) f(\lum),\eqn(nnf)
\eeq
where $f(\lum)$ is a normalized power law with lower limit $\lum_l$ and
upper limit $\lum_u$, proportional to $\lum^{-p}$.  If 
$\dot n(r)\propto r^{a-3}$ (as assumed in eqn.~\ceq(dRdr-a)) 
inside $r_c$, but vanishes
beyond $r_c$, then using \ceqn(dR-ndot)\ it is straightforward to show that
\beq
{dR \over d\F} \propto \cases{
   \F^{-p}, &for $\F_l < \F < \F_b$,\cr
   \F^{{a\over 2}+1}, &for $\F > \F_b$,\cr
   0, &for $\F < \F_l$;\cr
}
\eeq
where $\F_b = \lum_u/4\pi r_c^2$ and $\F_l = \lum_l/4\pi r_c^2$.
Thus, when the luminosity function is a power law, the differential
rate below the break directly mimics the luminosity function, potentially
allowing us to infer its logarithmic slope.  As shown by
Wasserman (1992), similar behavior
arises for cosmological bursters with a power-law luminosity function, except
that the break is smoothed, $c/H_0$ playing the role of $r_c$
(see also M\'esz\'aros and M\'esz\'aros 1995).

To summarize, the logarithmic slope of the flux distribution contains
information about the geometry of the burster distribution and the
flux-distance relationship, folded together.  A break in the
distribution contains information about characteristic distance and
luminosity scales in the burster distribution.  Finally, the
shape of the flux distribution below the break contains information
about the luminosity function; for power-law luminosity functions,
the differential flux distribution directly mimics the luminosity function.
With this as motivation, we now turn to the analysis of phenomenological
power law models for the 3B data.

%-------------------------------------------------------------------------------
\subsection{Power-Law Models}

First, we consider a simple power-law model, $M_1$, with
$dR/d\F = A\,\F^{-\gamma}$.
This model has one shape parameter, the power-law index, $\gamma$. 
Figure 3 shows the log posterior as a function of $\gamma$ using the
64~ms data (solid curve) and the 1024~ms data (dashed curve).  The
curves are very nearly parabolic, corresponding to nearly Gaussian
posteriors. For the 64~ms data, 
$\gamma=2.1\pm0.12$; for the 1024~ms data,
$\gamma=1.9\pm0.1$ (here and elsewhere we provide the mode and 95\%
credible region as parameter summaries); these best-fit parameter
values appear in Table 1.  The
index corresponding to an isotropic distribution ($\gamma=2.5$) is 
outside even the ``$5\sigma$'' range for both data sets.

The difference between the best-fit $\gamma$ values indicates that the
distribution of 64~ms fluxes falls off more quickly than does that of 
1024~ms peak fluxes.  Superficially,
the discrepancy between the inferred slopes for the data sets
appears only marginally significant, since the curves in Figure 3 overlap
just inside their ``2$\sigma$'' boundaries. 
However, these data sets are not independent since many bursts are
common to both, so the difference is likely to be 
more significant than
simple consideration of the sizes of credible regions for $\gamma$
would imply.  We will find that inferences based on these data sets
differ for every model we investigate.  This indicates that it is probably
necessary to consider the temporal behavior of bursts explicitly
in modeling and analysis of the flux distribution, a possibility we
discussed in some detail in LW95.  We
discuss this further in \S~6, and defer most discussion of the 
differences between the data sets to that section.

Figure 4a provides a graphical portrayal of how the best-fit power law
model compares with the 64~ms data.  Plotted are the normalized
effective cumulative rate distribution,
\beq
F(>\F) = { \int_\F^\infty d\F'\, \eta(\F')\,\dRdF \over
   \int_0^\infty d\F'\, \eta(\F')\,\dRdF},
  \eqn(F-def)
\eeq
as a smooth curve, and a cumulative histogram of estimated burst fluxes,
$\F_i$.  The dotted curve, associated with the right vertical axis,
shows the negative logarithmic slop of $F(>\F)$ as a function of $\F$.
Figure 4b is a similar plot based on the 1024~ms data.  In LW95 we
discuss using such curves to graphically indicate the goodness of
fit of a model; we emphasize that our analysis is not based on
comparison of the cumulative histograms shown here.

%-------------------------------------------------------------------------------
\subsection{Smooth Broken Power-Law Models}

It is apparent from Figure 4b that a single power law does not describe
the entire 1024~ms flux distribution very well; the distribution of bright
bursts seems to fall off more rapidly with $\F$ than does that of dim
bursts.  (Similar behavior is apparent in Figure~4a, but it is less
decisive.)
Also, {\it Pioneer Venus Orbiter} (PVO) burst observations imply
that at bright fluxes ($\F > 20$--50~cm$^{-2}$~s$^{-1}$), 
$dR/d\F \propto \F^{-2.5}$ for fluxes measured on a 256~ms time scale
(Fenimore et al.\ 1992).  Accordingly, we investigate a smooth broken power 
law model, $M_2$, with
\beq
{dR \over d\F} 
   = A\,{ (\F/\Fb)^{-\gamma_1} \over 1+(\F/\Fb)^{\gamma_2-\gamma_1}},
  \eqn(dR-sbpl)
\eeq
with $\gamma_2 \equiv 2.5$ (thus fixing the logarithmic slope at large
fluxes to that expected from a homogeneous source distribution). 
This model has two shape parameters, the logarithmic
slope a low flux $\gamma_1$, and the break flux $\F_b$.

Figure 5a shows joint credible regions for $\Fb$ and $\gamma_1$, based
on the 64~ms data.  
The best-fit parameter values are listed in Table 1, as are three
quantities useful for comparing this model to others we have
studied: (1) the
ratio, $R_{21}$, of the maximum likelihood for this model and that
for the single power law model; (2) the
asymptotic significance level, $p(>R_{21})$,
associated with this likelihood ratio; and (3) the Bayes factor, $B_{21}$
in favor of this model over the single power law model.
(The significance level is the approximate long-run probability that
one would falsely reject the simpler model if one were to reject it
when the likelihood ratio is at least as large as that observed.
Small values indicate high confidence in the complicated model.  
Asymptotically, $-2\log(R_{ij})$ is distributed as $\chi^2_\nu$,
where $\nu$ is the number of additional parameters in the more
complicated model.)
The Bayes factor indicates that the simpler single power law model
is favored.  The credible region is unbounded
at large values of $\Fb$ (the single power law model is the
$\Fb\rightarrow\infty$ limit of this model).
Figure 6a shows the best-fit model; very little curvature is
apparent.
No 3B bursts have 64~ms peak fluxes above 200~cm$^{-2}$~s$^{-1}$,
where the best-fit value of $\F_b$ lies.  
Thus the location and unboundedness of the credible region indicate
that these data do not provide significant evidence for a break in the 
logarithmic slope.  This is further borne
out by the maximum likelihood values:  the maximum likelihood for the
smooth broken power law model is only 1.3 times larger than that for
the single power law model, not enough to justify its greater
complexity.  We
conclude that there is no significant large-scale structure in the 
64~ms flux distribution.
Of course, these models may not be able to detect significant small scale
structure (``bumps and wiggles'') in the distribution; but we do
not know of any a priori reasons for such structure, and thus have
not attempted to model and detect it.

Figure 5b shows joint credible regions for $\Fb$ and $\gamma_1$ for the
1024~ms data.
The best-fit parameter values, likelihood ratio, and Bayes factor
appear in Table 1.
% are $\gamma_1 = 1.7$ and 
%$\Fb=12$~cm$^{-2}$~s$^{-1}$. 
The 68\% credible region is bounded at large values of $\Fb$, and
the Bayes factor in favor of this model over the single power law model
is 20, indicating
a significant preference for the smooth broken power law model.
But the 95\% credible region extends to $\Fb$ values well above
1000~cm$^{-2}$~s$^{-1}$.  We interpret this as implying
that the data require curvature in $\log R$ vs.\ $\log \F$, but not necessarily
a sharp break.  The allowed values of $\gamma_1$ are systematically smaller
than the best-fit $\gamma$ for a single power law, revealing that
the 1024~ms data favor even more flattening at low $\F$ than is
implied by fits of single power law models.
Figure 6b shows the best-fit model, illustrating 
how allowing curvature improves the fit, particularly at large flux
values.

As a final simple phenomenological model, $M_3$, we considered the smooth
broken power law model, but we allowed the upper power law index
to vary.  However, we changed parameters from the power law slope,
$\gamma_2$, to the angle, $\theta=\arctan(\gamma_2)$, and we used
a uniform prior from $\theta=1.1$ (corresponding to $\gamma_2\approx 2$)
to $\theta = 1.5$ (an 85$^\circ$ angle, corresponding to $\gamma_2\approx 15$).
We introduced this reparameterization to facilitate exploring steep
power laws for which a small change in angle results in a large change
in power law index.  Our uniform prior for $\theta$ corresponds to a
bounded Cauchy distribution prior for $\gamma_2$.  This change has little effect
on parameter estimates.  However, it probably leads to a larger Bayes factor
in favor of this model than would result from use of a flat prior on $\gamma_2$,
simply because the size of the $\theta$ parameter space is smaller.

Table 1 presents the best-fit parameter values and model comparison
statistics for $M_3$ (the $\gamma_2$ value corresponding to the best-fit
$\theta$ is quoted to facilitate comparison with other models).  
There is a mild preference for this model
over a single power law model for the 64~ms data (the Bayes factor
is 4.7).  There is a more significant, although not decisive, preference
for this model for the 1024~ms data.  Furthermore, the best-fit value
of $\theta$ is 1.5, the highest value we considered.  The 1024~ms data
fall off significantly more quickly at large fluxes than is expected
for a $\gamma=-2.5$ powerlaw, and the cutoff is quite sharp.  We
discuss some possible effects that could give rise to such an apparent
cutoff in the 1024~ms data, but not in the 64~ms data, in \S~6.

Finally, we note that our ability to detect changes in the logarithmic
slope of the 1024~ms data are not simply due to the 1024~ms data set
being larger and extending to lower fluxes than the 64~ms data
set.  If we raise the threshold
for the 1024~ms detection efficiency from 0.4 to 1.5~cm$^{-2}$~s$^{-1}$,
only 156 bursts remain in the data set (much less than the 279 in
the 64~ms data set).  Yet evidence for structure remains.
Although we have not performed a full Bayesian analysis, the
the maximum likelihood ratio in favor of $M_2$ is 6.4, corresponding to 5\% significance (only suggestive evidence for a break to $\gamma_2=2.5$);
and the maximum likelihood 
ratio in favor of $M_3$ is 138, corresponding to 0.7\% significance
(strong evidence for a cutoff, with $\Fb = 45$~cm$^{-2}$~s$^{-1}$ and
$\theta=1.5$).

%===============================================================================
\section{Simple Cosmological Models}

We now consider simple cosmological models.  To calculate $dR/d\F$ from a
cosmological source distribution, we must adapt \ceqn(dR-ndot)\ to
a cosmological setting, and then integrate over direction to
calculate $dR/d\F$.  We provide a detailed derivation of the cosmological
counterpart to \ceqn(dR-ndot)\ in Appendix A.  The result can be
written
\beq
{dR \over d\F d\drxn} =  \int dV(z,\drxn) \int d\lum\;
    (1+z)^2 \dot n_c(z,\lum) \,
      \delta[\F - \Fobs(z,\lum,\spars)].\eqn(dR-cosmo)
\eeq
This equation
differs from \ceqn(dR-iso)\ in three respects.  First,
the Euclidean volume element, $r^2 drd\drxn$, has been replaced
by a differential whose functional dependence accounts for
spacetime curvature; additionally, we choose to parameterize it in terms
of redshift and direction, the redshift here playing the role of a
radial coordinate.  Second, $\dot n(\rvec,\lum)$ has been replaced
with $(1+z)^2 \dot n_c(z,\lum)$.  The $\dot n_c(z,\lum)$ function is
the burst rate density per unit $\lum$
per comoving volume element for bursts
at a redshift of $z$ with a maximum photon emission rate of $\lum$
(hereafter the ``(peak) photon number luminosity'').
The $(1+z)$ factors arise from accounting for the 
redshift of the burst rate per unit time, and the
difference between proper volume and comoving volume.  Finally,
the observed flux of a burst from a source with luminosity $\lum$
at redshift $z$, $\Fobs(z,\lum,\spars)$, differs from its Euclidean 
counterpart.  It is given by
\beq
\Fobs(z,\lum,\spars) = {\lum \over 4\pi (1+z) d^2(z)}
           K_0(z,\spars),\eqn(Fobs-cos) 
\eeq
where $d(z)$ is the proper distance (at the current epoch)
to a redshift of $z$, and $K_0(z,\spars)$ 
is a spectral correction function (similar to the optical ``$K$-correction'') 
that depends on the shape of the burst spectrum through the spectral
shape parameters, $\spars$.  

We provide a detailed derivation of equations \ceq(dR-cosmo)\ and
\ceq(Fobs-cos)\ in Appendix A.
Here we remark only on the features of these equations necessary for understanding
the inferences we will make, and in particular on the parameters
one must specify to allow calculation of \ceqn(dR-cosmo).

The volume element, $dV(z,\drxn)$, depends on the cosmology adopted.  We study
cosmologies with zero cosmological constant, for which
$dV(z,\drxn)$ is specified by the Hubble constant, 
$H_0 = 100 h$~km~s$^{-1}$~Mpc$^{-1}$, 
and the density in terms of the critical density, $\Omega_0$ 
(or alternatively the deceleration parameter, $q_0=\Omega_0/2$).

To calculate the spectral correction factor $K_0(z,\spars)$, we
must specify the shapes of burst spectra.  We presume that all bursts
have a common spectral shape:  a power law proportional to $E^{-\alpha}$.
The spectral parameters, $\spars$, are then the power law index $\alpha$,
and the lower and upper limits of the burst spectrum in the rest
frame of the source.  The photon number luminosity, $\lum$, is the
{\it total} luminosity across the entire spectrum (i.e., not just
that in the trigger range, since the trigger range corresponds to
different rest frame energy ranges for sources at different $z$).
In this study, we simply fix $\alpha=1.5$ and fix the lower and
upper limits of the spectrum at 50~keV and 100~MeV;
we henceforth drop the $\spars$ argument from $K_0(z)$.  We chose the
energy range
values so that the redshifted lower limit is always at or below the lower 
limit of the trigger range, and the
upper limit is never redshifted into the trigger range
(since no bursts have sharp breaks or cutoffs observed in this range);
our results are not sensitive to these limits.
This constant value of $\alpha$ is the same assumed for the calculation
of the detection efficiency in the 3B catalog.  In fact, although
the median spectral index for 3B bursts is $\alpha\approx 1.5$,
the distribution of spectral indices is quite broad.
Figure 9 shows a histogram of approximate spectral indices in the
nominal trigger range of 50 -- 300 keV, derived from the ratios of
reported burst {\it fluences} just below and above this range 
(i.e., subtracting one from the fluence spectral index, since it
is for the energy spectrum rather than the number spectrum).
Although this is not a distribution of peak flux spectral indices
(no spectral information is tabulated for peak fluxes in the 3B catalog),
it does imply that bursts exhibit a wide variety of spectral
slopes.  We have
performed analyses using several values of $\alpha$ and verified
that our conclusions are not excessively sensitive
to our choice of $\alpha$ (best-fit parameters for $\alpha=1$, for
example, lie inside the 68\% credible region based
on $\alpha=1.5$).
A more rigorous
analysis would use measured values of $\alpha$ for each burst, but
our analysis gives us some confidence that the simplification of treating
$\alpha$ as a constant does not significantly corrupt
our findings.  More worrisome is the spectral dependence of
the efficiency, about which little information is available.  But
hopefully this dependence is weak above the fluxes where we have
truncated the efficiency for our analysis.  LW95 discusses incorporation
of spectral information into a Bayesian analysis in some detail.

We derive the detailed forms of $dV(z,\drxn)$ and $K_0(z,\spars)$ in
Appendix A.  Once these forms are specified, the isotropy of
cosmological models makes integration over $d\drxn$ trivial,
and the presence of the $\delta$-function in
\ceqn(dR-cosmo)\ permits us to perform one of the remaining two
integrals (over $z$ and $\lum$) analytically.  For standard
candle models, for which $\dot n_c$ contains a $\delta$-function in
$\lum$, both integrals are analytic.  More details about
computational methods appear in Appendix A.

The models we investigate differ with respect to the choice of
functional form for the comoving burst rate density, $\dot n_c(z,\lum)$.
We always use the total burst rate per comoving volume at $z=0$,
denoted by $\dot n_0$, as the amplitude parameter.
The shape parameters of the models are the parameters defining the
$z$ and $\lum$ dependence of $\dot n_c$,
and $\Omega_0$.  Hubble's constant appears as a scale
factor in the inferred luminosity and in
the amplitude parameter.  Since its value is highly uncertain, we
can only infer $h^{-3} \dot n_0$, and the product of luminosity and $h^2$.
For simplicity, we adopt $h=1$ below.

%-------------------------------------------------------------------------------
\subsection{Homogeneous Standard Candle Models}

First we consider a standard candle model, $M_4$, with
a comoving burst rate that is independent of redshift;
\beq
\dot n_c(z,\lum) = \dot n_0\,\delta(\lum - \lum_c).\eqn(ndot-sc)
\eeq
The shape parameters for this model are $\Omega_0$ and the standard
candle luminosity $\lum_c$. 
The luminosity is most conveniently written in terms of a
dimensionless photon number luminosity, $\nu_c$, according to
\beq
\lum_c = \nu_c\; {4\pi c^2 \over H_0^2 K_0(0)} \F_{\rm fid},\eqn(nu-sc)
\eeq
where $\F_{\rm fid}$ is a fiducial value of the observed flux, which
we set equal to 1~cm$^{-2}$s$^{-1}$ (near the triggering threshold
for BATSE).  For  $H_0=100$ km~s$^{-1}$~Mpc$^{-1}$ and the spectral
parameters given above, 
$\nu_c=1$ implies  $\lum_c = 2.2\times10^{57}$s$^{-1}$.  Since our
assumed spectrum has a mean rest frame photon energy of 2.2~MeV,
this corresponds to an energy luminosity of 
$L_c = 7.5 \times 10^{51}$ erg s$^{-1}$.  The inferred values of $\lum_c$
and $L_c$ depend more sensitively on our spectral assumptions than
does the inferred value of $\nu_c$, since the latter is defined 
with respect to the part of the spectrum in the trigger passband while
the former involve integrals over the entire spectrum.

For a flat universe ($\Omega_0=1$), Figure 10 shows the marginal posteriors for
the photon luminosity resulting from consideration of the 64~ms (solid
curve) and 1024~ms (dashed curve) data.  Properties of the best-fit models
appear in Tabel 2.
For the 64~ms data, the best-fit value is $\nu_c=0.37$, with $\nu_c=0.1$ to 1
at the 95\%  level; the best-fit model has a likelihood 4.5
times {\it smaller} than that of the single power law model.
For the 1024~ms data, the best-fit value is $\nu_c=0.44$, with $\nu_c=0.2$ to
0.8 at the 95\% level; the maximum likelihood is 7.2 times 
larger than that
of the single power law model (only a fraction of the improvement offered by
the smooth broken power law model). 
Figure 11 shows the shape of the flux distributions of the best-fit models.

Figure 12 shows contours of the joint posterior for $\dot n_0$ and
$\nu_c$.  The upper contours are from analysis of the 64~ms data;
the lower are from analysis of the 1024~ms data.  The inferred burst
rate density and luminosity are very strongly correlated in the
sense one would expect:  the burst rate density must be higher for
models with less luminous (i.e., closer) bursts.  But the inferred
rate falls more slowly with luminosity than the $\dot n \propto \nu_c^{-3/2}$
behavior one would expect in Euclidean space due to cosmological effects.
Figure~12 also makes it clear
that the two data sets imply significantly different burst rates.
Since their implied luminosities are similar, this is due simply
to the fact observed in \S~3.3:  the 1024~ms catalog has far
fewer bursts in the flux range where it overlaps the 64~ms catalog.

In Appendix A we describe how to calculate the redshift distribution
of burst sources once the parameters of a model are fully specified.
Figure 13 shows the burster redshift distributions for the best-fit
models.  The solid curves show the intrinsic distribution of sources
(top curve for 64~ms, bottom for 1024~ms).  The dashed curves,
corresponding to the right vertical axis, show the observable redshift
distribution (the cutoffs at large $z$ are due to the rapidly vanishing
efficiency for detecting weak bursts).  
The best-fit parameter values from both data sets
imply typical burst redshifts $z\approx 1$, 
although the typical redshifts of {\it observed} 1024~ms bursts
is about twice that of 64~ms bursts.

The calculations just discussed presumed $\Omega_0=1$.
Allowing $\Omega_0$ to vary, however, broadens the
allowed range of $\nu_c$, as shown by the joint posterior for
$\nu_c$ and $\Omega_0$, based on the 64~ms data, shown in Figure 14a.
Figure 14b shows that similar results arise from analysis of the
1024~ms data.
Although in principle we could hope to infer the cosmological
parameter $\Omega_0$ from the burst data, these Figures show that
this hope is forlorn; the posterior is very broad, spanning the
range $\Omega_0\approx 0.1$ to $2$ (i.e., the range we would have
considered reasonable a priori).  In practice, the data are thus too sparse to
usefully constrain $\Omega_0$.
This being the case, and since
the inferred value of $\nu_c$ is not strongly correlated with $\Omega_0$
(the contours are nearly vertical),
we concentrate on flat cosmologies ($\Omega_0=1$) in the
remainder of this paper.  Allowing $\Omega_0$ to vary will somewhat
broaden posteriors based on $\Omega_0=1$.

%Since $\Omega_0$ is one of the parameters of this (and any) cosmological
%model, we can in principle infer it from the burst data.  In practice,
%however, the data are too sparse to usefully constrain $\Omega_0$.
%This is illustrated in Figure 8, which shows the log posterior for
%$\Omega_0$, maximized with respect to $\nu$ for each value of
%$\Omega_0$.  The plotted curve is based on the 64~ms data; similar
%results hold for the 1024~ms data.  Clearly, the data provide no
%useful information about $\Omega_0$.  This being the case, and since
%the inferred value of $\nu$ is not strongly correlated with $\Omega_0$
%(CHECK!), we consider only flat cosmologies ($\Omega_0=1$) in the
%remainder of this paper.

%-------------------------------------------------------------------------------
\subsection{Standard Candle Models With Density Evolution}

Next we consider a standard candle model with power-law density evolution,
$M_5$, for which
\beq
\dot n_c(z,\lum) = \dot n_0\,(1+z)^{-\beta}\,\delta(\lum - \lum_c).
   \eqn(ndot-scd)
\eeq
The shape parameters are now $\Omega_0$, $\lum_c$, and $\beta$.
As before, we replace $\lum_c$ with the dimensionless luminosity $\nu_c$.
The  homogeneous model studied above corresponds to $\beta = 0$.

For a flat universe, 
Figure 15 shows joint credible regions for $\beta$ and $\nu_c$
for each data set; the best-fit values are listed in Tabel 2.
Note that the 95\% credible regions include
$\beta=0$ models for both data sets.  These are the homogeneous standard
candle models considered above (model $M_4$).
This implies that models with density
evolution are not significantly more probable than homogeneous models.
This is also borne out by model comparison calculations, as revealed
by the Bayes factors and likelihood ratios listed for this model
in Tabel 2, none of which significantly favor models with density
evolution over homogeneous models.
Figure 16 shows the cumulative flux distributions of the best-fit models.

Although allowing density evolution does not significantly improve the
fit to the data, 
it greatly weakens the ability of the data to
constrain the burst luminosity: the allowed range of $\nu_c$ now spans
over four decades.  

We can easily understand the shape of the credible regions plotted in
Figure 15 by analysis of the behavior of $dR/d\F$.
Large values of $\beta$ are associated with small values of
$\nu_c$ because when $\beta$ is large, most burst sources are 
nearby (with $z\lesssim \beta^{-1}$) and therefore must have small luminosities to
be able to account for the observed bursts.  This part of
parameter space has low probability because the observable bursts
are close enough that they may be considered to be a homogeneous
Euclidean population for which $dR/d\F \propto \F^{-5/2}$, which
we know is ruled out from our analysis of power law models.
On the other hand, negative values of $\beta$ are paired with large
values of $\nu_c$ because for these $\beta$ values, most sources
are at large redshifts and thus must be highly luminous in order
to be observed.  In Appendix B, we show that in this regime,
$dR/d\F \propto \F^{2\beta/3}$ (for an $E^{-3/2}$ burst spectrum).  
Since we know that the logarithmic
slope of the flux distribution is $\approx -2$, such models can
fit only if $\beta\approx -3$.  This is just where the ridge of
high probability is located in Figure 15.

Figure 17 shows contours of the joint posterior density for
$\dot n_0$ and $\nu_c$, conditional on $\beta=-2.5$, to illustrate
the behavior of the posterior for the burst rate density
in the vicinity of the most probable value of $\beta$.  We emphasize that this
figure is {\it conditional} on $\beta=-2.5$ and is not the marginal
distribution for $\dot n_0$ and $\nu_c$, which one could calculate
by averaging many such conditional distributions (with different
values of $\beta$).  In particular, the best-fit $(\nu_c,\dot n_0)$
points do not lie in the contours because the conditional density
varies strongly with $\beta$.  However, the conditional density
better illustrates the discrepancy between inferences based on the
two data sets.  As with homogeneous standard candle models, the
two data sets imply significantly different burst rate densities,
and there is a strong correlation between
inferred values of the burst rate density and luminosity.
But compared to the homogeneous case, allowing for density evolution 
significantly increases our uncertainty
about the burst rate density, so that even for the 1024~ms data,
the 95\% credible region spans nearly two orders of magnitude
in $\dot n_0$ (it is presumably larger for the marginal distribution).

The large uncertainty in burster luminosity implied by models with
density evolution leads to a
large uncertainty in burster redshifts.  Figure 18 shows the
burster redshift distribution for 
some representative models that lie in the 95\% credible regions for 
$\beta$ and $\nu_c$ for both
data sets (with parameter values corresponding to the dots in
Figure 15).    For positive values of
$\beta$, the observed bursts have typical redshifts $\lesssim 0.5$;
but for negative values of $\beta$, the observed bursts could extend
to redshifts of 30.

Models with large burst source luminosities and redshifts
are allowed only because the data are insufficient to constrain the
logarithmic slope of distribution of
bright bursts to the $-5/2$ value expected
for bursts originating from redshifts $\lesssim 1$.  Including
data from the burst detector on the Pioneer Venus Orbiter (PVO)
is likely to strengthen the constraint, since the flux distribution
of PVO bursts is consistent with $dR/d\F \propto \F^{-5/2}$.
A preliminary joint analysis of BATSE and PVO data by 
Fenimore and Bloom (1996a, b) indicates that the PVO data imply that
the dimmest observed bursts must have redshifts $z < 6$.
Cohen and Piran (1995) supplemented the BATSE data with a set
of burst fluxes comparable in size to those observed by PVO, and
drawn randomly from a $\F^{-5/2}$ distribution.  Their 
calculation seems to verify that
such data can improve constraints on the redshifts of the dimmest
bursts, but their analysis is only illustrative, since it presumes
that the entire PVO catalog samples the $\F^{-5/2}$ part of the
flux distribution.  

We expect that allowing $\Omega_0$ to vary would further weaken the
constraints on the burster luminosity, although we have not
explored the resulting higher dimensioned parameter space.
We do not expect introduction of $\Omega_0$ to affect the allowed
luminosity range as drastically as does introduction of $\beta$.

%-------------------------------------------------------------------------------
\subsection{Models With Power-Law Luminosity Functions}

Finally, we consider a model with no density evolution, but with a power-law
photon number luminosity function.  For this model, $M_6$, we set
\beq
\dot n_c(z,\lum) = \dot n_0 \, A\, \lum^{-p}\eqn(ndot-cpl)
\eeq
over a finite range, $\lum_l$ to $\lum_u$.  The parameter
$A$ is a normalization constant for the
$\lum$ power law whose value is fixed
by the other parameters.  The shape parameters for
$M_5$ are thus $\Omega_0$, $p$, $\lum_u$, and the dynamic range,
$\rho= \lum_u/\lum_l$.  As shown by Wasserman (1992), the luminosity
function allows the distribution to flatten from a power law with
index $5/2$ at large $\F$ to one with 
index $p$ at low $\F$, improving the fit in much the same manner as the
smooth broken power law model $M_2$ (see also M\'esz\'aros and 
M\'esz\'aros 1995).
Solar flares, which have similar temporal properties to bursts, have
power law distributions of peak intensity and fluence (Dennis 1985); such
power law distributions are general features of phenomena that involve
energy transfer via a cascade over a broad range of spatiotemporal
scales (Bak, Tant, and Wiesenfeld 1988; Press 1978; Lu and Hamilton 1991).  
In addition, relativistic beaming can produce
power law luminosity functions with a finite dynamic range, 
as we discuss in \S~7.  Thus power law distributions are natural
and obvious candidates for burst source luminosity functions.

As with the previous model, we analyze only the flat, $\Omega_0=1$ case;
this leaves three shape parameters.  In Figure 19 we show the profile
posteriors for $\rho$ based on each data set, that is, the posteriors
maximized with respect to $\lum_u$ and $p$ as a function of $\rho$.  
These curves can be considered to be
approximate marginal distributions for $\rho$.  We have normalized them
so that values of unity correspond to likelihoods equal to the maximum
likelihood found for homogeneous standard candle models.
If $\rho$ is smaller than the dynamic range of the data 
($\sim 10^2$), the luminosity function has negligible width, and
these models
resemble homogeneous standard candle models and thus have profile
posteriors equal to unity with this normalization.
As $\rho$ increases, the profile posterior increases indicating that
a broad luminosity function improves the fit.  Once $\rho$ significantly
exceeds the dynamic range of the data, the profile posterior varies
very weakly with $\rho$.  This indicates that the data cannot constrain
the width of a power law luminosity function.
Accordingly, we simply fix
$\rho$ at a large value $(10^4)$ to explore this model, rather than
incur the computational and graphical burden of keeping this
unconstrained parameter in the parameter space.  We discuss the
absence of a constraint on the luminosity function width further below.

With $\Omega_0$ and $\rho$ fixed,
two shape parameters remain, $p$ and $\lum_u$.
As with standard candle models, we write $\lum_u$ in terms of a
dimensionless parameter, $\nu_u$, so that
\beq
\lum_u = \nu_u\; {4\pi c^2 \over H_0^2 K_0(0)} \F_{\rm fid}.\eqn(nu-cpl)
\eeq
Figure 20 shows contours of the joint posteriors for $p$ and $\nu_u$,
and Figure 21 shows the best-fit models.  The best-fit parameter values
appear in Table 2, with the Bayes factors and likelihood ratios for
the models.
For the 64~ms data,
the best-fit point is at $p=2.0$, $\nu_u=23.6$; but $\nu_u$ is largely 
unconstrained over a dynamic range of the order of $\rho$ (here equal to
$10^4$).  Although the best-fit power law index and upper luminosity
limit are notably different for the 1024~ms data, they, too, imply
a broad posterior distribution that has significant overlap with
that based on 64~ms data.

The peculiar shape of the contours is easy to understand.  For
large negative values of $p$, the luminosity function is highly
concentrated in the vicinity of $\lum_u$, and the model thus has the
likelihood of a homogeneous standard candle model with $\nu_c=\nu_u$, 
independent of the value of $p$.  Thus the contours become vertical at
a value of $\nu_u$ near the credible range of $\nu_c$ found earlier for
homogeneous standard candle models.  Similarly, for large positive
values of $p$, the luminosity function is highly concentrated in the
vicinity of $\lum_l = \lum_u/\rho$, leading to a likelihood equal
to that of a homogeneous standard candle model with $\nu_c=\nu_u/\rho$.
Thus the contours become vertical at $\nu_u$ values that are a factor
of $\rho$ times the credible range of $\nu_c$ values found for homogeneous standard
candle models.  In between these limiting regimes, there is a nearly
horizontal ridge of high probability for models with $p$ nearly equal to
the value of the low flux power law index, $\gamma_1$, of the
phenomenological smooth broken power law model.  This is because, as
noted in \S~4, when there is a spatial scale (here $c/H_0$) and a
power-law luminosity function (with $1< p < 5/2$), the flux distribution will have the same
logarithmic slope as the luminosity function below some flux value.

For both data sets,
the 95\% contours fail to close at the large $\nu_u$, large 
positive $p$ end, or at the small $\nu_u$, large negative $p$ end.
This indicates that this model does not greatly improve upon the
standard candle model.  This is also clear from Figure 19, where it
is apparent that the likelihood for power law models is at most
nine times that of standard candle models.  This modest increase
in likelihood is insufficient to justify the larger parameter space
of these models.  Indeed, the model comparison
results in Table 2 indicate no significant preference for models
with power law luminosity functions over homogeneous standard
candel models.

Figure 22 shows contours of the joint posterior density for
$\dot n_0$ and $\nu_u$, conditional on $p=1.9$, to illustrate
the behavior of the posterior for the burst rate density
near the most probable value of $p$.  As was true with
models incorporating density evolution (see Figure 17), the
additional degree of freedom associated with the luminosity
function significantly increases our uncertainty about the
burst rate density; the 95\% credible region again spans
orders of magnitude.
As with the previously studied cosmological models, the
two data sets imply significantly different burst rate densities.

Figure 23 shows burster redshift distributions for 
some representative models in the 95\% credible regions.
Although the inferred luminosity of the brightest burters varies strongly
across the credible regions,
the redshift distributions do not vary as strongly; the
observed bursters always have redshifts of a few tenths, and
the characteristic redshift of the intrinsic distribution
is of order a few.  We elucidate the reasons for this behavior below.

%-------------------------------------------------------------------------------
\subsubsection{Width of the Luminosity Function}

A number of earlier investigations claim that the BATSE data constrain
the width of the luminosity function to be relatively narrow, with
most studies finding that the dynamic range must be $\lesssim 10$
(Horack, Emslie, and Meegan 1994; Cohen and Piran 1995;
Woods and Loeb 1995; Ulmer and Wijers 1995; 
Ulmer, Wijers, and Fenimore 1995), and a recent study finding instead
that it is constrained to $\lesssim 10^3$ (Hakkila et al.\ 1996).  
This is a surprising result in light of the wide diversity
evident in other burst characteristics (e.g., burst durations span
several orders of magnitude).
If true, this would be an important conclusion, potentially offering
important constraints on physical scenarios for bursts.  For example,
the high luminosities required in scenarios that place burst sources
at cosmological distances makes it likely that the emitting material
is accelerated to relativistic velocities.  Due to relativistic beaming,
an isotropically emitting ``blob'' with a particular rest frame luminosity will
have an apparent luminosity that varies strongly with the angle between
the observer and the velocity vector.  Thus even if sources were standard
candles in their own rest frame, the distribution of observing angles
gives rise to an apparent luminosity function.  We show in \S~7
that the resulting luminosity function is a bounded power law spanning
a range of luminosities differing by a factor of 
$\approx(2\gamma)^{2\alpha+4}$,
where $\gamma$ is the Lorentz factor and $\alpha$ is the spectral index
(typically $1.5\pm1$).  In this scenario, even with very modest Lorentz
factors, one thus expects the luminosity function to have a large dynamic
range ($> 10^4$ for $\gamma = 2$).  Evidence that the luminosity function
is narrow would thus place strong constraints on relativistic motion in
such a scenario.

Our results differ from those of previous studies, instead
demonstrating that the BATSE 3B data cannot constrain the width of power-law
luminosity functions.  This important difference requires explanation.
Before examining some specific ways that our work improves on previous studies, 
we offer some further calculations that clarify why broad luminosity
functions are compatible with the 3B data.

One might worry that our broad luminosity functions are {\it effectively}
narrow, because the best-fit models have luminosity functions that
fall fairly quickly with luminosity.  Although they extend to
large luminosities, the bright bursts are rare and perhaps can be considered
to be effectively absent from observations spanning only four years.
To address this, we explored the power law luminosity function model with
$p=0$, so the luminosity function becomes a flat ``top hat'' function.
Figure 24 shows the profile likelihood function for the dynamic
range spanned by the ``top hat;'' for each value of $\rho$, we
maximized the likelihood with respect to $\nu_u$ to get the plotted
function.  The profile likelihoods based on both the 64~ms and 1024~ms
data are nearly flat, showing that even the width of a {\it flat}
luminosity function cannot be constrained.

Figure 25 elucidates the reason the data cannot constrain the 
width of the luminosity function.  Shown are the
differential burst rates
for models with ``top hat'' luminosity functions with
three different dynamic ranges: $\rho=1$ (standard
candles), 2, and $10^4$, with $\nu_c$ and $\dot n_0$ set at their
best-fit values based on the 1024~ms data.  We have
plotted the rates over a much
broader range of flux values than that spanned by the data.  In
all three cases, the best-fit value of $\nu_u$ is of order unity.
This implies that the break from $dR/d\F \propto \F^{-2.5}$
to $dR/d\F \propto \F^{-p}$ will take place in the vicinity of
$\F=1$~s$^{-1}$~cm$^{-2}$, as is evident in the figure.  The extent
of the luminosity function affects the behavior of the rate below
this value; but the useful 
data extend only to $\F\approx 0.4$~s$^{-1}$~cm$^{-2}$.
There is simply no data in the region where rates from populations
with different values of $\rho$ distinguish themselves.

As already noted, our conclusion that the data cannot constrain the 
width of the luminosity function for cosmological burst sources differs
strongly from the findings of earlier studies.
In \S~10 of paper I we discussed a variety of ways that our methodology
improves on that of earlier studies.  Some of these improvements are
especially apparent in the analysis of the width of the burster
luminosity function; we elaborate on them here.

An important and general advantage of Bayesian methods over more
traditional frequentist methods is the objectivity of the Bayesian
approach.  Ironically, Bayesian methods have the reputation of
being subjective because of the presence of explicit prior probabilities
in Bayesian calculations.  Although in some problems objective priors
exist from analyses of previous measurements,
in cases where we start from ``ignorance,''
the need for a prior does impart subjectivity to the final result.
It is tempting to adopt some simple form for an ``ignorance prior,'' such
as a flat prior, but subjectivity arises because adopting a flat prior
in a problem with a parameter $\theta$ does not produce the same results
as adopting a flat prior in the same problem, reparameterized with
$\theta' = \theta^2$, say.  
But in practice, if the data provide
significant new information about a phenomenon, the choice of prior
has a negligible effect on one's final inferences.  Indeed, if the final
results change significantly due to a small change of the prior (or
of the parameterization), one
has quantitatively demonstrated that the data are not informative---surely
a useful capability.  Further, traditional frequentist methods are not
free of the subjectivity that arises from the choice of parameterization
of a problem.  For example, an unbiased estimator for $\theta$ will
not in general be an unbiased estimator for $\theta'=\theta^2$.

More importantly, Bayesian methods are far {\it more} objective than
their frequentist counterparts in terms of specifying how the data
should affect inferences.  Bayes's theorem uniquely
identifies the likelihood function as the relevant function of the data, and
the rules of probability theory dictate uniquely and mechanically how
to manipulate it to make inferences.  In frequentist statistics there
is considerable freedom in choosing the statistic one will use to
address a particular problem; in complicated problems different analysts
studying the same model are likely to make different choices and can
reach significantly different conclusions as a result.  Even when the
likelihood function is used (e.g., in maximum likelihood parameter 
estimation), considerable freedom remains regarding how to use it to
make inferences (particularly if there are nuisance parameters present).
Bayesian and frequentist analyses using likelihoods can produce
different results, as we discussed at length in \S~9 of paper I.

This point of contrast between Bayesian and frequentist approaches is
evident in analyses of the BATSE data.  A variety of statistics
have been used to study the same or similar models.  For example, 
studies of isotropic models have used various
estimates and summaries of the $V/V_{\rm max}$ distribution, 
$\chi^2$ fitting of binned flux data, the Kolmogorov-Smirnov statistic,
and various moments of the observed intensity distribution.
Besides the possibility of differing studies reaching different
conclusions, difficulties also arise if one examines several
statistics in the course of a study, choosing one as being
``best'' a posteriori.
Such difficulties arise in assessing the work of
Horack, Emslie, and Meegan (1994), one of the several studies claiming
that the BATSE data require a luminosity function with a dynamic
range less than 10.  Their study
used relationships among various 
integral moments of the best-fit peak fluxes to constrain
the luminosity function.  They examined many such relationships before
choosing one that provided a strong constraint, but their calculation
of the significance of their result did not
account for the number of statistics they examined.  Such an accounting
will weaken their constraint; possibly seriously.  
%***********[Also, the moments may be wrong, and
%***********the cosmology is a bit wrong.  Reference Petrosian???]
In any case, as we noted in LW95, the Pitman-Koopman theorem (Jeffreys 1961)
guarantees that if there is a set of averages
or moments of the data that contain all the information relevant to
assessing the considered hypothesis, then the 
likelihood will depend only on those averages, and they can be identified
by examining its functional form.  No set of such averages appears in
our likelihood function, so any calculation considering only a few
ad hoc moments of the data is discarding relevant information that the
likelihood function takes into account.

A further advantage of our methodology is that it straightforwardly
accounts for the uncertainty in the measured parameters of bursts
(e.g., their peak flux and direction).  This is a stumbling block for
frequentist studies because such parameters are technically ``nuisance
parameters,'' and there is no satisfactory method for handling
nuisance parameters in frequentist statistics.  In the Bayesian approach,
one simply uses Bayes's theorem to move them to the left of the 
conditional, and then integrates them out of the problem.
In analyses of isotropic models, the relevant uncertainties are the
peak flux uncertainties.  No previous study has rigorously accounted
for them.  The study of Horack, Emslie, and Meegan (1994) attempted
to account for them in
an ad hoc manner, but they performed no studies of simulated data to verify
or calibrate their procedure.  All other studies ignore the uncertainties,
implicitly assuming that the best-fit peak flux is the true peak flux.

Most studies analyzing the peak flux data use the 1024~ms data, 
for which the peak flux 
uncertainty is relatively small---at most $\sim 15$\% for the dimmest
bursts, compared to $\sim25$\% for the 64~ms data.  The 1024~ms
peak flux uncertainties
may be small enough to ignore (although we demonstrate below that the
distribution of 1024~ms peak fluxes is likely to be less suitable for
analysis than the 64~ms data for other reasons).  
However, two studies (Cohen and Piran 1995; Ulmer and Wijers 1995) instead analyzed the distribution
of peak {\it count rates} (with units of s$^{-1}$) rather than peak fluxes
(with units of s$^{-1}$~cm$^{-2}$).  The reported peak
count rates are the peak rates in the second most brightly illuminated
detector.  To compare them with burst distribution models, one must
convert from incident fluxes (the quantity one can predict with
a burster distribution model) to count rates,
taking into account the highly uncertain angle of incidence to
the second most brightly illuminated detector.  The resulting
uncertainty in the peak flux is {\it not} negligible; the uncertainty
in angle imparts a $\approx 16$\% uncertainty to the fluxes of
{\it all bursts}, to which must be added (in quadrature) the additional
uncertainty due to counting statistics (which can be relatively large
since the counts in only a single detector are used).
But this uncertainty was
ignored in both studies (both studies use equations that
incorrectly equate peak count rates with
peak fluxes, despite the dimensional inconsistency).
In our earlier work (Loredo and Wasserman 1993) we 
discussed proper fitting of the 1B peak count rate data; our analysis
found no constraint on the dynamic range of the luminosity function.
More importantly, since the release of the 1B catalog in 1993 
(Fishman et al.\ 1994), burst peak fluxes have been
available for analysis. 
The peak flux estimates not only incorporate burst direction
information; they also take into account the burst spectrum in the
conversion from count rates to fluxes.
They are thus far superior to the peak counts data
provided prior to the 1B catalog, and are the basis of our analysis here.

A further advantage of the Bayesian approach is that it produces
very straightforward inferences:  one calculates directly the
probabilities for the hypotheses under consideration.  To constrain
parameters, one simply plots contours of (possibly multidimensional)
posterior distributions.  When parameters are highly correlated,
this is evident in plots of the joint distribution; and even when 
inferences are summarized for a subset of the parameters (via
marginalization), all correlations are properly taken into account.
Similarly, to compare models, one simply calculates ratios of
the probabilities for competing models.

In the frequentist approach, one instead calculates values of the
chosen statistic, and then takes a further step of converting these
to probabilities, sometimes by appealing to a simple asymptotic limit,
or often by using Monte Carlo simulations.  A subtlety arises in that
one must assume a particular model is true with a particular set of
parameters in order to perform the probability calculation, but the resulting
probability may be interpreted as applying to a region of parameter
space (in calculating confidence regions), or to a space of models
(in calculating significances of goodness-of-fit tests).  This
complicated and indirect line of reasoning has led to shortcomings
in several studies of the BATSE data.  For example, 
Woods and Loeb (1995) found their constraints
on the width of a ``log-normal'' luminosity function (not of the standard
lognormal form) by fixing the parameter specifying the
most probable luminosity to its best-fit value, and varying§
only the width.  For the power-law models we studied, this would
correspond to fixing $\nu_u$, and varying only $\rho$ (for a particular $p$).
This strongly and artificially constrains the dynamic range, because
models with a large dynamic range are permitted only when we shift the bulk
of the luminosity
function to low fluxes, so that only the brightest bursts are visible
from redshifts of order unity.  This problem is evident in our approach
because Bayes's theorem gives us no alternative but to begin our
calculation by studying the full joint posterior, which displays
all correlations.

In addition, Woods and Loeb confused the roles of the probabilities
that appear in goodness-of-fit tests with those that appear in
calculating confidence regions:
they used the significance level assigned
by a goodness-of-fit test to define allowed parameter regions, when
instead a confidence interval should have been calculated.  Several
other investigators have made this same error, particularly in regard
to analyses of anisotropic models.  We thus discuss it in detail in
Appendix~A of Paper~III.  Such confusion cannot arise in the Bayesian
approach because one is always explicitly calculating probabilities
for the hypotheses of interest (statements about parameter values
for credible region calculations; statements about models in model
comparison calculations), rather than calculating probabilities
for ensembles of data conditional on a single point hypothesis that
may actually be representing an entire family of hypotheses of interest.

Finally, we went to great lengths in this work to ascertain how
approximations adopted in the preparation of the 3B catalog affect
one's inferences, particularly in regard to the accuracy of the
reported detection efficiency (see \S~3).  We found that a
self-consistent analysis must omit a significant amount of low-flux
data.  Most other analyses of the peak flux data omitted even more low-flux
bursts than we did, and thus are probably not affected by
inaccuracies in the reported efficiency.  
An exception is the work of Hakkila et al.\ (1996) who analyze a 
combination of data from BATSE and
PVO, studying the distribution of peak {\it energy} flux,
rather than peak photon number flux.  Unfortunately, they provide insufficient
details to allow duplication of their analysis or elucidation of
possible problems with their methods.  In particular, it is
not clear whether or how they took into account the detection efficiency.
Their analysis includes bursts with 1024~ms peak photon fluxes down to
$0.42$~cm$^{-2}$~s$^{-1}$.  The 3B efficiency is still varying
significantly with peak flux at these low fluxes, and must be
taken into account.  But the 3B efficiency is available only as a 
function of peak photon number flux, and they offer no discussion
of how it can be used to analyze the distribution of peak energy flux,
and no analysis of how accurately this can be done given the
approximations adopted in constructing the 3B catalog (for example, their
cutoff in peak number flux almost certainly does not correspond to
a simple cutoff in peak energy flux).  Their
study also exhibits some of the problems associated with choice of statistic
and accounting for uncertainty:  they analyzed a binned
flux distribution but did not discuss how bin boundary selection affects
their results ($\approx 20$\% of all bursts are in a single bin
at low flux in their analysis), 
and their binning implicitly presumes that the fluxes are
known without uncertainty.  Of course, the most obvious difference
between their analysis and that reported here is their inclusion of PVO data.
But we do not believe PVO data play an important role in constraining
the width of the luminosity function.  PVO predominantly detected
bright bursts, with fluxes well above the BATSE threshold; as
shown in Figure~25, models with different luminosity function widths
differ substantially only at {\it low} fluxes, near or below the
BATSE threshold.  It is possible that the BATSE
data are capable of constraining the width of the peak energy
luminosity function but not the peak photon number luminosity
function; but this would require a strong correlation between spectral
hardness and intensity.  The differences between our findings are thus
likely due to methodological differences.

%-------------------------------------------------------------------------------
\subsubsection{Effective Luminosity Function}

Our analysis so far has concerned the distribution of the luminosities
of {\it all} burst sources (the {\it intrinsic} luminosity function).  
The distribution of the luminosities of
the sources of {\it observed} bursts (the {\it effective} luminosity function) 
is in general different, and the difference between these functions has
been confused in some analyses.  

Ulmer and Wijers (1995) and Ulmer, Wijers, and Fenimore (1995) have
correctly distinguished the intrinsic and effective luminosity functions.
Unfortunately, the former
study incorrectly analyzed the BATSE peak count rates, treating them as
peak photon number fluxes as we discussed above.
The latter study combined BATSE data from two
timescales (256~ms and 1024~ms) with PVO data and found that 90\% of
observed bursts have peak luminosities within a range of 10.  We show
below that our results imply effective luminosity functions that
can span a much larger range.  The brief report of Ulmer, Wijers and Fenimore
(1995) does not provide sufficient detail for us to ascertain the origin
of our differing conclusions.

Other analyses recognize that the intrinsic and effective luminosity
functions are different, but go on to presume that the luminosity function
of observed sources is simply the 90\% most probable part of
the intrinsic luminosity function (Horack, Emslie, and Meegan 1994; 
Hakkila et al.\ 1996; Horack et al.\ 1996).  These studies also conclude that
the effective luminosity function must span a dynamic range $<10$.
Indeed, our best-fit {\it intrinsic} luminosity functions fall quickly
enough with $\lum$ that the 90\% most probable portion spans a range $\sim 10$.
However, the effective luminosity function has a different, flatter shape
than does the intrinsic luminosity function, giving it a much larger 90\% 
range.

In \S~4 of Appendix A we derive the effective luminosity function for cosmological
sources.  But the most important differences between intrinsic
and effective luminosity functions are apparent in the simpler
Euclidean case we now discuss.  Consider a population of sources with burst rate per
unit volume $\dot n(r)$ and intrinsic luminosity function $f(\lum)$.
The effective burst rate per unit peak flux is found by multiplying
\ceqn(dR-iso)\ by the detection efficiency.  The effective burst rate
per unit luminosity is found simply by exchanging the roles of $\F$ and
$\lum$.  The resulting integral over $\F$ is analytic, giving
\beq
{dR_{\rm eff} \over d\lum} = f(\lum) \int_0^\infty dr 4 \pi r^2 \dot n(r)\,
  \bar\eta'\left(\lum \over 4\pi r^2\right).\eqn(Euc-elf)
\eeq
This is the effective luminosity function, up to a factor converting the
normalized effective luminosity function into the effective burst rate
calculated here.
If all bursts are detectable, then $\bar\eta' = 1$ and the integral over
$r$ is simply a constant equal to the total burst rate.  In this case, 
the effective luminosity
function has the same shape as the intrinsic luminosity function.  But
selection effects can cause the shapes to differ drastically in more
realistic situations.  For example, if bursts are detectable only if their
peak flux exceeds a threshold value $\F_{\rm th}$, then the integral
over $r$ above is truncated to $r < (\lum/4\pi\F_{\rm th})^{1/2}$, and is
no longer a constant; its functional form will depend on the burst rate
density.  If $\dot n(r) \propto r^\delta$, then
\beq
{dR_{\rm eff} \over d\F\lum} \propto f(\lum) \lum^{\delta + 3/2}.\eqn(trunc-elf)
\eeq
It is apparent that even simple truncation can produce an effective
luminosity function drastically different in shape from $f(\lum)$, with
the tendency being toward increasing the probability for seeing luminous
burst sources (if $\delta > -3/2$).

Horack et al.\ (1996) recognized that the effective and intrinsic luminosity
functions could differ in shape, but never actually calculated the effective
luminosity function.  They maintained that the difference was
inconsequential because only burst rate densities that increased with $r$
could make bright bursts appear significantly more probable than is implied
by the best-fit $\lum^{-2}$ intrinsic luminosity functions.  However, 
their argument ignores the increase with $r$ of the volume element.  
The probability for seeing luminous burst sources
can be enhanced even for rate densities that fall with $r$, simply because
the volume at large $r$ is large enough that it is likely that rare, luminous
bursts actually occur within the sample volume.

In Figure~26 we show the effective luminosity functions for the best-fit
models for the 64~ms and 1024~ms data; also shown are the logarithmic
slopes of the functions (these calculations use the full cosmological expression).  
The effective luminosity functions are significantly
flatter than the intrinsic functions (which have logarithmic slopes $\approx 2$).
The dots indicate the luminosities bounding the 90\% most probable luminosities.
The 90\% ranges span over three orders of magnitude.  Figure~27 shows the
effective luminosity function for the best-fit 1024~ms model, along with 
functions corresponding to two other models lying in the 68\% credible
region shown in Figure~20(b) that have luminosity upper limits $\nu_u$
significantly smaller and larger than the best-fit value.  Again, dots
indicate the boundaries of regions containing the 90\% most probable luminosities.
It is clear that the shape of the effective luminosity
function is not well-determined;
the sign of its slope is not even determined.  The sizes of the 90\% regions
are very large, spanning two to three orders of magnitude.  Interestingly,
although the three curves correspond to $\nu_u$ values spanning a factor
of 400, the upper limits of the 90\% regions differ by only a factor of 3,
and the 90\% regions overlap substantially.  Thus although widely disparate
values the luminosity of the most luminous sources are allowed by the data, the
luminosity of the most luminous {\it observed} sources
is constrained to be of order $\nu \sim 0.1$ to 1.

Note that the 90\% ranges are larger than the ranges of best-fit peak fluxes in the
data sets we have analyzed.  Geometry is to blame for this somewhat counterintuitive
result.  Underluminous bursters are visible only nearby, in a volume small 
enough that
few if any luminous bursts occur during the duration of the observations.
But luminous bursters are visible through a much larger volume, large enough
that a significant number occur at large distances (and hence at relatively
low fluxes) during the observation time.
In this way burst sources with very disparate luminosities are sampled from
differing depths such that their observed fluxes span a range smaller than is
spanned by the luminosities.

%===============================================================================
\section{A Phenomenological Model\\ With Duration Dependence}

As we have seen time after time, inferences based on an analysis
of the 64~ms data are somewhat different from those based on analysis
of the 1024~ms data.  The complicated nature of the dependence of
these data sets on one another makes it difficult if not impossible
to assess the significance of the discrepancy quantitatively, and also
to determine the extent to which the discrepancy is 
merely a consequence of the different sizes of the
data sets.  Nevertheless, even the simplest phenomenological models
we have studied indicate that there are important systematic differences
between the data sets.  This conclusion is supported by the fact, alluded
to above, that analyses of the 256~ms data always lead to inferences
intermediate to those found with the 64~ms and 1024~ms data.

This leads us to conclude that explicit consideration of the temporal
behavior of bursts is necessary for understanding the flux distribution,
a possibility we discussed at some length in LW95.  This finding is
somewhat ironic in view of the fact that the the gamma ray burst
community moved from quantifying burst intensity with fluences to
peak energy fluxes and finally to peak photon number fluxes in an effort
to avoid the effects of differing burst light curves and spectra on the
shape of the intensity distribution.  But as we emphasized in LW95, use
of peak photon number flux may weaken, but cannot eliminate, the effects
of differing burst light curves and spectra on the shape of the 
distribution of measured fluxes.

To see how the temporal properties of bursts can affect the observed
flux distributions, consider a ``top hat'' model for burst
light curves that presumes the burst emission maintains its peak value
over some time scale, $\tau$, and is significantly smaller for times outside
of the peak.  Denote the actual peak flux by $\F_a$.  If $\tau$ is
longer than the trigger time scale $\delta t$, a somewhat subtle
{\it peak counts bias} results in overestimation of $\F_a$, if one
does not carefully account for the peak duration.  This is because
one is taking several independent samples of counts during the peak,
and is thus likely to identify an upward fluctuation in the counts
as the peak.  We analyzed this bias in detail in LW95, and pointed
out that since its size depends nonlinearly on $\F_a$, it can
distort the shape of the observed flux distribution.  Interestingly,
the peak counts bias causes the
the observed distribution to steepen as $\delta t$ is decreased (provided
$\delta t < \tau$).  Thus
it is possible that some of the steepening of the 64~ms peak flux
distribution relative to the 1024~ms distribution (at low flux values)
is due to this effect.  Lamb, Graziani, and Smith (1993) discuss
the effects of peak counts bias on burst classification.

On the other hand, if $\tau < \delta t$, {\it peak dilution} results.
If one does not account for the fact that the peak is narrower than the
measuring interval, $\F_a$ can be underestimated by a factor as small
as $\tau/\delta t$.  If peak duration is not correlated with peak flux,
this dilution simply shifts the entire flux distribution downward
without altering its shape (but possibly broadening it if there is
a distribution of peak durations).  But if peak duration and peak
flux are correlated, peak dilution can easily distort the shape of
the observed flux distribution.  Some of the effects of peak dilution
have been previously discussed by Lamb, Graziani, and Smith (1993),
Mao, Narayan and Piran (1993), and
Petrosian, Lee, and Azzam (1994).  The fact that the observed 64~ms peak flux
distribution extends to about 150~cm$^{-2}$~s$^{-1}$, but the
1024~ms peak flux distribution extends only to
about 40~cm$^{-2}$~s$^{-1}$ could be taken as evidence that peak
dilution is important in the 1024~ms data.

In LW95 we indicated some ways in which effects like peak counts bias
and peak dilution could be
incorporated into an analysis of the burst peak flux and direction data.  
We additionally discussed incorporation of spectral information.
The 3B catalog provides
extremely limited information about the temporal and spectral properties
of bursts, and thus severely limits the possibilities for an analysis
more sophisticated than that described in the previous sections.  In this
and the following section, we analyze two simple models that illustrate
how temporal and spectral information can play a role even in an analysis
of peak flux data alone.
Neither model is completely successful at explaining the patterns we have
uncovered in the data; but they remain useful as illustrations of the
principles described in LW95.
Underlying both models is the ``top hat'' light curve model just
discussed.  Although it is a highly limited caricature of the shape of actual burst light curves, it has the virtue of simplicity, and it can be 
analyzed approximately using only the limited information available in the 
3B catalog.  As part of the approximation, we neglect peak counts
bias (as we have tacitly done throughout this work;
no raw count data is available in the catalog), and consider
only the effects of peak dilution.

We begin with a purely phenomenological model that presumes
that bursts have
an intrinsic time scale $\tau$ that is a deterministic function of the
actual peak flux of the bursts, which we denote by $\F_a$ to distinguish
it from the measured value, $\F$.  That is, we take the $\tau$ distribution
to be a $\delta$-function whose location is a function of $\F_a$.  
If we let $\tau(\F_a)$ denote the duration of a burst with actual
flux $\F_a$, then the $\delta t$-averaged peak flux is
\beq
\F_{\rm eff}(\F_a) = \cases{
  \F_a, &for $\tau(\F_a) \ge \delta t$,\cr
  \F_a{\tau(\F_a)\over\delta t}, &for $\tau(\F_a) < \delta t$.\cr
}\eqn(F-tau)
\eeq
Once $\tau(\F_a)$ is specified, the observable burst flux distribution
can be calculated from the actual flux distribution $dR/d\F_a$ according to
\beq
{dR \over d\F} = \int d\F_a\; {dR\over d\F_a}
   \,\delta[\F - \F_{\rm eff}(\F_a)].\eqn(dRdF-dFa)
\eeq

For our phenomenological model, we use a power law form for $\tau(\F_a)$,
writing
\beq
\tau(\F_a) = \tau_0 \left(\F_a\over\F_0\right)^{-\sigma},\eqn(tau-F)
\eeq
where $\tau_0$ is the peak duration for bursts with some fiducial
actual peak flux $\F_0$, and $\sigma$ is the power law index.
If $\sigma>0$, bright bursts have shorter peaks than dim ones; if
$\sigma<0$, dim bursts have shorter peaks than bright ones.  For
$\sigma > 0$, this model
can qualitatively mimic the behavior of a simple physical model:
a homogeneous standard candle distribution of bursters that are also
``standard clocks.''  For such a model, dim bursts must originate from
larger redshifts than bright ones, and will therefore have longer
observed durations.  This is the same qualitative behavior as 
our $\sigma>0$ models.

To complete specification of the model, we let the actual peak flux
distribution be a power law, writing
\beq
{dR \over d\F_a} = A 
   \left(\F_a \over \F_0\right)^{-\gamma_1}.\eqn(dRdFa-pl)
\eeq
We can now use \ceqn(dRdF-dFa)\ to calculate the observable flux
distribution.  For $0 < \sigma < 1$, the result can be written as
\beq
{dR \over d\F} = A' \times
  \cases{
     \left(\F\over\F_\tau\right)^{-\gamma_1},&for $\F \le \F_\tau$,\cr
     {1\over 1-\sigma}
         \left(\F\over\F_\tau\right)^{-\gamma_2},&for $\F > \F_\tau$;\cr
}\eqn(dRdF-pdbpl)
\eeq
where $\gamma_2 = (\gamma_1-\sigma)/(1-\sigma)$, and $\F_\tau$ is the
flux where $\tau(\F_a)=\delta t$,
\beq
\F_\tau = \F_0 \left(\tau_0 \over \delta t\right)^{1/\sigma}.\eqn(Ftau-def)
\eeq
The observable flux distribution is thus a broken power law with a
discontinuity at $\F_\tau$, where the power law index changes from
its low flux value of $\gamma_1$ (the index for the underlying
actual distribution) to its high flux value of $\gamma_2$.  
The cumulative distribution is continuous.

Note that $\F_\tau$ is a decreasing function of $\delta t$ as long
as $\sigma>0$.  In our study of smooth broken power law models above, we found
that the break flux inferred from the 1024~ms data is lower than
that inferred from the 64~ms data, so the positive $\sigma$ regime is
the regime of interest.  Note also that 
\beq
\gamma_2 - \gamma_1 = {(\gamma_1 - 1)\sigma \over 1 - \sigma},\eqn(g2g1)
\eeq
so that $\gamma_2 > \gamma_1$ as long as
$\gamma_1 > 1$ and $\sigma < 1$.  
From our earlier analyses, we know $\gamma_1\approx 2$;
thus for this model the flux distribution will steepen at large
fluxes.  This is the behavior exhibited by the data and expected
for observational and theoretical reasons (the PVO data for bright
bursts has a differential distribution like that expected from a
homogeneous Euclidean distribution, $\propto \F^{-2.5}$).  When
$\sigma > 1$, the distribution flattens rather than steepens if
$\gamma_1 > 0$.  Thus the $0<\sigma < 1$ regime described above is
the only regime of interest.

This discontinuous broken power law model has three shape parameters:
$\gamma_1$, $\sigma$, and $\tau_0$ ($\F_0$ is an arbitrary fiducial
flux that we set equal to 1~cm$^{-2}$~s$^{-1}$).  If the differences
between the 64~ms and 1024~ms data are due to simple duration effects
of the kind built into this model, then parameter estimates from
each data set should agree, even though the shapes of the flux
distributions in the two data sets differ.  For example, Figure~28
shows the differential flux distribution for a representative choice of
$\gamma_1$, $\sigma$, and $\tau_0$, plotted for both time scales,
illustrating that the differing values of $\delta t$ lead to
different break locations.

Table 3 lists the best-fit parameter values and model comparison 
statistics for this model ($M_7$).  For the 64~ms data, the best-fit
model has a burst peak duration of $\tau_0=0.16$~s, and
fits the data substantially better than a single power law
model without any duration dependence ($M_1$), although not so much better
that it justifies the additional parameters of this model.  
In contrast, for the 1024~ms data, the timescale is 40~s, and
the improvement of the fit is substantial enough to favor this model
over $M_1$, although not decisively.  

Note that of all the models
studied in this work, this simple model has by far the highest
likelihood.  Figure~29 displays the cumulative flux
distributions for the best-fit models, elucidating the reason
for their large likelihood.  Comparing with Figure 8, 
we find that this model accounts for the data in much the same
way as our broken power law model ($M_3$); but it does so presuming
an intrinsically {\it un}broken power law distribution of burst
fluxes, the apparent break resulting entirely from peak dilution.
The curvature and cutoff at large fluxes in the 1024~ms model
arise because the best-fit model 
has a strong correlation between $\F_a$ and peak duration, the peak duration
being inversely proportional to $\F_a$.  Thus although $\tau_0=40$~s,
the predicted peak duration for the brightest bursts is significantly
smaller than 1~s, so the observed 1024~ms peak flux seriously
underestimates the actual peak flux for bright bursts.

As a test of this feature of this model, we have examined the
publicly available raw OSSE light curves of the five brightest bursts
in the 1024~ms data set (Matz 1996).  These bursts all have peak 1024~ms
fluxes between 20 and 30 ct~s$^{-1}$~cm$^{-2}$, and span the
brief range where the observed flux distribution suddenly
cuts off (see the corresponding region of the cumulative
histogram in Figure 29b).  For one of these bursts, the BATSE and
OSSE detectors triggered on a faint precursor, and the main component
of the burst lies beyond the 60~s segment in the public OSSE
catalog.  In Figure~30 we display the light curves for the remaining 
four bursts; the insets magnify the region of the peak, and include
``root-$n$'' error bars to help distinguish the boundaries of the
peak from mere statistical fluctuations.  Also shown are the BATSE
$T_{50}$ and $T_{90}$ burst duration measures.
For all of these bursts, the peaks are
less than $0.5$~s in duration; for three of them the peaks are
shorter than $0.1$~s.    
Although we have not performed any
rigorous peak fitting, it is clear that the true peak intensity
is larger than a 1024~ms average by factors that could easily be
much larger than one.  
The peak durations show no obvious correlation
with the $T_{50}$ and $T_{90}$ measures of the total burst duration,
preventing any simple rescaling of the peak intensity using the
timescales reported in the 3B catalog.  
The underestimation of peak intensity evident for these bursts
is consistent with our simple model, although
we have not determined whether the data indicate that the true
peak duration is correlated with burst intensity.  A rigorous
analysis of this type requires BATSE light curve and background
data that is not part of the 3B catalog.  A number of investigators have
undertaken analyses of this data in an effort to detect such a
correlation (which is expected for cosmological models), but they
reach conflicting conclusions (cf.\ Mitrofanov 1996, Fenimore 1996, 
and Norris and Nemiroff 1996).

% values of $\gamma_1=1.8$, $\sigma=0.99$ (near the maximum value of 1),
%and $\tau_0=40.1$~s; the maximum likelihood is
%$6\times 10^3$ times larger than that for a single power law.

Our motivation for considering this model was not only to attempt to explain
the cutoff in the distribution of 1024~ms fluxes, but also to attempt to
reconcile the 64~ms and 1024~ms distributions.
The best-fit parameter values for the two data sets
are clearly inconsistent with each
other, and in that sense this model fails to reconcile them.
The inconsistency is not as great as it may appear,
however.  Figure~31 shows contours of the joint posterior distribution
for $\tau_0$ and $\sigma$, conditional on the best-fit values of
the underlying flux distribution power law index, $\gamma_1$.  The
68\% (dashed) and 95\% (solid) contours are shown.  They are
highly structured, the structure resulting from the discontinuity
in the differential rate for this model (which causes the likelihood
to vary greatly as the break point passes through the best-fit flux
value for an observed burst).  The 1024~ms data
constrain $\tau_0$ to have large values.  But the 64~ms data
does not strongly constrain $\tau_0$; small values are preferred,
but the profile posterior is relatively flat, and large values are
thus not strongly ruled out.  Despite the disparity between the best-fit
parameter values, the 95\% credible regions thus have significant regions
of parameter space in common.  

We conclude that although this model is not entirely successful in
explaining the discrepancy between the two data sets, it does
indicate that duration effects can strongly distort the observed
flux distribution (especially on the 1024~ms time scale) in a manner
that can account for the salient features of the data, and that
it is likely that duration effects {\it must} be taken into 
account in order to understand the shapes of the observed flux
distributions.

%===============================================================================
\section{Cosmological Models With Beamed Sources}

The final model we discuss here is a physical model that
combines aspects of many of the models
previously discussed.  It is a cosmological model with a power-law
luminosity function, but it also incorporates temporal and spectral
information about bursts.  It is of intrinsic physical interest, but
it also serves to illustrate how the many characteristics of bursts---not
just their peak fluxes and directions, but also their temporal and
spectral characteristics---can influence an analysis of the 
distribution of peak fluxes from bursts.

We consider a homogeneous distribution of burst sources in a flat
($\Omega_0=1$) cosmology.  We presume the sources are standard candles
and clocks {\it in their rest frame}, but that all the sources are
in relativistic motion with common speed $v\sim c$ (relative to the
rest frame of the cosmic background radiation).  We also presume
that they emit gamma rays isotropically in their rest frame with
a common power-law spectrum.  As a
consequence of their relativistic motion, these sources will appear to
be highly beamed to observers at or near rest with respect to the
cosmic rest frame.  Thus, even though the sources are standard candles
intrinsically, sources at a common redshift will have a {\it distribution}
of apparent luminosities, depending on whether they happen to be beamed
toward or away from an observer.  The shape and width of the
luminosity distribution
depends both on the source speed and on spectral index.
In addition, the sources will appear
to have a distribution of peak durations due to the relativistic
Doppler shift and cosmological redshift.  Since the peak duration affects
the measurement of the peak flux, relativistic beaming can
make the flux distribution 
take on different shapes when measured using different integration time scales.

There are two ways we could proceed to calculate the flux distribution
from this model.  We could generalize \ceqn(dR-cosmo), introducing the beam
direction (with respect to the line of sight) as an additional
integration variable.  The comoving burst rate formally gains the
beam direction as an argument (although the distribution of beam directions
is presumed isotropic).  The $\F_{\rm obs}$ function also gains this
argument, and is complicated by it.  The standard candle
assumption introduces a $\delta$-function in $\lum$ that trivially eliminates
one integration dimension, leaving integrations over beam direction and
redshift.  Of course, $\F_{\rm obs}$ must additionally be modified to take into
account duration effects.

We here adopt an alternative approach that builds upon our existing
intuition about models with luminosity functions; it also leads to
somewhat simpler numerical calculations.  
We use \ceqn(dR-cosmo), but with $\lum$ taken to denote the peak {\it apparent} 
luminosity along the null ray to Earth, that is, the luminosity one
would infer presuming the flux reaching Earth is from an isotropically
emitting source.  The
distribution of beam directions results in an easily calculated
distribution of apparent luminosities.  With this interpretation of
$\lum$, the only modification of  $\F_{\rm obs}$ required is that
needed to account for duration effects.  We denote the apparent
luminosity by $\luma$ to emphasize that it is different from the
actual (rest-frame) luminosity of the burst sources.

In Appendix B we carry out a straightforward exercise in relativistic
kinematics that shows that the apparent luminosity distribution for
a population of beamed sources is a bounded power law,
\beq
f(\luma) \propto \cases{
  \left(\luma\over \lum_0\right)^{-p}, 
       &for $\lum_l < \luma < \lum_u$;\cr
  0, &otherwise.\cr
}\eqn(beam-lfunc)
\eeq
The power law index is related to the spectral index of the sources
according to
\beq
p = {3+\alpha\over 2+\alpha};\eqn(beam-pl)
\eeq
and the lower and upper bounds of the distribution are related to
the rest-frame luminosity, $\lum_0$, according to
\beqa
\lum_l &= \lum_0 \Dop_b^{-(\alpha+2)},\\
\lum_u &= \lum_0 \Dop_b^{\alpha+2}.\eq(beam-bounds)
\eeqa
In these equations $\Dop_b$ is the relativistic Doppler factor along the beam 
direction.  In terms of the beam velocity parameter, $\beta=v/c$, and
$\gamma\equiv 1/\sqrt{1-\beta^2}$, 
\beq
\Dop_b = {1 \over \gamma(1 - \beta)}.\eqn(Dop-def)
\eeq
From \ceqn(beam-bounds), the dynamic range of the power law is
$\Dop_b^{2\alpha+4}$, or $\Dop_b^7$ for $\alpha=1.5$.  For $\beta\approx 1$,
$\Dop_b \approx 2\gamma$.  Thus for beaming with even moderate values
of $\gamma$, a very broad luminosity function results.

Since the luminosity function is a power law, in the absence of duration
effects this model is identical to the cosmological model with power law
luminosity function considered above, with two exceptions.
First, the power law range is
parameterized in terms of the physical parameters
$\lum_0$ and $\Dop_b$, rather than by its upper limit and dynamic range.
More importantly, the power law index
is determined by the burst spectrum, and thus is not a free parameter
to be inferred separately from the shape of the peak flux distribution.
Some of these features of a population of beamed sources have been invoked in
analyses of active galaxies (Urry and Shafer 1984, Urry and Padovani 1991).  
Krolik and Pier (1991) noted other benefits
of beaming for modeling burst sources, and Yi (1993, 1994) performed
a rough statistical analysis of early BATSE data using models with
beamed sources.

Since the sources are standard candles and clocks all moving with the
same speed, the observed duration of any particular source 
is a deterministic function of $\luma$.
Put another way, $\luma$ is a measure of the beam direction, and thus
can be used to specify the Doppler shift.  In Appendix B we show that
the observed peak duration of a beamed source with apparent luminosity
$\luma$ at redshift $z$ is
\beq
\tau(\luma,z) = \tau_0  (1+z)
     \left(\luma\over \lum_0\right)^{1-p},
     \eqn(tcrf-def)
\eeq
where $\tau_0$ is the peak duration in the source's rest frame.

With the apparent luminosity and duration of beamed sources specified,
we can now specify a model based on a population of such sources.
We will calculate the differential burst rate using \ceqn(dR-cosmo),
but with $\lum$ replaced by the apparent luminosity $\luma$, and
with $\Fobs(z,\lum,\spars)$ generalized to be a function of $\luma$
and to take into account durations effects.
We take the comoving burst rate density to be constant with redshift,
so that $\dot n_c(z,\lum) = \dot n_0 f(\luma)$.  The observed flux
function is
\beq
\Fobs(z,\luma,\spars) = \cases{
    {\luma \over 4\pi (1+z) d^2(z)} K_0(z,\spars), 
       &if $\tau(\luma,z) \ge \delta t$,\cr
    {\luma \over 4\pi d^2(z)} K_0(z,\spars)
     \left(\luma\over \lum_0\right)^{1-p} {\tau(\luma) \over \delta t},
       &if $\tau(\luma,z) < \delta t$.\cr
}\eqn(Fobs-tau)
\eeq
The shape parameters of this model are $\lum_0$, $\Dop_b$ (determined
by $v$), and $\tau_0$.
As noted above, the power law index $p$ is a function of the spectral
index of bursts, $\alpha$.  We could consider this to be a free parameter
(since it influences the shape of the observed flux distribution), but
instead we simply take $\alpha=1.5$, the value we adopted for the
cosmological models discussed above (and the value adopted by the BATSE
team for their detection efficiency calculations).  If spectral indices
were available for each burst, we could use that information.
In fact, this model
predicts that the flux distributions of bursts with different $\alpha$
will have different slopes at low $\F$, so that including such information
in the analysis could strengthen or weaken our preference for this
model.  But such information is not available in the 3B catalog.

Evaluating the integral in \ceqn(dR-cosmo)\ for these beamed models
is significantly more complicated than for the cosmological models
considered earlier.  The $\Fobs(z,\luma,\spars)$ function has two cases,
depending in a nontrivial way on $\F$, $\luma$, and $z$.  Breaking
the integral into a sum of integrals separating each case as a function
of $\F$ is tedious but straightforward.  The resulting differential
rate resembles that of the power-law luminosity function models of
\S~5.3, except that there is a ``kink'' where the logarithmic slope
of the rate changes from $p$ at low fluxes to $2.5$ at large fluxes.
At fluxes below the break, the rate is somewhat elevated due to
bursts with actual peak fluxes larger than the break flux but with brief
peaks being displaced to below the break.
The location and shape of the kink depends on $\delta t$, so that
the same underlying distribution of true peak fluxes will produce
observed distributions with differing shape for data obtained with
different trigger time scales.

To study this model, which we denote by $M_8$, we fixed the
line-of-sight Doppler factor to $\Dop_b=4$.  As just noted, this parameter
specifies the dynamic range of the apparent luminosity function, much
as does the $\rho$ parameter for our simple cosmological model with
a luminosity function ($M_6$).  Recall that $\rho$ is
essentially unconstrained by the data (see Figure 19); the same
holds true for $\Dop_b$.  Taking $\Dop_b=4$ results in an apparent
luminosity function with a dynamic range $\approx 1.6\times 10^4$,
about equal to the $\rho=10^4$ value we adopted in our analysis
of $M_6$.  We have verified that varrying $\Dop_b$ does not greatly
affect inferences of the remaining shape parameters, $\lum_0$ and
$\tau_0$.  As with our earlier cosmological models, we parameterize
the luminosity with a dimensionless parameter, $\nu_0$, defined by
\beq
\lum_0 = \nu_0\; {4\pi c^2 \over H_0^2 F(0)} \F_{\rm fid}.\eqn(nu-beam)
\eeq

Table 3 lists the best-fit parameters and model comparison statistics
for this model.  The Bayes factors indicate ambivalence between
this model and a simple standard candle model ($M_4$).  The best-fit
parameter values based on the 64~ms and 1024~ms data are discrepant.
Figure~32 shows contours of the joint posterior for $\nu_0$ and
$\tau_0$.  The posterior is bimodal.  For $\tau_0>\delta t$, there
is an uncorrelated ridge in the posterior.  In this part of parameter
space, there is no peak dilution, so the models correspond to
simple cosmological models with a power law luminosity function of
fixed slope ($p\approx 1.3$).  This ridge thus corresponds to a
$p=1.3$ ``slice'' of the posterior plotted in Figure~20 (with
the $\nu_u=1$ corresponding to $\nu_0=1/128$).
For $\tau_0<\delta t$, peak dilution leads to a strong correlation
between the inferred luminosity and duration.

Not surprisingly, the best-fit 64~ms model lies in the ``simple''
part of parameter space (recall that this data is well-modelled by
a single power law), and the best-fit 1024~ms model lies in 
the part of parameter space where duration effects are important.
Figure~33 shows the cumulative flux distributions for the best-fit
models, and shows that the duration effects increase the slope of
the best-fit 1024~ms flux distribution at large fluxes.  But this
model is not capable of producing as drastic a change in slope as
is the phenomenological model ($M_7$), and so is not as strongly 
favored.

Despite the disparate best-fit parameter values, the credible
regions calculated from the two data sets have significant overlap
for rest frame durations $\tau_0\gtrsim 2$~s.  Unfortunately,
this is in the ``simple'' region of parameter space, corresponding
to the overlapping parts of the credible regions for the
luminosity function model plotted in Figure~20.  Thus the duration
effects incorporated in this beamed ``top hat'' model do not account for the
differing shapes of the 64~ms and 1024~ms flux distributions.
The model remains of interest, however, both as an interesting
physical model for burst sources, and as an illustration of how
both duration and spectral information can be incorporated into
analyses of the flux distribution.

%===============================================================================
\section{Summary and Discussion}

We have analyzed the 64~ms and 1024~ms peak flux data in the 3B catalog
using the Bayesian method described in detail by Loredo and Wasserman (1995).
The method identifies several shortcomings of the summaries of the data
comprising the 3B catalog that prevent consistent analyses of the entire
catalog.  In particular, counting uncertainties and atmospheric scattering
were omitted from the calculation of the detection efficiencies reported
in the 3B catalog, requiring that the dimmest 38\% of the 64~ms bursts, and
the dimmest 16\% of 1024~ms bursts, be omitted from any analysis of
the peak flux distribution.

We have used the resulting self-consistent data sets to analyze a
variety of phenomenological and physical (cosmological) models for
burst sources that presume burst sites are distributed isotropically.
A companion paper presents analyses of anisotropic models that
associate some or all bursts with an extended Galactic halo.

%-------------------------------------------------------------------------------
\subsection{Simple Phenomenological Models}

Our analysis of phenomenological models based on power laws
and broken power laws verifies that neither the 64~ms nor the 1024~ms
data is consistent with a homogeneous (Euclidean) distribution of sources,
for which the differential burst rate obeys $dR/d\F \propto \F^{-5/2}$.
There is no significant evidence for a break in the logarithmic slope
of the distribution of 64~ms peak fluxes, but there is moderately
significant evidence for such a break to the homogeneous $\gamma=2.5$
slope in the 1024~ms data, and stronger evidence for a steep cutoff
in the distribution of bursts with 
$\F_{1024} \gtrsim 40$~cm$^{-2}$~s$^{-1}$.
Also, the power law indices that best describe
the low flux portion of each data set differ with at least moderate
significance, the low-flux distribution of 1024~ms peak fluxes being
somewhat flatter than those favored for the 64~ms peak fluxes.
The different inferred shapes
of the two data sets are not simply due to their different sizes and dynamic
ranges.  This argues that a full understanding of the shape of the observed
peak flux
distribution requires explicit consideration of the temporal structure of
burst light curves.  A simplified analysis (summarized below) indicates
that the structure in the 1024~ms flux distribution is an artifact
of its longer measurement time scale, so that the shape of the 64~ms
flux distribution is more representative of the shape of the distribution
of instantaneous peak fluxes.

Several physical models for the distribution of burst sources in space and 
luminosity predict flux distributions that are well approximated by
power laws and broken power laws, as noted in \S~4.
Quite generically, information about characteristic length and luminosity
scales in such models is revealed by a change in the logarithmic slope
of the flux distribution from a relatively flat differential distribution for
dim bursts to a steeper $\F^{-5/2}$ distribution for bright bursts.
That there is no evidence for such a change
in the 64~ms data, and only moderate evidence for a change to a
$\F^{-5/2}$ power law in the 1024~ms data, presaged the conclusions
we found in our analyses of physical models:  the data are unable
to constrain properties of cosmological populations of burst sources.

%-------------------------------------------------------------------------------
\subsection{Simple Cosmological Models}

We analyzed three simple cosmological models in an effort to determine
whether the 3B data could detect or rule out evolution of the burst
rate density with redshift, and whether the data could constrain the
width of the burster luminosity function.  The data are unable to
discriminate among homogeneous standard candle models and models with
strong density evolution or broad luminosity functions.  As a consequence,
the luminosity of burst sources is uncertain over many orders of
magnitude, and the typical redshifts of observed bursts can be as
small as a few tenths or $\gtrsim 20$.  The upper limit could
almost certainly be reduced by considering PVO data, since it depends
on locating the flux where the flux distribution steepens to 
$dR/d\F \propto \F^{-5/2}$.  A stronger constraint may arise from
the absence of large time dilation in burst lightcurves, although
the wide variety of temporal behavior exhibited by bursts severely
complicates the modeling and detection of such time dilation
(see, e.g., Mitrofanov 1996, Fenimore 1996, and Norris and Nemiroff 1996,
who reach conflicting conclusions regarding the presence of 
``time stretching'' in BATSE data).  A lower limit on the redshift
of observed sources may be sought more effectively from the absence of
anisotropy in the distribution of burst directions (as would appear
if many bursts were visible from within the local supercluster,
for example) than
from the flux distribution (see, e.g., Quashnock 1996).  Since such
anisotropy should correlate with burst intensity, our Bayesian
methodology is an ideal tool for rigorously studying it.

Even in the absence of 
strong density evolution (in which case the observed bursts have
typical redshifts of a few tenths), the width of the luminosity 
function for burst sources is unconstrained and could span several
orders of magnitude.  Unfortunately, we find that the uncertainty in the 
luminosity of the brightest burst sources is comparable in size to the
range of the luminosity function, and thus is largely
unconstrained by the BATSE data.  The addition of PVO data is
unlikely to strengthen this constraint, because models with luminosity
functions of very different widths distinguish themselves at
low fluxes rather than at large fluxes.  
Without a vastly larger data set, the best hope
for constraining the width of the luminosity function of cosmological
burst sources is to obtain data on the distribution of bursts with fluxes
well below the threshold of the BATSE detectors.

\clearpage
%-------------------------------------------------------------------------------
\subsection{Consideration of Temporal\\ and Spectral Properties of Bursts}

Analyses with sufficiently flexible models reveal systematic differences
between the shapes of the distributions of 64~ms and 1024~ms peak fluxes.
Parameters inferred from the two data sets differ with moderate
significance presuming they are independent.  But since the
two data sets are not independent (well over half of the 1024~ms bursts
triggered on the 64~ms timescale) one would expect close agreement between
the inferred values; the discrepancy between the inferences is thus
probably very significant.
In addition, the normalizations of the two distributions are extremely
different, the number of 1024~ms bursts with fluxes above
1.5~cm$^{-2}$~s$^{-1}$ (the cutoff for 64~ms bursts) being only 56\% of
the number of 64~ms bursts.  In LW95 we argued that explicit consideration
of the temporal properties of bursts would be necessary for understanding
the distribution of measured burst fluxes.  The disparity between the
64~ms and 1024~ms data supports this claim.

We therefore attempted to model the data in a manner that crudely accounts
for ``peak dilution'':  the underestimation of peak flux that occurs when
estimating peak flux with data from a time interval longer than the peak
duration.  If peak duration is correlated with peak flux, peak dilution
can result in an observed peak flux distribution that is different in shape
from the underlying actual peak flux distribution.  Unfortunately, the
3B catalog contains no direct information about the peak durations of
bursts.  Thus we have been able to perform only illustrative calculations
that show how explicit consideration of temporal properties of bursts
might enter an analysis of the flux distribution.  Should peak duration
measurements become available, more reliable and definitive analyses
will be possible.

We analyzed a purely
phenomenological model in which bright bursts were presumed to have shorter
peak durations than dim bursts (the qualitative behavior expected in
cosmological models).  Thus bright bursts have their peak fluxes
systematically underestimated when long measuring time scales are used,
steepening the flux distribution at bright fluxes.
This model is moderately successful in reconciling
the shapes of the two data sets, and in particular is capable of producing
a cutoff in the distribution of observed peak fluxes, as is seen in
the 1024~ms data.  Additionally, we analyzed a physical model
in which a cosmological population of relativistically beamed sources
that are standard candles and clocks produces an apparent distribution of
sources with a broad luminosity function and distribution of peak
durations, due to the distribution of the angle between the source velocity
and the line of sight.  Besides correlating burst duration and peak flux,
this model also correlates the burst spectrum with peak flux
and duration.  However, it does not successfully account for the differences
between the two data sets.  Nevertheless, it is of intrinsic physical
interest, and further, it is 
the simplest model illustrating how the spectral and temporal properties of 
bursts can enter the analysis of the flux distribution.

We thus conclude that the BATSE peak flux data cannot usefully
constrain cosmological models for burst sources.  Useful constraints
from peak fluxes alone will result only from consideration of
data about the infrequent bright bursts that BATSE has not yet seen
(to constrain the redshifts of the most distant observed sources),
and about bursts dimmer than BATSE is capable of seeing (to constrain
the width of the luminosity function).  Joint analyses of
temporal and spectral properties of bursts with their peak fluxes
may well provide more useful constraints on cosmological models.
The Bayesian methodology adopted here is the ideal tool for such a joint
analysis.

%===============================================================================
\acknowledgments

This work was supported in part by NASA grants
NAG 5-1758, NAG 5-2762, NAG 5-3097, and NAG 5-3427; 
and by NSF grants AST91-19475 and AST-93-15375.

%===============================================================================
\appendix
\section{Derivation and Properties of the Cosmological Burst Rate}

We present here a detailed derivation of the expressions referred to in the text
for calculating the differential burst rate from cosmological models
(eqn.~\ceq(dR-cosmo)\ and the various functions that appear in that equation).
Much of our treatment follows that of Weinberg (1972); the most important
differences between our analysis and his result from our interest in sources
detected by measurements of peak photon flux, rather than energy flux or 
fluence, and from our interest in inferring a rate density rather than
a number density.
We also derive here expressions for the distribution of
observable sources with redshift and luminosity, 
and we find expressions for the logarithmic
slope of the differential rate in various regimes.
We use these latter expressions in the main text to motivate our study of
power laws and to explain the shape of some of our
posterior distributions.  

Our task is to generalize \ceqn(dR-ndot)\ to a cosmological setting.  
This involves
finding an expression for the volume element in terms of convenient
coordinates for events in spacetime, and taking into account cosmological
effects in expressions for the burst rate density, and for
the observed peak flux from a cosmological source.  We address these
tasks in turn.

\subsection{The Volume Element}

Our starting point is the Robertson-Walker metric, which specifies the
interval of proper time, $d\tau$, between events separated by an
infinitesimal cosmic time interval $dt$ and by
infinitesimal comoving spherical coordinate intervals $dr$, $d\theta$, and 
$d\phi$:
\beq
d\tau^2 = dt^2 - {a^2(t)\over c^2} \left[{dr^2 \over 1-kr^2} + 
     r^2(d\theta^2 + \sin^2\theta d\phi^2)\right].\eqn(metric)
\eeq
The time dependence of the scale factor, $a(t)$, will be determined by the 
Einstein equations.
We assign units of length to $a(t)$, so that the comoving coordinate $r$
is a dimensionless coordinate label.  The dimensionless constant $k$ is
the curvature constant whose sign determines the sign of the spatial
curvature of three-dimensional spaces of constant $t$.  We take the
origin of the coordinate system ($r=0$) to be at Earth.

Equation \ceq(dR-ndot)\ requires the volume element for three-dimensional
spaces of constant $t$.  The metric for such spaces can easily be read
off of \ceqn(metric); the square root of its determinant gives
\beq
dV(r,\theta,\phi) 
  = {a^3(t) \over \sqrt{1-kr^2}} r^2 dr d\drxn, \eqn(volume)
\eeq
where $d\drxn = \sin\theta d\theta d\phi$ is the familiar solid
angle element in spherical coordinates.  Besides the three spatial coordinates,
the cosmic time, $t$, appears in $dV$ in the scale factor.  It must be
set equal to the coordinate time of the
observed event.  Since we are interested only in events observed with
light, there is a one-to-one correspondence between $t$ and $r$
determined by the condition that events be connected to the origin
by radial null rays.  In addition, it proves most convenient to parameterize
such rays by the redshift, $z$, rather than by $t$ or $r$, where
$z$ is given by
\beq
1 + z = {a_0 \over a(t)},\eqn(z-def)
\eeq
where $a_0 = a(0)$.
For null rays, \ceqn(volume)\ thus can be written as
\beq
dV(z,\theta,\phi) = c a_0^2 { r^2(z) \over (1+z)^3 H(z)} 
     dz d\drxn,\eqn(dV-z-1)
\eeq
where $r(z)$ is the radial coordinate of a source at redshift $z$, and
$H(z)$ is the Hubble factor defined by $H(z) = \dot a(z)/a(z)$.

To calculate $H(z)$ and $r(z)$ we need to solve the Einstein equations for the
evolution of the scale factor.  The energy-momentum tensor in
a Robertson-Walker universe necessarily takes the form of that of a perfect
fluid, and thus can be characterized by the fluid density $\rho$
and pressure $p$.  The Einstein equations based on this metric
and energy-momentum tensor
reduce to two second order differential equations relating $\rho(z)$,
$p(z)$, and $a(t)$.  
Since we are concerned
with events at epochs when the universe is matter-dominated, we
simply set $p=0$.  We also set the cosmological constant equal to zero
throughout this paper.  
The Einstein equations can then be easily solved, giving
\beq
H(z) = H_0 (1+z) \sqrt{1 + 2 q_0 z},\eqn(H-z)
\eeq
where $H_0 =H(0)$ is the Hubble constant and $q_0$ is the deceleration
constant.  The deceleration constant is related to the density at
the current epoch, $\rho_0$,
according to $\Omega_0 = 2 q_0$, where $\Omega_0$ is the ratio
of $\rho_0$ to the critical density corresponding to a $k=0$ universe,
\beq
\Omega_0 = {8 \pi G \rho_0 \over 3 H_0^2}.\eqn(O-def)
\eeq
Since $H(z)$ specifies the redshift dependence of the scale factor,
we can use it to evaluate $r(z)$ (Weinberg 1972,
eqn.~15.3.23).  The result is
\beq
r(z) = {c \over H_0 a_0} {1 \over q_0^2 (1+z)}
  \left[q_0 z + (1-q_0)\left(1-\sqrt{2q_0 z+1}\right) \right].\eqn(r-vs-z)
\eeq
With $H(z)$ and $r(z)$ now available, 
the volume element finally takes the form,
\beq
dV(z,\theta,\phi) = {c a_0^2 \over H_0} 
         { r^2(z) \over (1+z)^4 \sqrt{1 + 2 q_0 z}} 
     dz d\drxn.\eqn(dV-z)
\eeq

%-------------------------------------------------------------------------------
\subsection{The Burst Rate}

The differential rate integrand includes the burst rate density,
denoted $\dot n(\rvec,\lum)$ in \ceqn(dR-ndot).  In a cosmological setting,
where there is a difference between proper volume and comoving volume,
and between local time intervals and time intervals observed at large
redshift, care must be taken in defining the burst rate. 

The observed time interval, $dt$, between events at redshift $z$
separated by a time interval $dt'$ is given by $dt = (1+z)dt'$.
Let $\dot n_p(z,\lum)$ denote the proper burst rate density, so that
the burst rate measured by observers
at redshift $z$ due to sources in a volume $dV$
and luminosity interval $d\lum$ at that redshift is
$\dot n_p(z,\lum) dVd\lum$.  Then the {\it apparent} burst rate from that
volume and luminosity interval, as measured by observers at $z=0$,
will be $\dot n_p(z,\lum) dV d\lum / (1+z)$.  Thus a factor of $1/(1+z)$
must be inserted into the integrand of \ceqn(dR-ndot)\ when cosmological
sources are considered.  This factor was neglected by Wickramasinghe
et al.\ (1993) and by Horack, Emslie, and Meegan (1994), who
used results appropriate for number densities rather than rate densities.

In addition, it is more natural to specify the burst rate per unit
{\it comoving} volume element (i.e., per unit volume element expanding
with the separations between galaxies) than per unit {\it proper}
volume element.  The  comoving burst rate density, $\dot n_c$,
is related to the proper burst rate density
according to $\dot n_c = \dot n_p a^3 = \dot n_p a_0^3/(1+z)^3$.
Note that since the comoving radial coordinate $r$ is dimensionless,
$\dot n_c$ has units of the product of inverse time and inverse luminosity, 
with no inverse volume dimensions.

These considerations, combined with the volume element 
calculated above, lead to a cosmological counterpart
to \ceqn(dR-ndot)\ that we can trivially integrate over $\drxn$
(presuming an isotropic burst rate density)
to obtain the counterpart to \ceqn(dR-iso),
\beq
{dR \over d\F} = {4\pi c \over H_0 a_0} \int dz \int d\lum\;
     { r^2(z) \over (1+z)^2 \sqrt{1 + 2 q_0 z}} \,
   \dot n_c(z,\lum) \, \delta[\F - \Fobs(\lum,\spars,z)]. \eqn(dR-zl-1)
\eeq
Here $\Fobs(\lum,\spars,z)$ is a function specifying the peak flux one would
observe from a source with peak photon luminosity $\lum$ at redshift $z$,
with a spectrum described by spectral parameters $\spars$.
The $4\pi$ factor out front came from performing the integration over 
$d\drxn$ (ignoring any apparent and, presumably, slight anisotropy 
induced by motion with respect to the cosmic rest frame).

%-------------------------------------------------------------------------------
\subsection{The Peak Flux}

To evaluate \ceqn(dR-zl-1), we must specify 
$\dot n_c(z,\lum)$ and $\Fobs(\lum,\spars,z)$.  The former depends on
the burst source model, but the latter depends primarily on the cosmology and
burst spectrum (it will also depend on the light curve, as noted in the text
and in Appendix B).
We evaluate $\Fobs(\lum,\spars,z)$ for bursts with power-law spectra here.

We will calculate $\Fobs$ simply by requiring that the number of photons that
pass through spherical surfaces centered at a burst site
be conserved.  Let $A$ denote the area of a spherical wavefront just reaching 
Earth from a source at redshift $z$.  We define the distance $d(z)$ traveled by 
the wavefront by writing
$A = 4\pi d^2(z)$; as shown by Weinberg (1972), this implies that
$d(z) = R_0 r(z)$.

Let $\lum dt'$ denote the number of photons emitted by the source in
a pulse of duration $dt'$ in its rest frame.  Focus attention on a
small group of the photons with rest-frame energies $\eps'$ in the interval
$[\epsilon',\epsilon'+d\epsilon']$, and let the fraction of
all photons with energies in this range be given by 
$\phi(\epsilon') d\epsilon'$.  The $\phi(\epsilon')$ function describes
the shape of the burst spectrum, and must be normalized so that
$\int d\epsilon'\phi(\epsilon') = 1$.

Now consider this pulse of photons as it reaches a sphere of radius $A$
at a redshift $z$ from the source.
Let $\xi(\eps)A d\eps dt$ denote the number of photons that pass through
the sphere in a time $dt$ and with energies in $[\epsilon,\epsilon+d\epsilon]$,
with all quantities measured on the sphere.  The function $\xi(\eps)$
is thus the photon number flux per unit energy through the spherical
surface.  Now focus attention on
the group of photons just described above.  Any such photon that
started with rest-frame energy $\epsilon'$ will be detected with
energy $\epsilon = \epsilon'/(1+z)$; and the energy interval spanned
by the photons will be $d\epsilon = d\epsilon'/(1+z)$.  Also, the
time spanned by the pulse will be $dt = (1+z)dt'$.  Requiring the
number of photons in these energy and time intervals to equal the
number emitted in the corresponding intervals at the source implies
\beq
4\pi d^2 \xi(\eps) d\eps dt = \lum \phi(\eps') d\eps' dt'.\eqn(xi-phi)
\eeq
Solving for $\xi(\eps)$, and casting all quantities in terms
of those measured at the sphere gives
\beq
\xi(\eps) = {\lum \phi[\eps(1+z)] \over 4 \pi d^2(z)}.\eqn(xi-def)
\eeq
The photon number flux measured by a detector with an energy-dependent
detection efficiency
$k(\eps)$ is $\Fobs = \int_0^\infty d\eps k(\eps) \xi(\eps)$.  Using \ceqn(xi-def),
and transforming the integral from $\eps$ to $\eps'$, we can write
this as
\beq
\Fobs(\lum,\spars,z) = {\lum \over 4 \pi (1+z) d^2(z)}
  K_0(z,\spars),\eqn(Fobs-def)
\eeq
where the $1/(1+z)$ factor arose from changing the integration variable
from $\eps$ to $\eps'$; and the spectral correction function $K_n(z,\spars)$ is
given by
\beq
K_n(z,\spars) =  {1 \over (1+z)^n} 
       \int_0^\infty d\eps' \, (\eps')^n k[\eps'/(1+z)]\phi(\eps').\eqn(Kn-def)
\eeq
This function specifies the gamma ray burst counterpart to the $K$-corrections
familiar from analyses of optical observations of cosmological sources.
Note that it depends not only on redshift, but also on the shape of the
burst spectrum and the efficiency function of the detector.  We have
defined it with a general index, $n$, for the sake of generality.
Had we been concerned with observations of a burst's {\it energy} flux
$F$, rather than its photon number flux, then the observed energy
flux would be $F_{\rm obs} = \int d\eps k(\eps) \eps \xi(\eps)$.  
The energy flux counterpart of \ceqn(Fobs-def)\ is then
\beq
F_{\rm obs}(\lum,\spars,z) = {\lum \over 4 \pi (1+z) d^2(z)}
  K_1(z,\spars).\eqn(EFobs-def)
\eeq
The energy flux thus has a different dependence on both $z$ and $\spars$ than
does the photon number flux.  Other measures of burst intensity (such as
the total (time-integrated) photon number per unit area, or the fluence) have 
yet different dependences (e.g., time integrals introduce further
$(1+z)$ factors when one transforms from the detector frame to the
source frame).

In this work, we calculate $K_0$ using a detection efficiency that
is constant (equal to unity) from $\eps_1$ to $\eps_2$ and that vanishes
outside this range.  We set $\eps_1=60$~keV and $\eps_2=300$~keV, the
nominal energy boundaries of the BATSE detectors during the observations
comprising the 3B catalog.  This ``top hat'' efficiency function
results in the integral over $\eps'$ in
\ceqn(Kn-def)\ having a $z$-dependent range, extending from
$\eps_1(1+z)$ to $\eps_2(1+z)$.
We also use a power-law photon spectrum,
with $\phi(\eps') \propto (\eps')^{-\alpha}$, with the spectrum extending
from $\eps'_l$ to $\eps'_u$.  The spectral parameters
are thus $\spars = \{\alpha,\eps'_l,\eps'_u\}$.  The resulting
spectral correction function is
\beq
K_0(z,\spars) = \cases{
  {\log(\eps_2/\eps_1) \over \log(\eps'_u/\eps'_l)}, &for $\alpha=1$,\cr
  \left(\eps_2\over \eps'_u\right)^{1-\alpha} (1+z)^{1-\alpha}
      {(\eps_2/\eps_1)^{\alpha-1} - 1 \over
             (\eps'_u/\eps'_l)^{\alpha-1} - 1}, &for $\alpha \ne 1$.\cr
}\eqn(K0-def)
\eeq
This form applies only for redshifts close enough that $\eps'_u$ is not
redshifted to within the detector passband; this assumption is justified
in that no bursts exhibit sharp spectral breaks within the passband.
Note that, for $\alpha=1$, the spectral
correction function is independent of $z$, and that it depends only
weakly on the spectral parameters $\eps'_u$ and $\eps'_l$.
We set $\eps'_l = 60$~keV, and $\eps'_u=X$~keV.

We now have all the ingredients needed to calculate the burst rate
according to \ceqn(dR-zl-1).
We can facilitate the differential rate calculation
by rewriting the $\delta$-function in \ceqn(dR-zl-1)\ 
so that its argument is an expression
for one of the integration variables, rather than for $\F$. 
For standard candle models, the burst rate density includes a
$\delta$-function in $\lum$, so the appropriate transformation
is from $\F$ to $z$ (so as to avoid a product of $\delta$-functions
with the same argument).  Let $z'(\F,\lum,\spars)$ be the redshift value
that solves \ceqn(Fobs-def)\ when the value of
the left hand side is given as $\F$.  
This function must be calculated
by numerically solving \ceqn(Fobs-def).  In terms of $z'(\F,\lum,\spars)$,
\beq
\delta\left[\F-\Fobs(\lum, \spars, z)\right]
 = \delta[z - z'(\F,\lum,\spars)]\;
  \left|{d\Fobs \over dz}\right|^{-1}.\eqn(delta-z-1)
\eeq
We can calculate the derivative from \ceqn(Fobs-def), giving
\beq
{d\Fobs \over dz} = \Fobs(\lum,\spars,z) 
   \left( {K'_0(z,\spars) \over K_0(z,\spars)} - 2 {r'(z)\over r(z)}
     - {1 \over 1+z}\right),\eqn(F-deriv)
\eeq
where $r'(z)$ and $K'_0(z,\spars)$ denote the derivatives of $r(z)$
and $K_0(z,\spars)$ with respect to $z$, which we can easily calculate
from their defining expressions above.  With these results, and a
standard candle luminosity of $\lum_c$, the
differential rate integral of \ceqn(dR-zl-1)\ becomes
\beq
{dR \over d\F} = {4\pi c \over H_0 a_0} \;
     { r^2(z') \over (1+z')^2 \sqrt{1 + 2 q_0 z'}} \,
   \dot n_c(z') \, 
  {1 \over \Fobs(\lum_c,\spars,z')
       \left( {K'_0(z',\spars) \over K_0(z',\spars)} - 2 {r'(z')\over r(z)}
     - {1 \over 1+z'}\right)}, \eqn(dR-sccosmo)
\eeq
where $z'$ is everywhere equal to $z'(\F,\lum_c,\spars)$, and
$\dot n_c(z)$ is the burst rate density without the $\delta$-function
luminosity factor.
 
For models with nondegenerate luminosity functions, calculations
may be facilitated by instead transforming from $\F$ to $\lum$,
in which case
\beq
\delta\left[\F-\Fobs(\lum, \spars, z)\right]
  = {4\pi (1+z) d^2(z) \over K_0(z,\spars)} \;
     \delta\left[\lum - {4\pi d^2(1+z)\F \over K_0(z,\spars)}\right].
      \eqn(delta-lum)
\eeq
Using this in \ceqn(dR-zl-1)\ gives this version of the differential
rate equation,
\beq
{dR \over d\F} = {16\pi^2 c a_0 \over H_0} \int dz \;
     { r^4(z) \over K_0(z,\spars) (1+z) \sqrt{1 + 2 q_0 z}}\, 
     \dot n_c(z,\lum'),      \eqn(dR-plcosmo)
\eeq
where $\lum'$ is given by
\beq
\lum'(\F,z,\spars) = {4\pi d^2(1+z)\F \over K_0(z,\spars)}.\eqn(lump-def)
\eeq

%-------------------------------------------------------------------------------
\subsection{Redshift and Luminosity Distributions of Observable Sources}

From the results already derived, it is straightforward to calculate
the redshift distribution of sources visible from Earth.  We first
note that the differential rate per unit $\F$ per unit $z$ is
available from inspection of \ceqn(dR-zl-1); we need only omit
the integration over $dz$ on the right hand side.  Integrating
the resulting expression over $\F$ is trivial due to the 
$\delta$-function; the result is
\beq
{dR \over dz} = {c \over H_0 a_0} \,
     { 4\pi r^2(z) \over (1+z)^2 \sqrt{1 + 2 q_0 z}} 
    \int d\lum\; \dot n_c(z,\lum). \eqn(dRdz-1)
\eeq
If we write $\dot n_c(z,\lum)$ as the product of the burst rate
per comoving volume, $\dot n_c(z)$, and a normalized conditional
luminosity function, $f(\lum\mid z)$, we can also easily perform
the integral over luminosity, giving
\beq
{dR \over dz} = {c \over H_0 a_0} \;
     { 4\pi r^2(z) \over (1+z)^2 \sqrt{1 + 2 q_0 z}} \,
   \dot n_c(z). \eqn(dRdz-2)
\eeq
This gives the redshift distribution visible to a {\it perfect} detector.
The {\it effective} differential rate---that visible to a detector with
limited detection efficiency---is simply $\bar \eta'(\F) dR/d\F$.
Repeating the above calculations, we find that the redshift
distribution of bursts visible to a real detector is given by
\ceqn(dRdz-1), but with a factor of $\bar \eta'[\Fobs(\lum,z,\spars)]$
inserted on the right hand side.  The luminosity integral no longer
gives a simple, general result.  But for standard
candle models with luminosity $\lum_c$, the integral can be
performed, giving a result like \ceqn(dRdz-2), but with a
factor of $\bar \eta'[\Fobs(\lum_c,z,\spars)]$ appearing on the
right hand side.  The efficiency function has the
effect of truncating the distribution at large $z$.

We can similarly find the effective luminosity function.  Multiplying
\ceqn(dRdz-1)\ by $\bar \eta'[\Fobs(\lum,z,\spars)]$, interchanging
the roles of $\F$ and $\lum$, and performing the integral over
$\F$ (made trivial by the $\delta$-function) gives
\beq
{dR_{\rm eff} \over d\lum} = {c \over H_0 a_0} \; \int dz\;
     { 4\pi r^2(z) \over (1+z)^2 \sqrt{1 + 2 q_0 z}} \,
   \dot n_c(z,\lum)\,
   \bar \eta'[\Fobs(\lum,z,\spars)]. \eqn(dReffdL)
\eeq
We discuss the properties of this function at the end of \S~5.3.

%-------------------------------------------------------------------------------
\subsection{Limiting Behavior of the Rate}

To understand some of the inferences found in the text, it is useful to 
know the behavior of the logarithmic slope of the differential rate as
as function of $\F$ for various choices of $\dot n_c(z,\lum)$.  We
collect some such results here.

First, if $\dot n_c(z,\lum)$ concentrates observable bursts to redshifts
significantly less than unity, it is clear that cosmological effects
are negligible, so that $dR/d\F$ shares the properties of rates derived
from Euclidean models.  This will be the case if the burst rate per unit
volume is concentrated at low $z$, or if it is spread out but sources
have luminosities that are visible only from low redshifts.

It follows that the brightest bursts (presumably from nearby sources
at $z<1$) should have a differential distribution proportional to
$\F^{-5/2}$.  Of course, the flux corresponding to sources at $z<1$
could be beyond the range of the BATSE data, so that the
$\F^{-5/2}$ regime is not yet evident in the BATSE data (although it
appears to have been detected by PVO; see Fenimore et al.~1993,
and Fenimore and Bloom 1996a).

The dimmest bursts could possibly be due to sources at large redshifts,
so it is interesting to know the behavior of the differential rate when
it is dominated by sources with $z\gg 1$.  Following along the lines of
the discussion at the start of \S~4, we first find the $z$ dependence
of $dR/dz$ at large $z$, then find the $z$ dependence of
$\Fobs(\lum,z,\spars)$, and finally change variables from $z$ to $\F$
to calculate $dR/d\F$.

Begin by noting that, from \ceqn(r-vs-z), we find that
$r(z) \rightarrow c/H_0R_0q_0$, a constant, at large $z$.
Now examine \ceqn(dRdz-1), and take $\dot n_c(z)\propto (1+z)^{-\beta}$,
as we did in the inhomogeneous standard candle model considered
in \S~5.  For $z\gg1$, we find that
\beq
{dR \over dz} \propto z^{-\beta-{5\over 2}}.\eqn(dRdz-lim)
\eeq
To find the large $z$ behavior of the flux, as given by
\ceqn(Fobs-def), we need to know the large $z$ behavior of
$K_0(z,\spars)$.  From \ceqn(K0-def)\ we find $K_0\propto z^{1-\alpha}$.
Since $d(z) \propto r(z)$, we thus find from \ceqn(Fobs-def)\ that
$\F\propto z^{-\alpha}$.  This allows us to change variables
in \ceqn(dRdz-lim), giving
\beq
{dR \over d\F} \propto \F^{{\beta\over\alpha}+{3\over2\alpha}-1}.
    \eqn(dRdF-lim)
\eeq
For $\alpha=3/2$, we
find $dR/d\F \propto \F^{2\beta/3}$, the power law
behavior noted in the main text.

%===============================================================================
\section{Sources With Relativistic Beaming}

In this Appendix we derive two important properties of a population of
standard candle, standard clock beamed sources used in \S~7:  
the power-law distribution of apparent
luminosities, and the relationship between apparent luminosity
and peak duration.

Consider a source emitting photons isotropically in its rest frame
with a photon number luminosity $\lum_0$, and moving with velocity
$\vvec$ with respect to a cosmologically local observer
at rest with respect to the cosmic background radiation (hereafter
the ``local observer'').  
We describe events in the source's rest frame (the ``source frame'') 
and the local observer's frame using
coordinate systems whose origins are
coincident at the moment of the events under consideration.
We will identify rest-frame quantities with a ``0'' subscript, with 
corresponding quantities in
the local observer's frame denoted with a prime.  Quantities measured
at Earth are denoted without a subscript or prime.  We use the
standard symbols $\beta=v/$ and $\gamma=1/\sqrt{1-\beta^2}$
where convenient.

Consider a pulse of photons emitted by the source in a time $dt_0$ and in
a narrow cone of solid angle $d\drxn_0$ along the line of
sight to Earth at an angle $\theta_0$ from the direction of motion
with direction cosine $\mu_0=\cos\theta_0$.  Denote the fraction
of photons emitted
with rest-frame energies in $[\eps_0,\eps_0+d\eps_0]$ by
$\phi(\eps_0)d\eps_0$.  Then the number of photons in the
pulse with energies in $d\eps_0$ is
\beq
dn = {\lum \over 4\pi} \phi(\eps_0) dt_0 d\eps_0 d\drxn_0.\eqn(dn-rest)
\eeq
The local observer sees this pulse over a time interval given
by the Doppler formula as $dt' = dt_0/\Dop(\beta,\mu)$, where
$\beta=v/c$ and $\Dop(\beta,\mu)$ is the Doppler factor,
\beq
\Dop(\beta,\mu) = {1 \over \gamma(1-\beta\mu)}.\eqn(Dop-def-app)
\eeq
In addition, the cone will appear at an angle $\theta'$ with
respect to the direction of motion of the source, with $\mu'=\cos\theta'$
given by
\beq
\mu' = {\mu_0 + \beta \over 1 + \beta \mu_0}.\eqn(mup-mu0)
\eeq
It will subtend a solid angle
$d\drxn' = d\drxn_0/\Dop^2(\beta,\mu)$.  Finally, the photons
will be observed at an energy $\eps'=\Dop(\beta,\mu')\eps_0$,
over an energy interval $d\eps' = \Dop(\beta,\mu')d\eps_0$.
These relationships between source and local observer quantities
lets us rewrite \ceqn(dn-rest)\ as
\beq
dn = \Dop^2 {\lum \over 4\pi} \phi(\eps'/\Dop) 
       dt' d\eps' d\drxn',\eqn(dn-local)
\eeq
where we have temporarily suppressed the arguments of $\Dop(\beta,\mu')$.

At Earth, the pulse will have duration $dt = (1+z) dt'$.  The photons
will have energy $\eps = \eps'/(1+z)$ and will span an energy interval
of size $d\eps = d\eps'/(1+z)$.  If the pulse is observed with a detector
of area $A$ normal to the line of sight, then the solid angle subtended
by the detector in the local observer frame is $d\drxn' = A/d^2(z)$.
Thus we can rewrite \ceqn(dn-local)\ as
\beq
dn = \Dop^2 {\lum \over 4\pi d^2(z)} \phi[\eps'(1+z)/\Dop] 
       A dt d\eps.\eqn(dn-Earth)
\eeq
The flux per unit energy is just $\xi(\eps) = dn/(A dt d\eps)$.
As in Appendix A, the photon number flux measured by a detector
with efficiency $k(\eps)$ is $\Fobs = \int d\eps k(\eps) \xi(\eps)$.
Using \ceqn(dn-Earth), and transforming the integration variable 
from $\eps$ to $\eps'$, this expression becomes
\beq
\Fobs(z,\luma,\spars) = \Dop^2 {\lum_0 \over 4\pi (1+z) d^2(z)} 
  \int d\eps'\; k[\eps'/(1+z)]\, \phi(\eps'/\Dop).\eqn(F-beam)
\eeq

Now define the apparent luminosity of the source, $\luma$, by writing,
in a manner analogous to \ceqn(F-def),
\beq
\Fobs(z,\luma,\spars)
  \equiv {\luma \over 4\pi (1+z) d^2(z)} K_0(z,\spars).\eqn(luma-def)
\eeq
Comparing this with \ceqn(F-beam), and using the definition of
$K_0$ in \ceqn(Kn-def), gives
\beq
\luma = \Dop^2 \lum_0 
   {  \int d\eps'\; k[\eps'/(1+z)]\, \phi(\eps'/\Dop) \over
      \int d\eps'\; k[\eps'/(1+z)]\, \phi(\eps')  }.\eqn(luma-gen)
\eeq
Although the ratio of integrals appearing here is not trivial in
general, for power-law spectra proportional to $(\eps')^{-\alpha}$
we easily find that
\beq
\luma =  [\Dop(\beta,\mu')]^{2+\alpha} \lum_0,\eqn(luma-lum0)
\eeq
provided the lower and upper limits of the spectrum do not enter
the detector passband at the redshifts of interest.

Equation \ceq(luma-lum0)\ reveals the apparent luminosity of
a beamed source to be strongly angle and velocity dependent.
Distributions of beaming angles and velocities thus produce
an apparent luminosity function.  In this work, we presume that
all sources have the same velocity.  The luminosity function is
thus that due to the distribution of beaming angles, which must
be isotropic in an isotropic cosmology.  If we denote the fraction
of sources with apparent luminosity in $[\luma,\luma+d\luma]$ by
$f(\luma)d\luma$, and if the range of beaming angle cosines
corresponding to $d\luma$ is $d\mu'$, then $f(\luma)d\luma=d\mu'/2$,
so that
\beq
f(\luma) = {1 \over 2} \left(d\luma \over d\mu'\right)^{-1}.
  \eqn(fluma-def)
\eeq
Calculating the derivative reveals the apparent luminosity function
to be a power law;
\beq
f(\luma) = \cases{
   {1 \over 2 \lum_0 \gamma \beta (2+\alpha)}
      \left(\luma\over \lum_0\right)^{-{3+\alpha \over 2+\alpha}}
      & for $\lum_l < \luma < \lum_u$,\cr
   0 &otherwise.\cr
}\eqn(fluma)
\eeq
The limits of the power law are
\beqa
\lum_l &= \lum_0 \Dop_b^{-(\alpha+2)},\\
\lum_u &= \lum_0 \Dop_b^{\alpha+2};\eq(luma-bounds)
\eeqa
where $\Dop_b=\Dop(\beta,1)$ is the relativistic Doppler factor along the beam 
direction (i.e., for $\mu'=1$).  For large $\gamma$ (so that $\beta\approx 1$),
$\Dop_b \approx 2\gamma$.  If we had considered a distribution of
velocities, the resulting $f(\luma)$ would consist of a superposition
of powerlaws of the same index but of differing dynamic range, smoothing
the cutoffs at low and high $\luma$.

If the duration of the burst peak becomes smaller than $\delta t$, the
effective peak flux is reduced from the value given by \ceqn(luma-def)\ by
the factor $\tau /\delta t$, where $\tau$ is the peak duration at
Earth.  If the rest-frame duration is $\tau_0$, then beaming and
redshift effects imply $\tau = \tau_0 (1+z) / \Dop$.  Using
\ceqn(luma-lum0), we can rewrite $\Dop$ in terms of $\luma$ to
reveal explicitly how the observed peak duration is correlated with
the luminosity (and through it, the peak flux).  The result is
\beq
\tau(\luma,z) = 
  \tau_0 (1+z) \left(\luma \over \lum_0\right)^{-{1 \over 2+\alpha}},
   \eqn(tau-luma)
\eeq
as quoted in \S~7.

%===============================================================================

\clearpage

%=================================================
% 1
\figcaption{True (solid) and approximate sky-averaged detection efficiencies
for the simulated observations described in the text.  Dashed curve
was calculated ignoring both counting uncertainties and atmospheric
scattering, as was the efficiency reported in the BATSE catalogs.  Dotted
curve incorporates counting uncertainties.  Solid curve additionally
incorporates atmospheric scattering.}

% 2
\figcaption{a--c.  Determination of cutoff flux for self-consistent
analysis of simulated 64~ms data with approximate detection efficiency.
Panels show scatterplots of the logarithm of the ratio of maximum 
likelihood for parameters of a broken power law model to the likelihood
for the true parameter values, calculated using the true detection
efficiency ($\Delta L_{\rm true}$) and the approximate one
($\Delta L_{\rm approx}$), analyzing all bursts (a), and only those
with peak fluxes $>1.2$ and 1.5~cm$^{-2}$~s$^{-1}$ (b, c).
Results are shown for 10 simulated data sets of 400 bursts.}

% 3
\figcaption{Posterior distributions for the power-law index, $\gamma$,
for phenomenological 
power law models, based on 64~ms (solid) and 1024~ms (dashed)
data.  Intersections with the horizontal dotted lines indicate
the 68.3\% (top), 95.4\% (middle), and 99.7\% (bottom) credible
regions.}

% 4
\figcaption{Cumulative peak flux distributions predicted
by best-fit simple power law models based on 64~ms (a) and
1024~ms (b) data.  Histogram shows
cumulative distribution of best-fit peak flux values of
detected bursts.} 

% 5
\figcaption{Joint credible regions for break flux $\Phi_b$ and 
low-flux power-law index $\gamma_1$ in
a simple broken power law model with high-flux power-law
index $\gamma_2\equiv 2.5$,
based on 64~ms (a) and 1024~ms (b) data.  Here and throughout
this work contours enclose
68.3\% (dotted),
95.4\% (dashed), and 99.7\% (solid) of the posterior probability;
crosses indicate best-fit parameter values.
}

% 6
\figcaption{Cumulative peak flux distributions predicted by
best-fit simple broken power law models
with $\gamma_2\equiv 2.5$, based on 64~ms (a) and
1024~ms (b) data.  Histogram shows 
cumulative distribution of best-fit peak flux values of
detected bursts.}

% 7
\figcaption{Joint credible regions for break flux $\Phi_b$ and 
inclination of the logarithmic differential rate at large flux, $\theta$,
conditional on the best-fit values of the low-flux index $\gamma_1$ in
the simple broken power law model,
based on 64~ms (a, $\gamma_1=2.04$) and 1024~ms (b, $\gamma_1=1.83$) data.}

% 8
\figcaption{Cumulative peak flux distributions predicted by
best-fit simple broken power law models, based
on  64~ms (a) and 1024~ms (b) data.  
Histogram shows cumulative distribution of best-fit peak flux values of
detected bursts.}

% 9
\figcaption{Distribution of approximate spectral indices for
burst photon number spectra, based on broadband fluence data.}

% 10
\figcaption{Posterior distributions for dimensionless
luminosity $\nu_c$ for simple,
homogeneous standard candle cosmological models, based
on  64~ms (solid) and 1024~ms (dashed) data.  
Intersections with the horizontal dotted lines indicate
the 68.3\% (top), 95.4\% (middle), and 99.7\% (bottom) credible
regions.}

% 11
\figcaption{Cumulative peak flux distributions predicted by
best-fit homogeneous standard candle cosmological models, based
on  64~ms (a) and 1024~ms (b) data.
Histogram shows cumulative distribution of best-fit peak flux values of
detected bursts.}

% 12
\figcaption{Joint credible regions for dimensionless luminosity $\nu_c$
and comoving burst rate density $\dot n_0$ for simple,
homogeneous standard candle cosmological models, based
on 64~ms (upper contours) and 1024~ms (lower contours)
data.}

% 13
\figcaption{Distribution of redshifts of burst sources
predicted by best-fit homogeneous standard candle cosmological models.
Solid curves (with left axis) show the
redshift distributions of all sources; dashed curves (with right axis)
show those of the sources visible to BATSE.  The uppermost curves
are based on 64~ms data, the lowermost on 1024~ms data.}

% 14
\figcaption{Joint credible regions for dimensionless luminosity
$\nu_c$ and cosmological density parameter $\Omega_0$ for
homogeneous standard candle cosmological models,
based on  64~ms (a) and 1024~ms (b) data.}

% 15
\figcaption{Joint credible regions for dimensionless luminosity
$\nu_c$ and density function power law index $\beta$ for
standard candle cosmological models with $(1+z)^{-\beta}$
density evolution,
based on  64~ms (a) and 1024~ms (b) data.  Crosses indicate best-fit
points, dots indicate representative points used for redshift
distributions shown in Fig.~18.}

% 16
\figcaption{Cumulative peak flux distributions predicted by
best-fit standard candle cosmological models with density evolution, based
on  64~ms (a) and 1024~ms (b) data.
Histogram shows cumulative distribution of best-fit peak flux values of
detected bursts.}

% 17
\figcaption{Joint credible regions for dimensionless luminosity $\nu_c$
and comoving burst rate density $\dot n_0$ for
standard candle cosmological models with $(1+z)^{-\beta}$
density evolution, based
on  64~ms (upper contours) and 1024~ms (lower contours)
data.  These are conditional on a power law index of
$\beta=-2.5$.}

% 18
\figcaption{Distribution of redshifts of burst sources
as predicted by representative standard candle cosmological models with
density evolution in
the 95.4\% credible regions.  Solid curves (with left axis) show the
redshift distributions of all sources; dashed curves (with right axis)
show those of the sources visible to BATSE.  The uppermost curves
are based on  64~ms data, the lowermost on 1024~ms data.  Models have
parameter values indicated by dots in Fig.~15.}

% 19
\figcaption{Profile likelihood functions as a function of
power-law luminosity function dynamic range, $\rho$, based on
64~ms (solid) and 1024~ms (dashed) data.}

% 20
\figcaption{Joint credible regions for maximum luminosity parameter
$\nu_u$ and power law index $p$ for
cosmological models with bounded power law luminosity functions,
based on  64~ms (a) and 1024~ms (b) data.  Crosses indicate best-fit
points, dots indicate representative points used for redshift
distributions shown in Fig.~23.}

% 21
\figcaption{Cumulative peak flux distributions predicted by
best-fit cosmological models with power law luminosity functions, based
on  64~ms (a) and 1024~ms (b) data.
Histogram shows cumulative distribution of best-fit peak flux values of
detected bursts.}

% 22
\figcaption{Joint credible regions for dimensionless luminosity $\nu_c$
and comoving burst rate density $\dot n_0$ for
cosmological models with power-law luminosity functions,
based on  64~ms (upper contours) and 1024~ms (lower contours)
data.  These are conditional on a power law index of $p=1.9$.}

% 23
\figcaption{As Fig.~18, but for representative cosmological models
with power law luminosity functions indicated by dots in Fig.~20.}

% 24
\figcaption{Profile likelihood functions as a function of
dynamic range, $\rho$,  of a ``top hat'' luminosity function, based on
64~ms (solid) and 1024~ms (dashed) data.}

% 25
\figcaption{Differential burst rate for best-fit homogeneous standard candle 
model (short dashed), and ``top hat'' luminosity function models with
a dynamic range of 2 (long dashed) and $10^4$ (solid); based on 1024~ms
data.}

% 26
\figcaption{Effective luminosity functions for best-fit cosmological
models with bounded power-law intrinsic luminosity functions.  Solid
curve (with left axis) shows luminosity function, dashed curve (right
axis) shows its logarithmic slope. (a) For 64~ms data; intrinsic
luminosity function has power law index $p=2.12$.  (b) For 1024~ms data;
intrinsic luminosity function has power law index $p=1.68$.}

% 27
\figcaption{Effective luminosity functions for 1024~ms data based on
cosmological models with bounded power-law intrinsic luminosity functions.
Shown are the effective luminosity functions for the best-fit model
(solid curve; $p=1.68$) and two other models in the 68.3\% credible
region of Fig.~20b: a low luminosity model with $p=1.0$ and $\nu_u=1$
(dotted curve) and a high luminosity model with $p=2.2$ and $\nu_u=300$.
Dots bound the regions containing the 90\% most probable luminosities;
the 90\% region is bounded on the left by the lower cutoff for the
solid and dashed curves.}

% 28
\figcaption{Differential burst rates for duration-dependent broken power
law model with intrinsic power law index $\gamma_1=1.9$, 
duration-flux power law index $\sigma=0.6$, and fiducial
duration $\tau_0=2$~s, for
64~ms (solid) and 1024~ms (dashed) measuring timescales, illustrating
effect of peak dilution.}

% 29
\figcaption{Cumulative peak flux distributions predicted by
best-fit duration-dependent broken power law models, based
on  64~ms (a) and 1024~ms (b) data.
Histogram shows cumulative distribution of best-fit peak flux values of
detected bursts.}

% 30
\figcaption{OSSE light curves for the brightest bursts in the 1024~ms
catalog used in these analyses.  Insets detail a 1.5~s duration
including the peak.}

% 31
\figcaption{Joint credible regions for $\tau_0$ and $\sigma$
conditional on the best-fit values of $\gamma_1$ in
the duration-dependent broken power law model,
based on 64~ms (a, $\gamma_1=2.1$) and 1024~ms (b, $\gamma_1=1.8$) data.
Only contours bounding the 95.4\% (dashed) and 99.7\% (solid) credible
regions are shown.}

% 32
\figcaption{Joint credible regions for dimensionless rest-frame 
luminosity $\nu_0$ and rest-frame duration $\tau_0$ of standard candle, 
standard clock cosmological models with relativistic beaming,
conditional on a Doppler factor of ${\cal D}_b =4$,
based on 64~ms (a, $\gamma_1=2.1$) and 1024~ms (b, $\gamma_1=1.8$) data.}

% 33
\figcaption{Cumulative peak flux distributions predicted by
best-fit standard candle, standard clock
cosmological models with relativistic beaming, based
on  64~ms (a) and 1024~ms (b) data.
Histogram shows cumulative distribution of best-fit peak flux values of
detected bursts.}

\clearpage

%---------------------- Simple Phen -----------------------------------------
\begin{deluxetable}{lrr}
\tablecolumns{3}
\tablewidth{0pc}
\tablecaption{Simple Phenomenological Models}
\tablehead{
\colhead{Quantity}    &   \colhead{64~ms Results}   &
\colhead{1024~ms Results} } 
\startdata
\cutinhead{$M_1$: Single Power Law} \nl
$\gamma$ & 2.11 & 1.90 \nl
$R_{11}$ & $\equiv 1.0$ & $\equiv 1.0$ \nl
$B_{11}$ & $\equiv 1.0$ & $\equiv 1.0$ \nl
\cutinhead{$M_2$: Broken Power Law, $\gamma_2\equiv 2.5$} \nl
$\gamma_1$ & 2.00 & 1.67 \nl
$\Phi_b$ (cm$^{-2}$~s$^{-1}$) & $1.6\times10^2$ &  12 \nl
$R_{21}$ & 1.3 & 48 \nl
$p(>R_{21})$ & 0.48 & $5\times10^{-3}$ \nl
$B_{21}$ & 0.54 & 20 \nl
\cutinhead{$M_3$: Broken Power Law} \nl
$\gamma_1$ & 2.04 & 1.83 \nl
$\Phi_b$  (cm$^{-2}$~s$^{-1}$) & $1.2\times10^2$ &  43 \nl
$\gamma_2$ & 4.3 & 15 \nl
$R_{31}$ & 3.6 & $1.7\times10^3$ \nl
$p(>R_{31})$ &  0.28 & $6\times10^{-4}$ \nl
$B_{31}$ & $4.7$ & 19 \nl
\enddata
\end{deluxetable}

%--------------------- Simple Cosmo ---------------------------------------
\begin{deluxetable}{lrr}
\tablecolumns{3}
\tablewidth{0pc}
\tablecaption{Simple Cosmological Models With $\Omega_0=1$}
\tablehead{
\colhead{Quantity}    &   \colhead{64~ms Results}   &
\colhead{1024~ms Results} } 
\startdata
\cutinhead{$M_4$: Homogeneous Standard Candles} \nl
$\nu_c$ & 0.37 & 0.44 \nl
$\dot n_0$ (yr$^{-1}$~Gpc$^{-3}$) & 53 & 24 \nl
$R_{41}$ & 0.22 & 7.2 \nl
$p(>R_{41})$ & \nodata & \nodata \nl
$B_{41}$ & 0.24 & 5.9 \nl
\cutinhead{$M_5$: Inhomogeneous Standard Candles} \nl
$\nu_c$ & $3.3\times10^2$ & $1.0\times10^1$ \nl
$\beta$ & -3.0 & -2.1 \nl
$\dot n_0$ (yr$^{-1}$~Gpc$^{-3}$) & $5.8\times10^{-3}$ & 0.40 \nl
$R_{54}$ & 5.6 & 3.7 \nl
$p(>R_{54})$ & $6.4\times10^{-2}$ & 0.11 \nl
$B_{54}$ & 1.3 & 1.0 \nl
\cutinhead{$M_6$: Power Law Luminosity Function} \nl
$\nu_u$ & 25.1 & 4.84 \nl
$p$ & 2.12 & 1.68 \nl
$\rho$ & $\equiv 10^4$ & $\equiv 10^4$ \nl
$\dot n_0$ (yr$^{-1}$~Gpc$^{-3}$) & $2.0\times10^3$ & $7.2\times 10^2$ \nl
$R_{64}$ & 5.5 & 4.4 \nl
$p(>R_{64})$ & $6.5\times10^{-2}$ & $3.0\times10^{-3}$ \nl
$B_{64}$ & 1.3 & 0.92 \nl
%   $\nu_u$ & 56 & 5.5 \nl
%   $p$ & 2.1 & 1.7 \nl
%   $\rho$ & $\equiv 10^5$ & $\equiv 10^5$ \nl
%   $R_{64}$ & 7.2 & 4.3 \nl
%   $p(>R_{64})$ & $4.7\times10^{-2}$ & $8.6\times10^{-2}$ \nl
%   $B_{64}$ & 1.4 & 1.1 \nl
\enddata
\end{deluxetable}

%------------------- Duration-Dependent Models -------------------------

\begin{deluxetable}{lrr}
\tablecolumns{3}
\tablewidth{0pc}
\tablecaption{Duration-Dependent Models}
\tablehead{
\colhead{Quantity}    &   \colhead{64~ms Results}   &
\colhead{1024~ms Results} } 
\startdata
\cutinhead{$M_7$: Broken Power Law/Top Hat} \nl
$\gamma_1$ & 2.1 & 1.8 \nl
$\sigma$ & 0.36 & 0.996 \nl
$\tau_0$ (s) & 0.16 & 40 \nl
$R_{71}$ & 18 & $6.0\times10^3$ \nl
$p(>R_{71})$ & $5.7\times10^{-2}$ & $1.7\times10^{-4}$ \nl
$B_{71}$ & $\approx 0.3$ & $\approx 5$ \nl
\cutinhead{$M_8$: Cosmological Beamed Sources} \nl
$\nu_0$ & $8.7\times10^{-3}$ & 0.60 \nl
$\tau_0$ (s) & 0.73 & 0.18 \nl
${\cal D}_b $ & $\equiv 4$ & $\equiv 4$ \nl
$R_{84}$ & 1.8 & 1.2 \nl
$p(>R_{84})$ & 0.27 & 0.52 \nl
$B_{84}$ & 1.1 & 1.3 \nl
\enddata
\end{deluxetable}

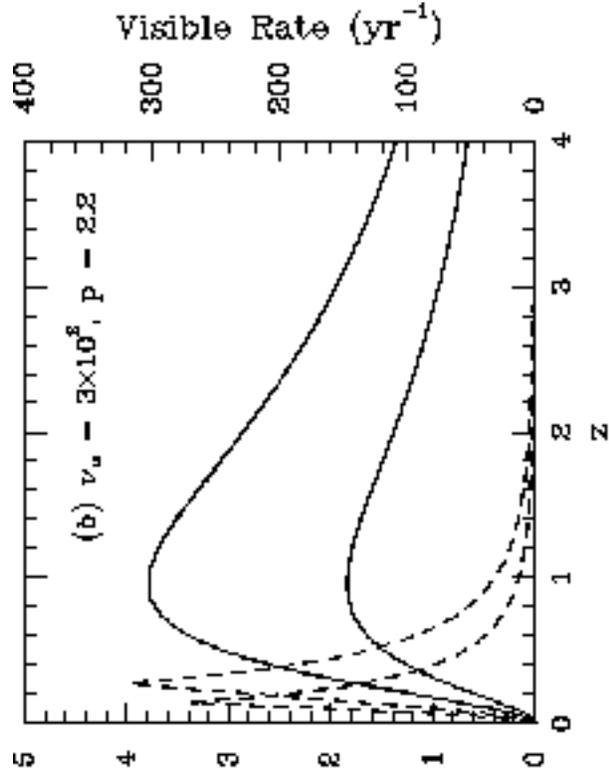
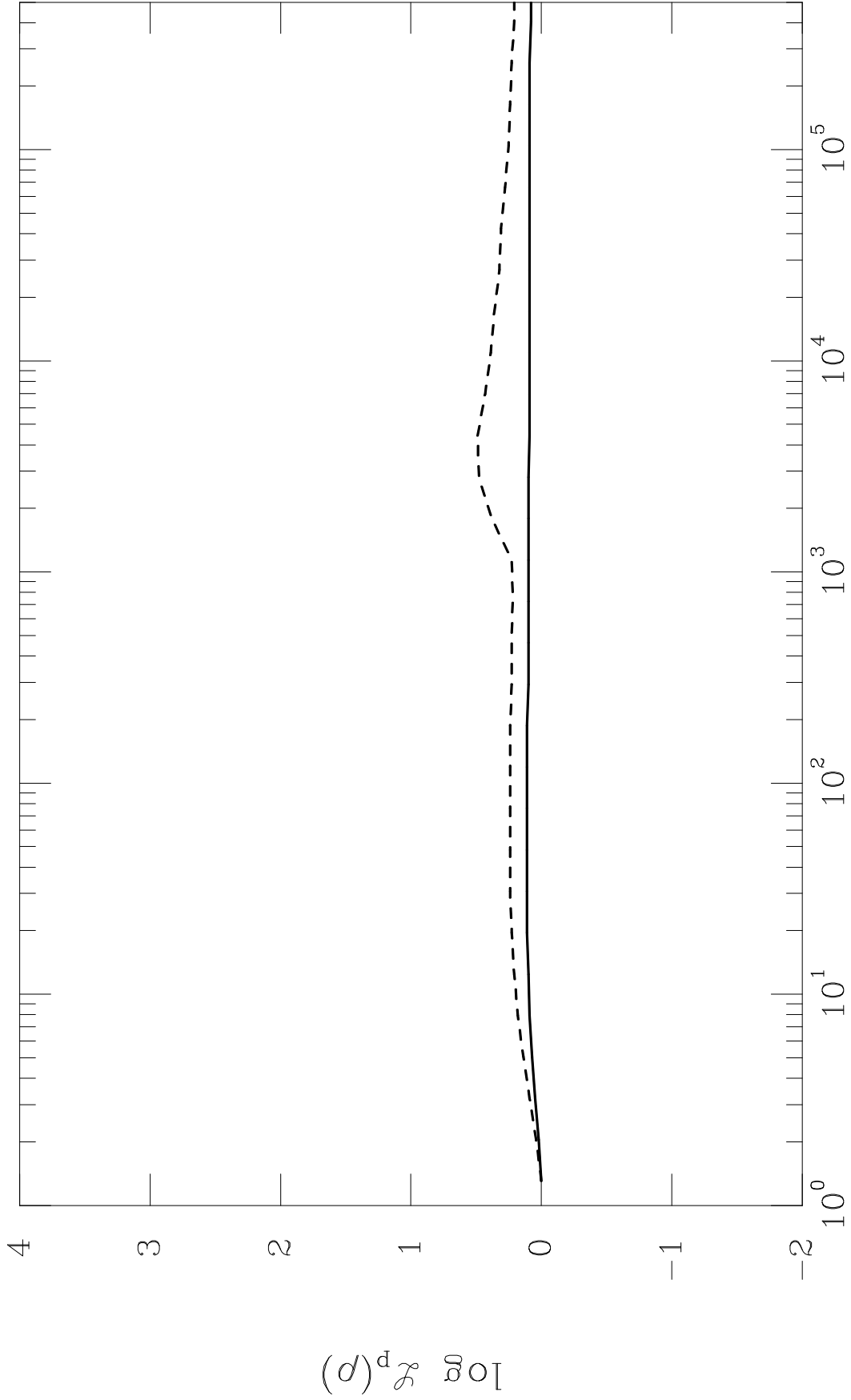
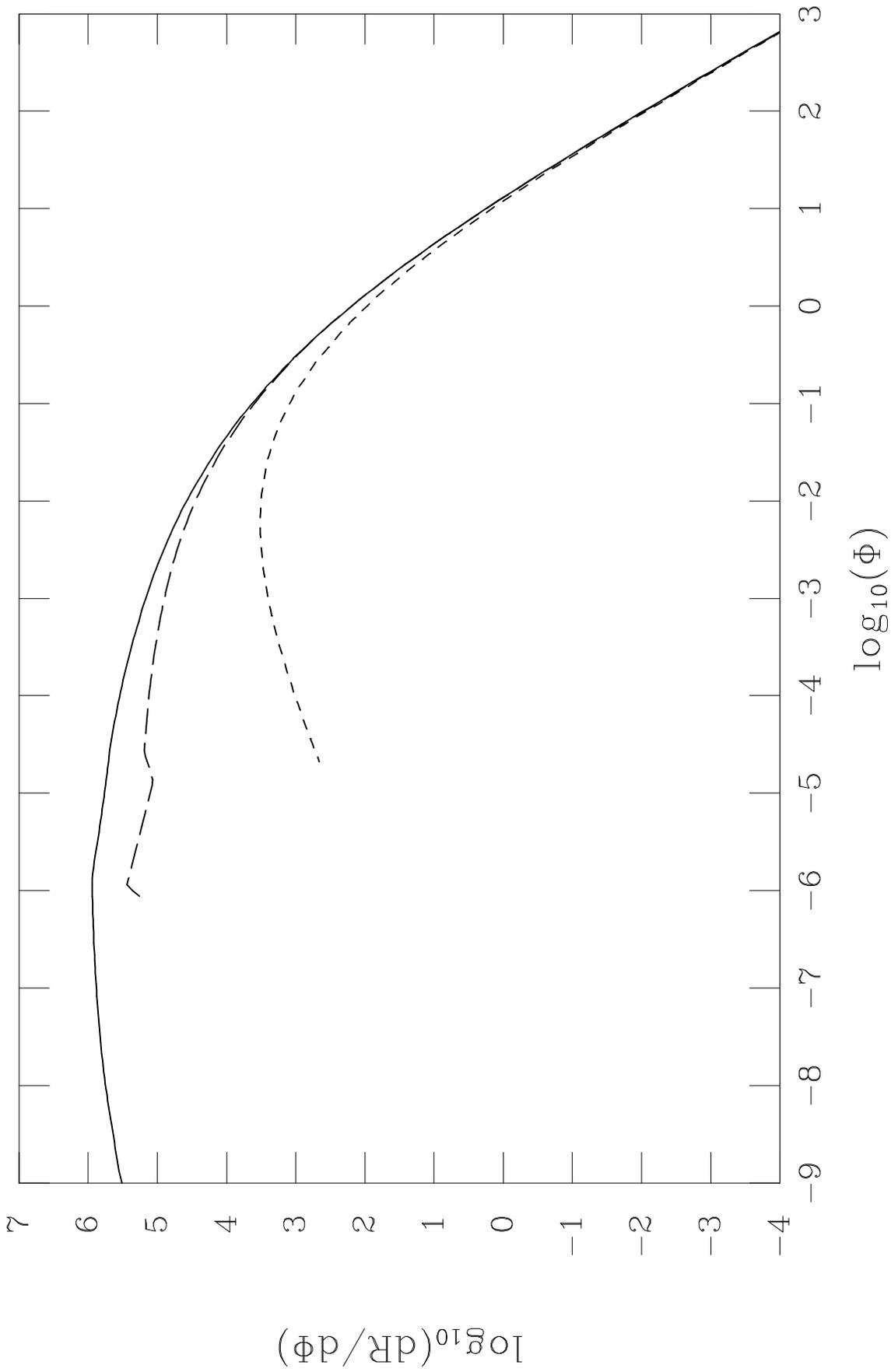
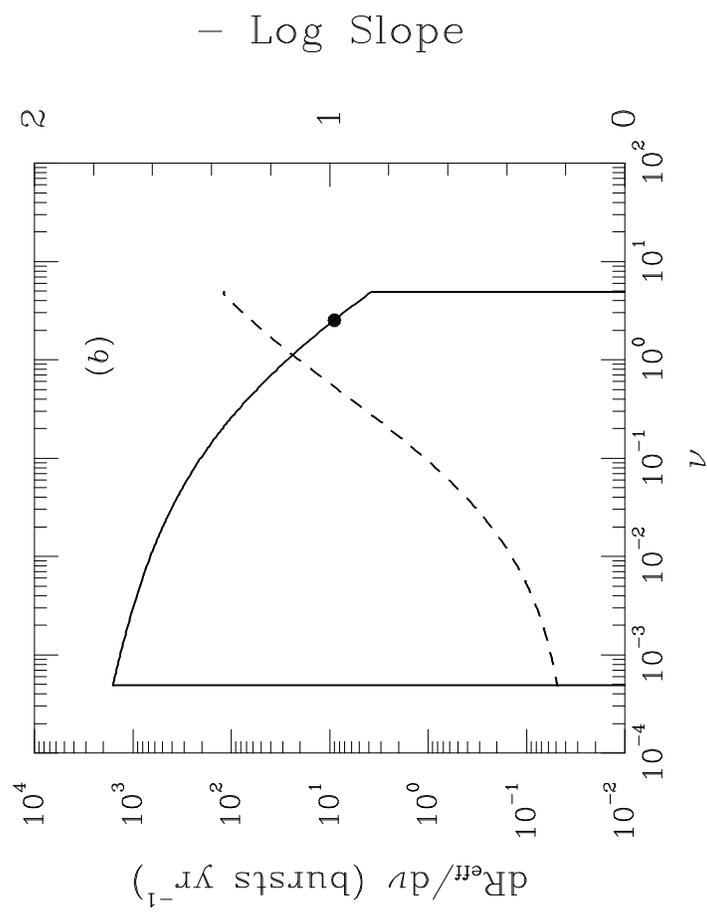
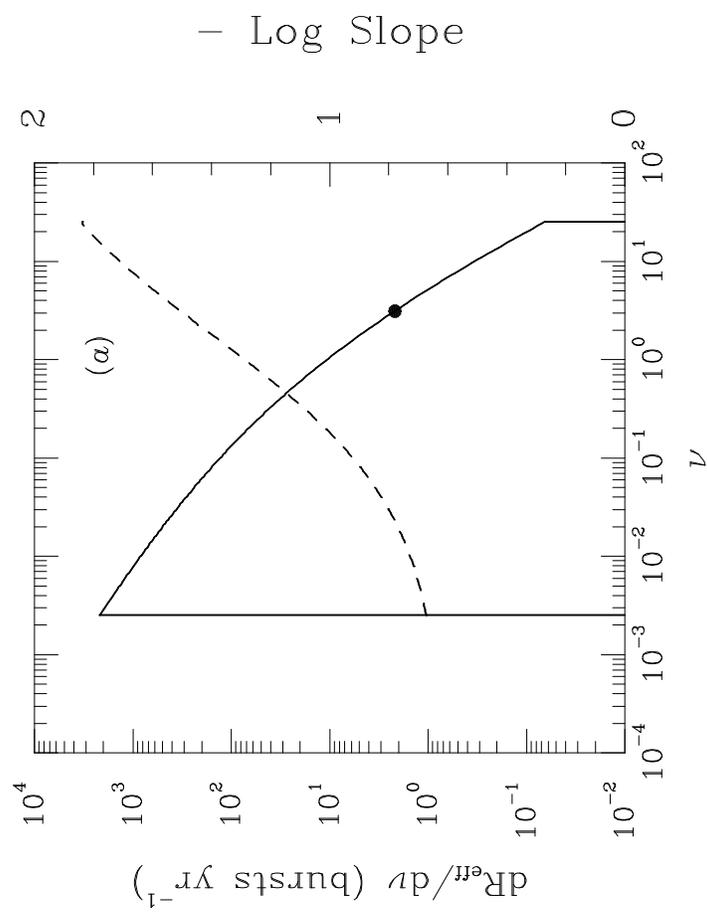
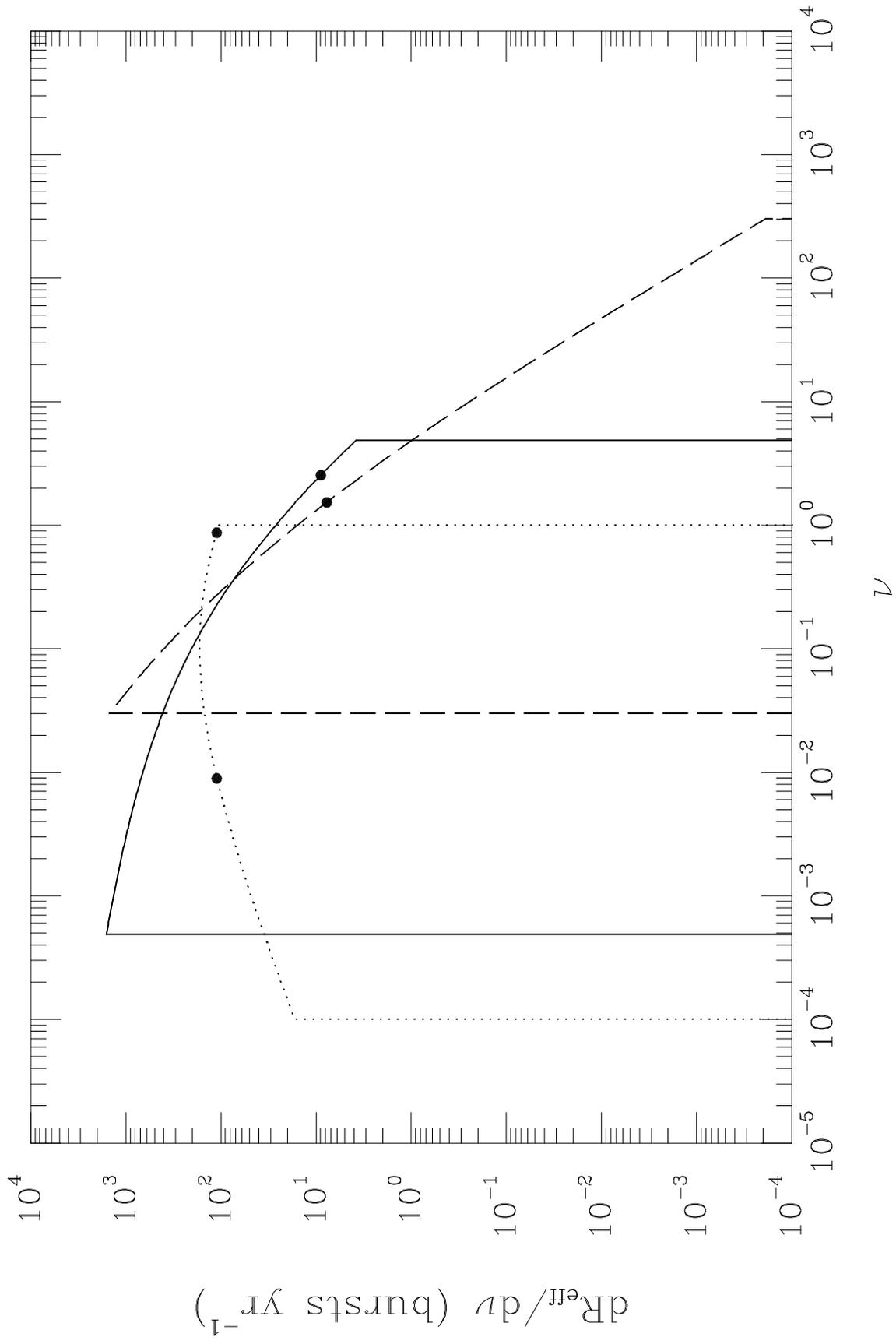
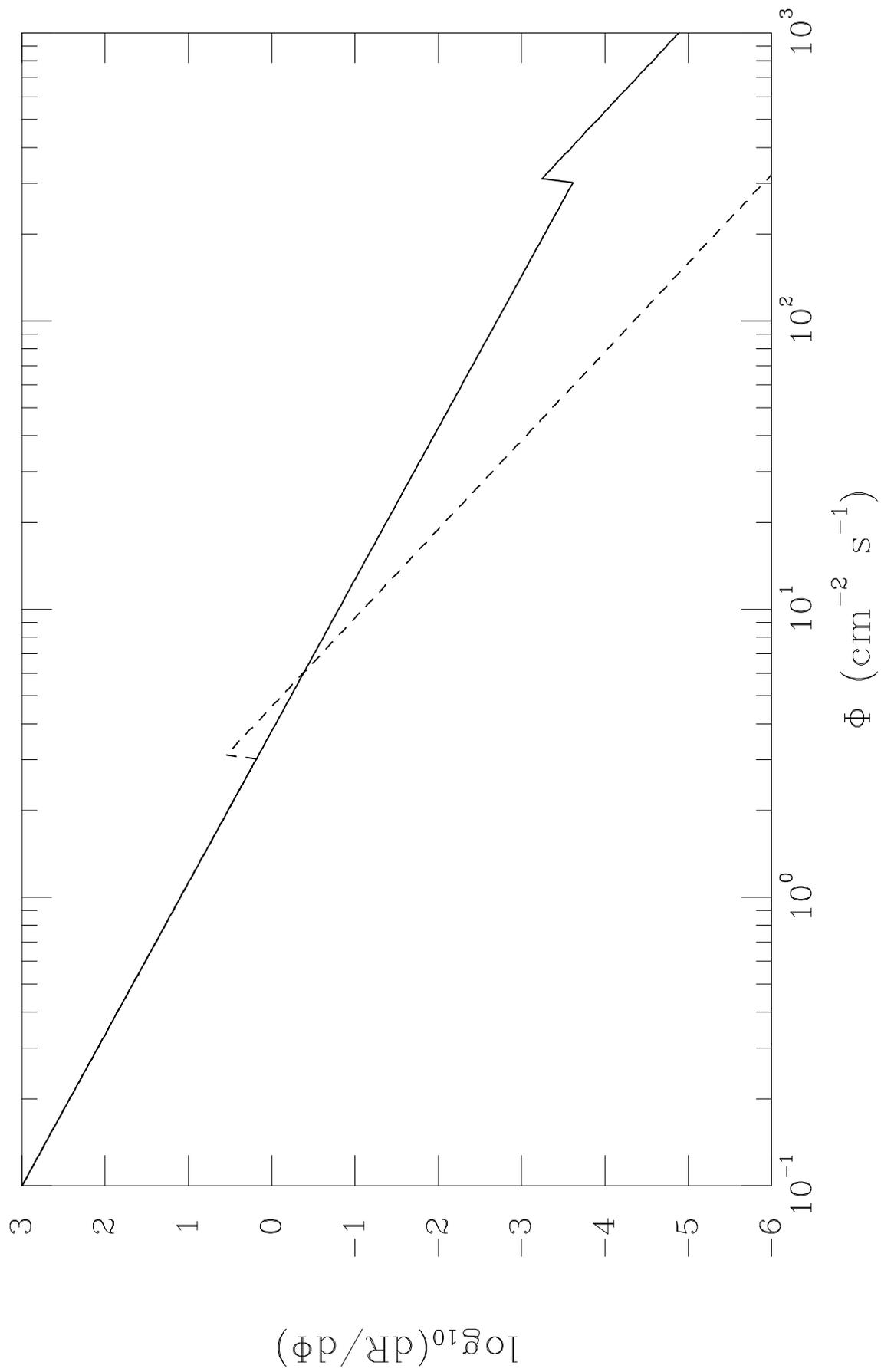
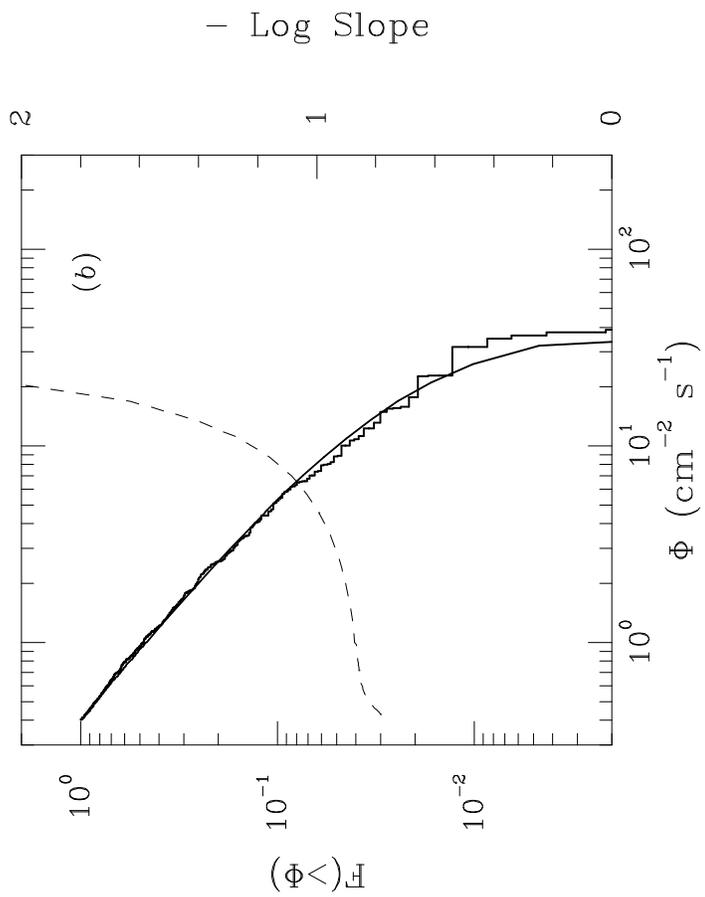
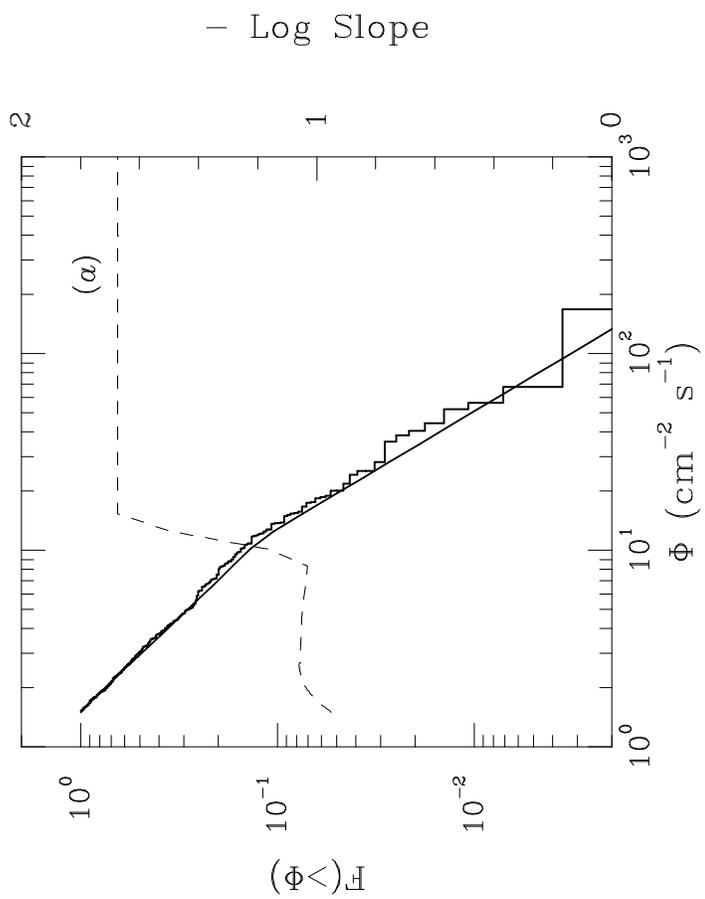
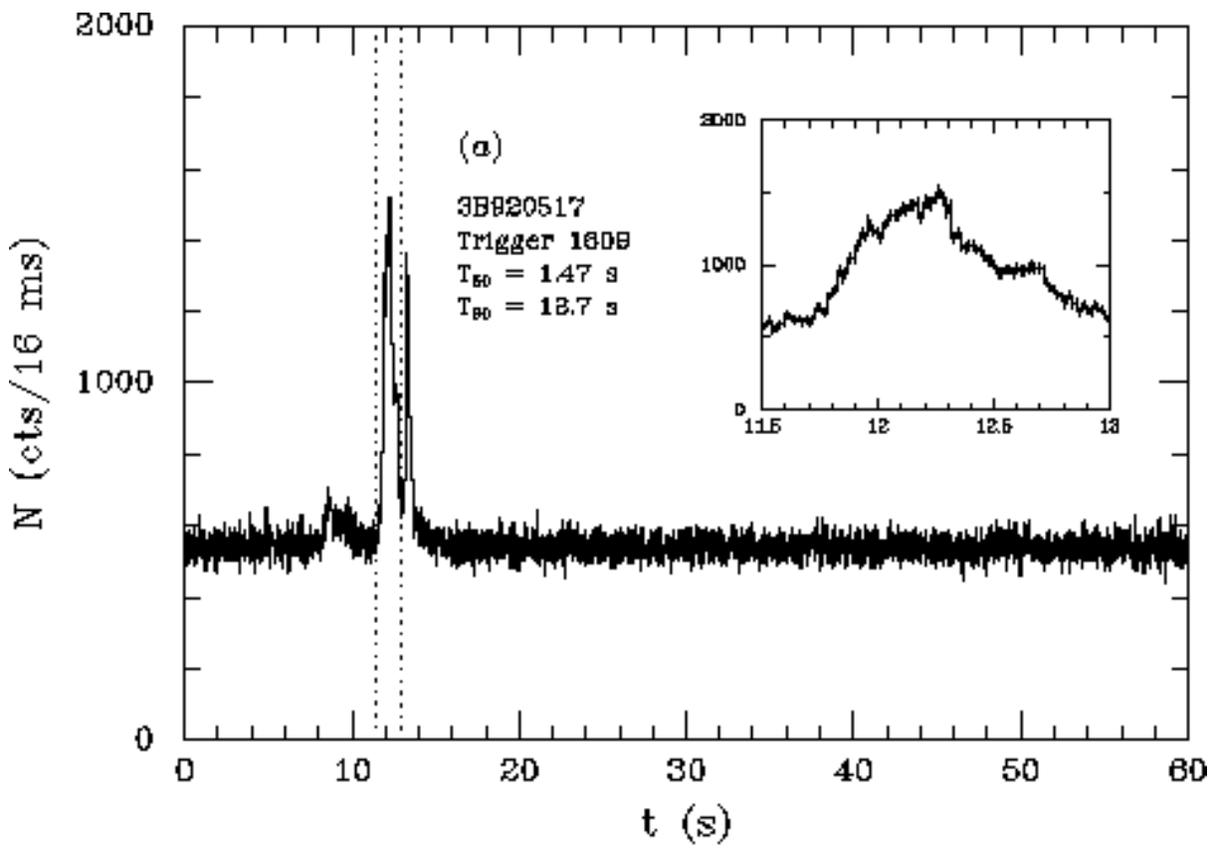
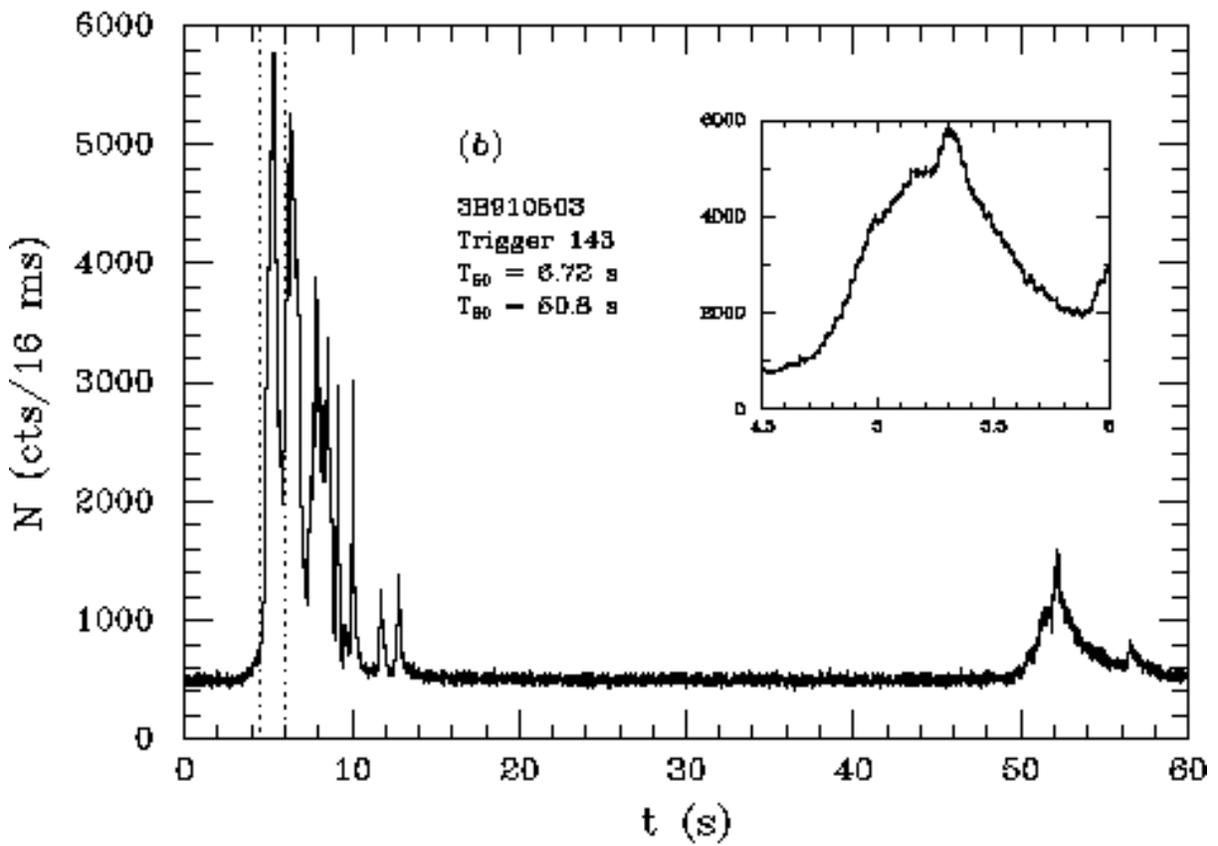
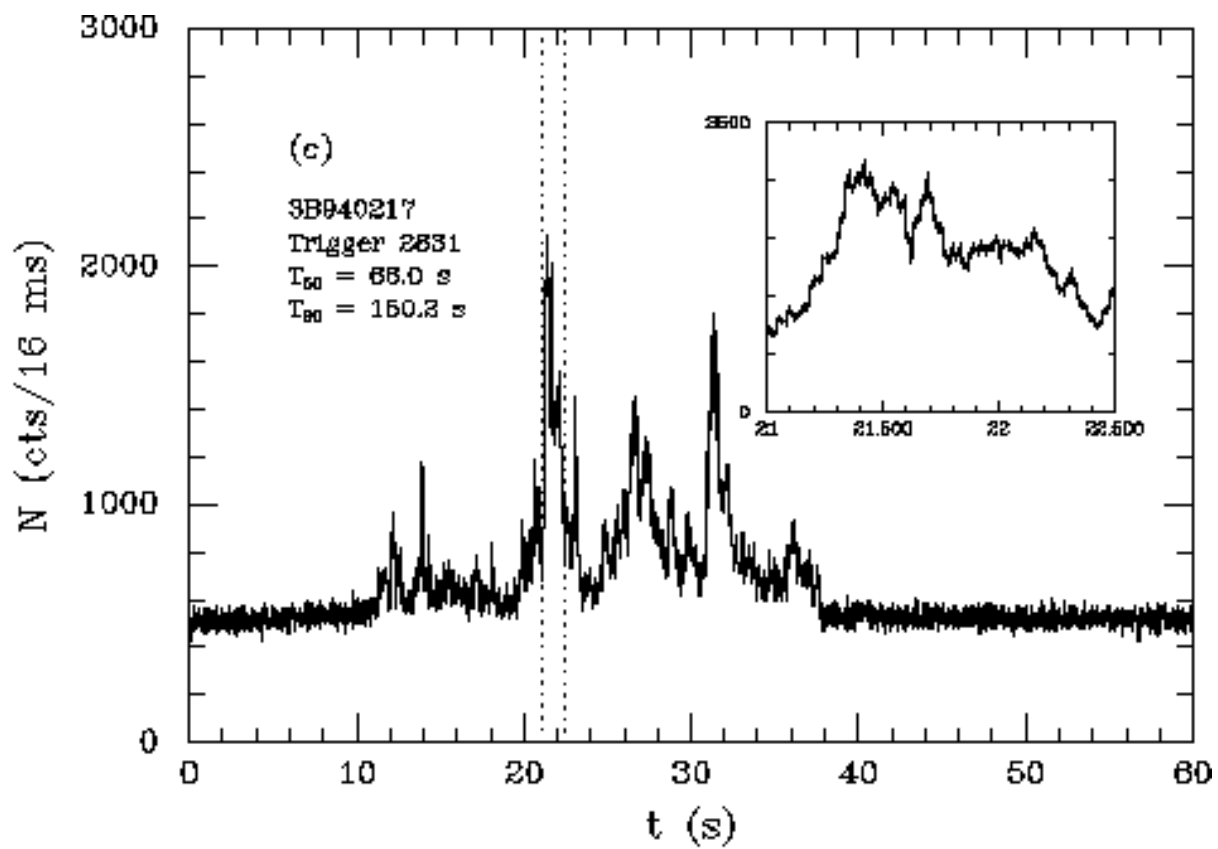
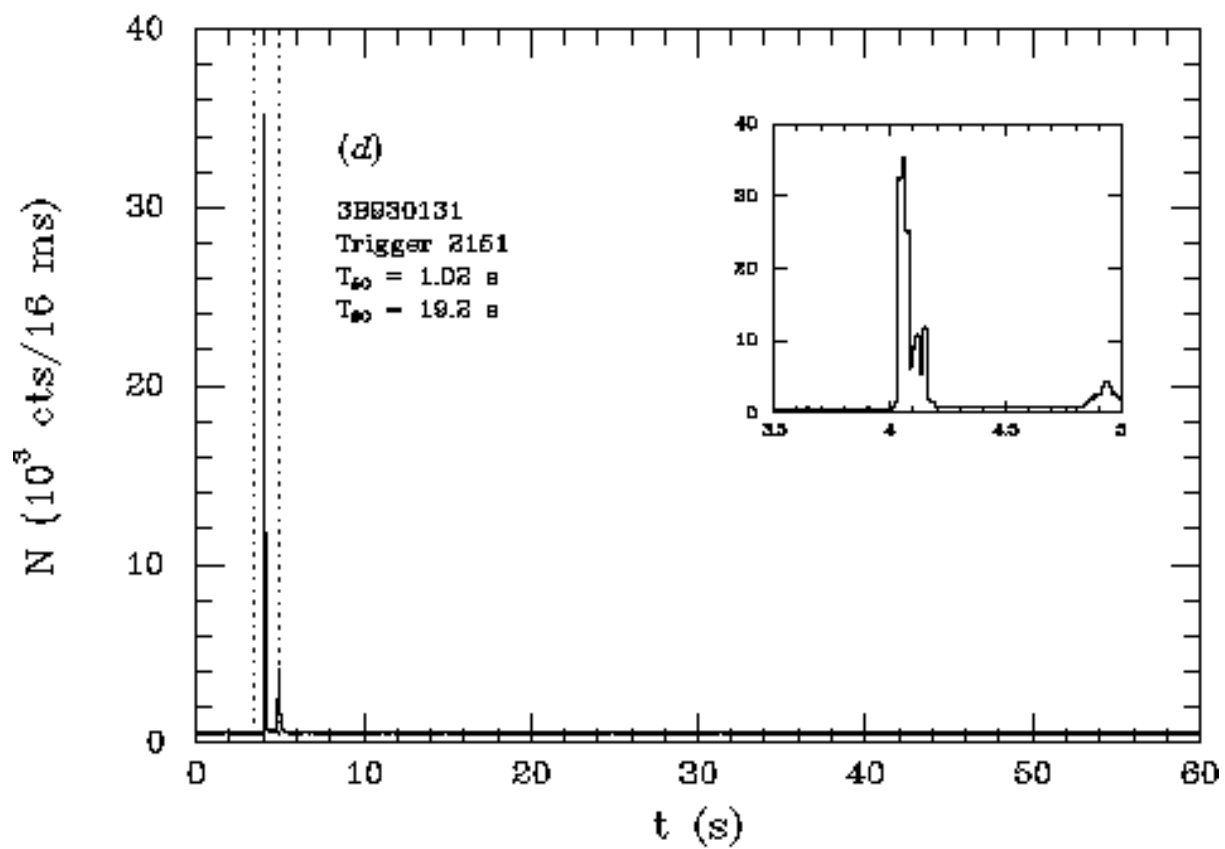
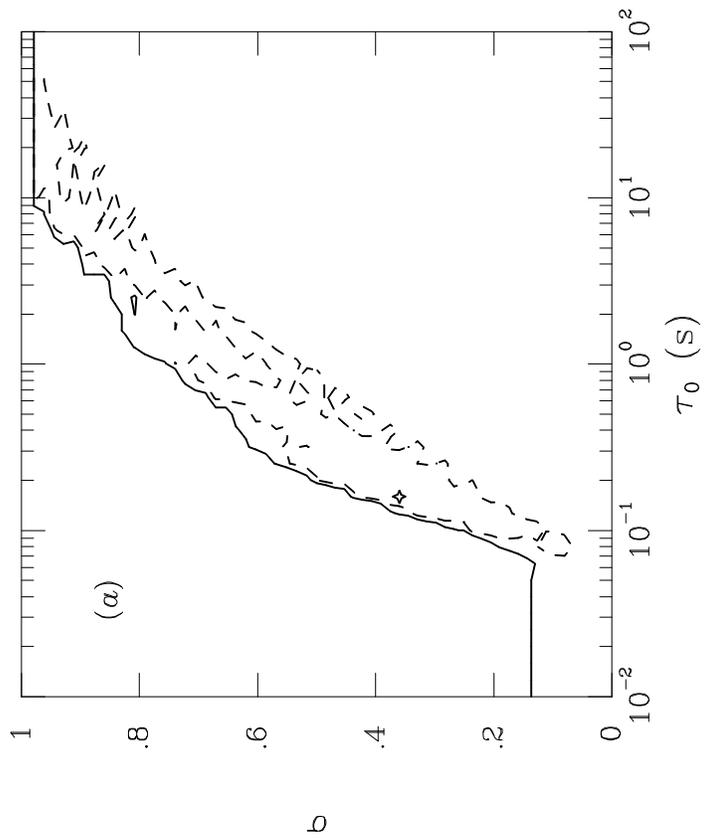
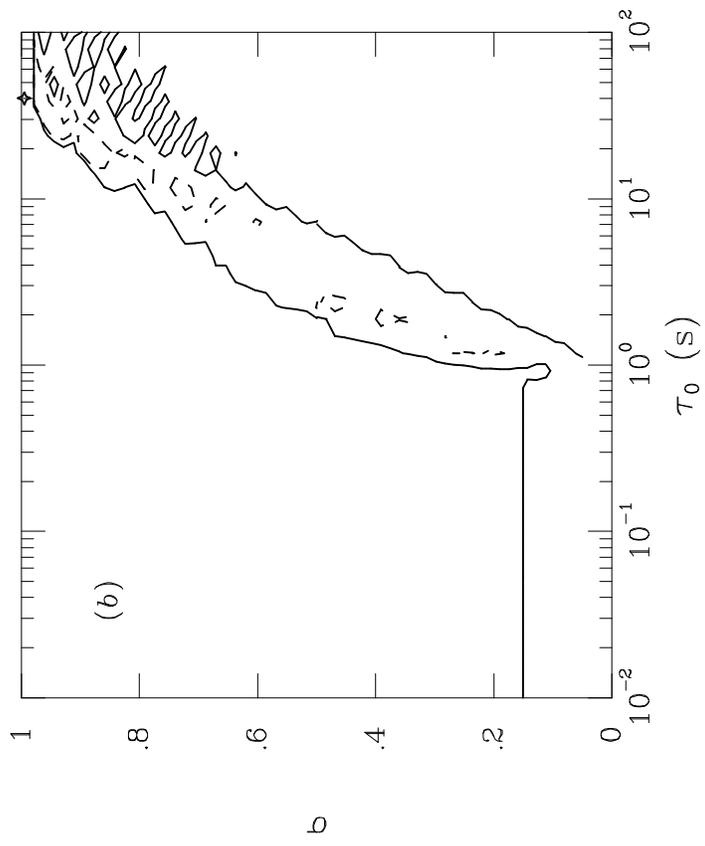
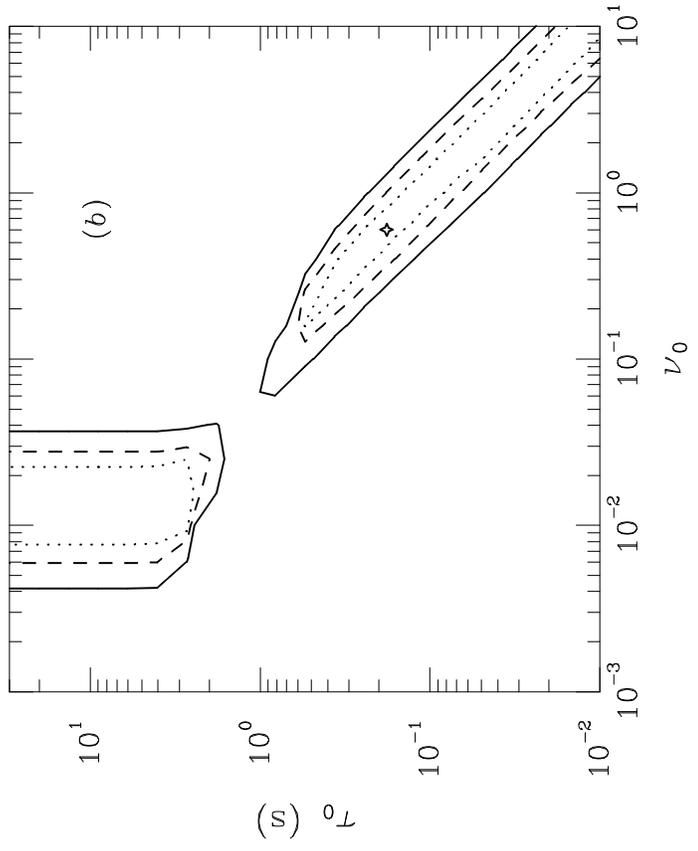
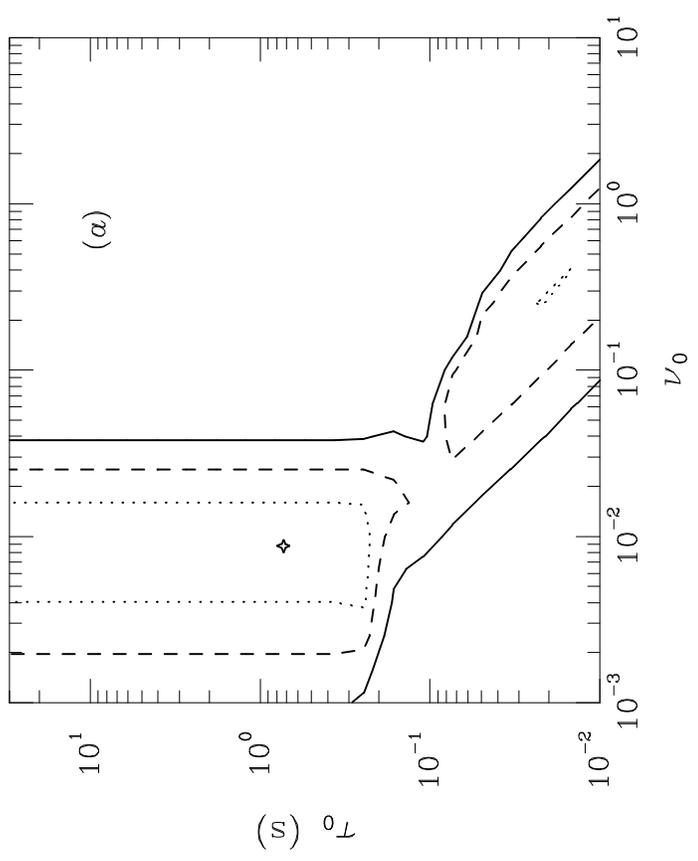
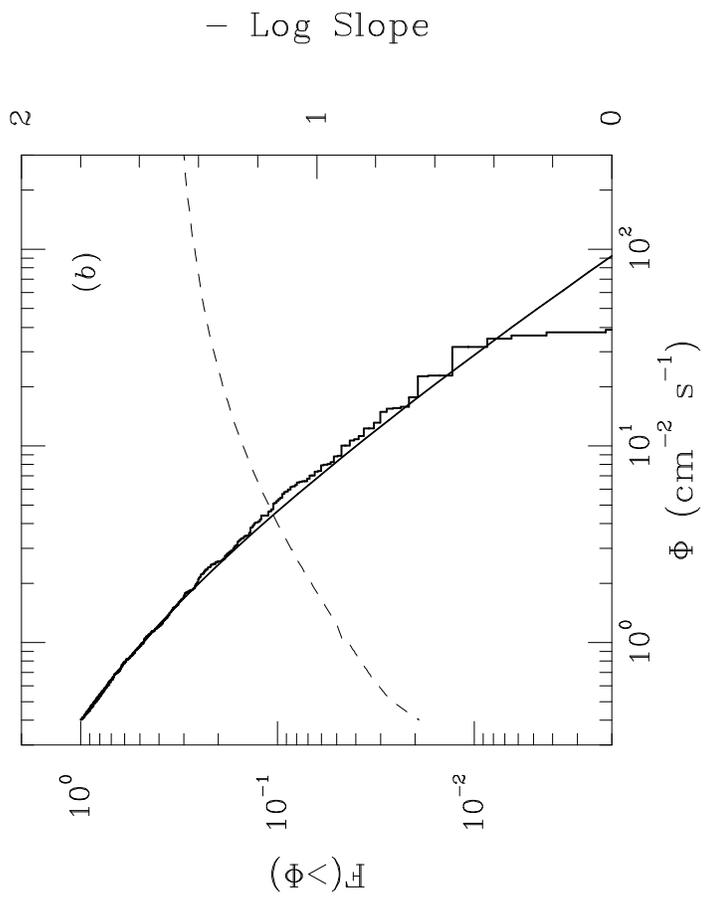
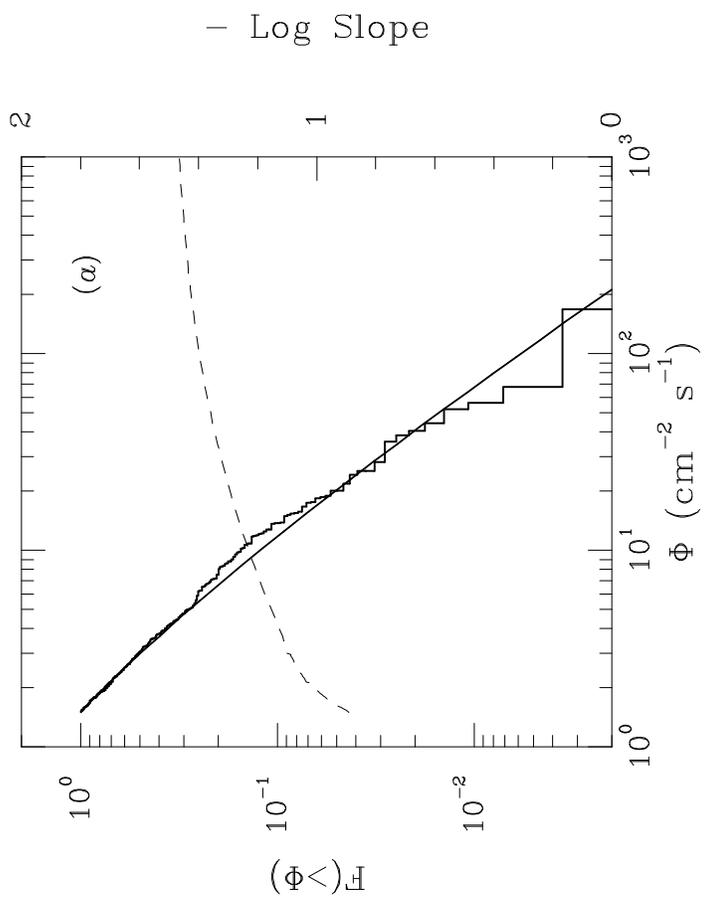
\end{document}